\documentclass[12pt]{article}
\usepackage{epsfig}
\usepackage{amsmath}
\usepackage{hhline}
\usepackage{amssymb}
\usepackage{times}

\newlength{\dinwidth}
\newlength{\dinmargin}
\begin{document}
% The rest
\newcommand{\pom}{{I\!\!P}}
\newcommand{\reg}{{I\!\!R}}
\newcommand{\slowpi}{\pi_{\mathit{slow}}}
\newcommand{\fiidiii}{F_2^{D(3)}}
\newcommand{\fiidiiiarg}{\fiidiii\,(\beta,\,Q^2,\,x)}
\newcommand{\n}{1.19\pm 0.06 (stat.) \pm0.07 (syst.)}
\newcommand{\nz}{1.30\pm 0.08 (stat.)^{+0.08}_{-0.14} (syst.)}
\newcommand{\fiidiiiful}{F_2^{D(4)}\,(\beta,\,Q^2,\,x,\,t)}
\newcommand{\fiipom}{\tilde F_2^D}
\newcommand{\ALPHA}{1.10\pm0.03 (stat.) \pm0.04 (syst.)}
\newcommand{\ALPHAZ}{1.15\pm0.04 (stat.)^{+0.04}_{-0.07} (syst.)}
\newcommand{\fiipomarg}{\fiipom\,(\beta,\,Q^2)}
\newcommand{\pomflux}{f_{\pom / p}}
\newcommand{\nxpom}{1.19\pm 0.06 (stat.) \pm0.07 (syst.)}
\newcommand {\gapprox}
   {\raisebox{-0.7ex}{$\stackrel {\textstyle>}{\sim}$}}
\newcommand {\lapprox}
   {\raisebox{-0.7ex}{$\stackrel {\textstyle<}{\sim}$}}
\def\gsim{\,\lower.25ex\hbox{$\scriptstyle\sim$}\kern-1.30ex%
\raise 0.55ex\hbox{$\scriptstyle >$}\,}
\def\lsim{\,\lower.25ex\hbox{$\scriptstyle\sim$}\kern-1.30ex%
\raise 0.55ex\hbox{$\scriptstyle <$}\,}
\newcommand{\pomfluxarg}{f_{\pom / p}\,(x_\pom)}
\newcommand{\dsf}{\mbox{$F_2^{D(3)}$}}
\newcommand{\dsfva}{\mbox{$F_2^{D(3)}(\beta,Q^2,x_{I\!\!P})$}}
\newcommand{\dsfvb}{\mbox{$F_2^{D(3)}(\beta,Q^2,x)$}}
\newcommand{\dsfpom}{$F_2^{I\!\!P}$}
\newcommand{\gap}{\stackrel{>}{\sim}}
\newcommand{\lap}{\stackrel{<}{\sim}}
\newcommand{\fem}{$F_2^{em}$}
\newcommand{\tsnmp}{$\tilde{\sigma}_{NC}(e^{\mp})$}
\newcommand{\tsnm}{$\tilde{\sigma}_{NC}(e^-)$}
\newcommand{\tsnp}{$\tilde{\sigma}_{NC}(e^+)$}
\newcommand{\st}{$\star$}
\newcommand{\sst}{$\star \star$}
\newcommand{\ssst}{$\star \star \star$}
\newcommand{\sssst}{$\star \star \star \star$}
\newcommand{\tw}{\theta_W}
\newcommand{\sw}{\sin{\theta_W}}
\newcommand{\cw}{\cos{\theta_W}}
\newcommand{\sww}{\sin^2{\theta_W}}
\newcommand{\cww}{\cos^2{\theta_W}}
\newcommand{\trm}{m_{\perp}}
\newcommand{\trp}{p_{\perp}}
\newcommand{\trmm}{m_{\perp}^2}
\newcommand{\trpp}{p_{\perp}^2}
\newcommand{\alp}{\alpha_s}

\newcommand{\alps}{\alpha_s}
\newcommand{\sqrts}{$\sqrt{s}$}
\newcommand{\LO}{$O(\alpha_s^0)$}
\newcommand{\Oa}{$O(\alpha_s)$}
\newcommand{\Oaa}{$O(\alpha_s^2)$}
\newcommand{\PT}{p_{\perp}}
\newcommand{\JPSI}{J/\psi}
\newcommand{\sh}{\hat{s}}
\newcommand{\uh}{\hat{u}}
\newcommand{\MP}{m_{J/\psi}}
\newcommand{\PO}{I\!\!P}
\newcommand{\xbj}{x}
\newcommand{\xpom}{x_{\PO}}
\newcommand{\ttbs}{\char'134}
\newcommand{\xpomlo}{3\times10^{-4}}  
\newcommand{\xpomup}{0.05}  
\newcommand{\dgr}{^\circ}
\newcommand{\pbarnt}{\,\mbox{{\rm pb$^{-1}$}}}
\newcommand{\gev}{\,\mbox{GeV}}
\newcommand{\WBoson}{\mbox{$W$}}
\newcommand{\fbarn}{\,\mbox{{\rm fb}}}
\newcommand{\fbarnt}{\,\mbox{{\rm fb$^{-1}$}}}

\newcommand{\vtab}{\rule[-1mm]{0mm}{4mm}}
\newcommand{\htab}{\rule[-1mm]{0mm}{6mm}}
\newcommand{\photoproduction}{$\gamma p$}
\newcommand{\ptmiss}{$P_{T}^{\rm miss}$}
\newcommand{\epz} {$E{\rm-}p_z$}
\newcommand{\vap} {  $V_{ap}/V_p$}
\newcommand{\Zero}   {{Z^{0}}}
\newcommand{\Ftwo}   {{\tilde{F}_2}}
\newcommand{\Ftwoz}   {{\tilde{F}_{2,3}}}
\newcommand{\Fz}   {{\tilde{F}_3}}
\newcommand{\FL}   {{\tilde{F}_{_{L}}}}
\newcommand{\wtwogen} {\tilde{W_2}}
\newcommand{\wlgen} {\tilde{W_L}}
\newcommand{\xwthreegen} {x\tilde{W_3}}
\newcommand{\Wtwo}   {{$W_2$}}
\newcommand{\Wz}   {{$W_3$}}
\newcommand{\WL}   {{$W_{_{L}}$}}
\newcommand{\Fem}  {{F_2}}
\newcommand{\Fgam}  {{F_2^{\gamma}}}
\newcommand{\Fint} {{F_2^{\gamma Z}}}
\newcommand{\Fwk}  {{F_2^{Z}}}
\newcommand{\Ftwos} {{F_2^{\gamma Z, Z}}}
\newcommand{\Fzz} {{F_3^{\gamma Z, Z}}}
\newcommand{\Fintz} {{F_{2,3}^{\gamma Z}}}
\newcommand{\Fwkz}  {{F_{2,3}^{Z}}}
\newcommand{\Fzint} {{F_3^{\gamma Z}}}
\newcommand{\Fzwk}  {{F_3^{Z}}}
\newcommand{\Gev}  {{{\rm GeV}}}
\newcommand{\Gevv}{{{\rm GeV}^2}}
\newcommand{\QQ}  {{{Q^2}}}
%\newcommand{\lapprox}{\stackrel{<}{_{\sim}}}
%\newcommand{\gapprox}{\stackrel{>}{_{\sim}}}

%
% Some useful tex commands
%
\newcommand{\qsq}{\ensuremath{Q^2} }
\newcommand{\gevsq}{\ensuremath{\mathrm{GeV}^2} }
\newcommand{\et}{\ensuremath{E_t^*} }
\newcommand{\rap}{\ensuremath{\eta^*} }
\newcommand{\gp}{\ensuremath{\gamma^*}p }
\newcommand{\dsiget}{\ensuremath{{\rm d}\sigma_{ep}/{\rm d}E_t^*} }
\newcommand{\dsigrap}{\ensuremath{{\rm d}\sigma_{ep}/{\rm d}\eta^*} }

\pagestyle{empty}
\begin{titlepage}

\noindent
DESY 00-187  \hspace*{8.5cm} ISSN 0418-9833 \\
December 2000

\vspace*{3cm}

\begin{center}
  \Large
{\bf 
Measurement of Neutral and Charged Current \\
 Cross Sections in Electron-Proton Collisions \\
at High {\boldmath{$Q^2$}} }

\vspace*{1cm}
    {\Large H1 Collaboration} 
\end{center}

\begin{abstract}
\noindent
The inclusive $e^- p$~single and double differential cross sections
for neutral and charged current processes are measured with the H1
detector at HERA, in the range of four-momentum transfer squared $Q^2$
between $150$ and $30\,000$ GeV$^2$, and Bjorken $x$ between $0.002$
and $0.65$. The data were taken in 1998 and 1999 with a centre-of-mass
energy of 320 GeV and correspond to an integrated luminosity of
16.4~${\rm pb}^{-1}$. The data are compared with recent measurements
of the inclusive neutral and charged current $e^+ p$ cross sections.
For $Q^2>1\,000$~${\rm GeV}^2$ clear evidence is observed for an
asymmetry between $e^+p$ and $e^-p$ neutral current scattering and the
generalised structure function $x\Fz$ is extracted for the first time
at HERA.  A fit to the charged current data is used to extract a value
for the $W$ boson propagator mass. The data are found to be in good
agreement with Standard Model predictions.
\end{abstract}
\vfill

\begin{center}
 {To be submitted to {\em Eur. Phys. J. C}}
\end{center}
\end{titlepage}

\pagestyle{plain}

%-- H1AUTS Author list by names 
%-- Status: Wed Sep 13 16:30:46 MET DST 2000  Number of authors = 338 

C.~Adloff$^{33}$,              %WUPP-ST        01/96           Adloff              
V.~Andreev$^{24}$,             %LPI -PD        8/88            Andreev             
B.~Andrieu$^{27}$,             %ECPL-PD        8/88            Andrieu             
T.~Anthonis$^{4}$,             %ANTW-ST        11/99           Anthonis            
V.~Arkadov$^{35}$,             %ZEUT-ST        10/96           Arkadov             
A.~Astvatsatourov$^{35}$,      %ZEUT-ST        02/98           Astvatsatourov      
I.~Ayyaz$^{28}$,               %PARI-LEFT      12/99           Ayyaz               
A.~Babaev$^{23}$,              %ITEP-PD        8/88            Babaev              
J.~B\"ahr$^{35}$,              %ZEUT-PD        8/88            Baehr               
P.~Baranov$^{24}$
% ...

%-- H1 Institutes 
\bigskip{\it
% ...
$ ^{2}$ III. Physikalisches Institut der RWTH, Aachen, Germany$^{ a}$ \\
% ...
$ ^{4}$ Inter-University Institute for High Energies ULB-VUB, Brussels;
          Universitaire Instelling Antwerpen, Wilrijk; Belgium$^{ c}$ \\
% ...
$ ^{23}$ Institute for Theoretical and Experimental Physics,
          Moscow, Russia \\
 $ ^{24}$ Lebedev Physical Institute, Moscow, Russia$^{ e,h}$ \\
% ...
$ ^{27}$ LPNHE, Ecole Polytechnique, IN2P3-CNRS, Palaiseau, France \\
 $ ^{28}$ LPNHE, Universit\'{e}s Paris VI and VII, IN2P3-CNRS,
          Paris, France \\
% ...
$ ^{33}$ Fachbereich Physik, Bergische Universit\"at Gesamthochschule
          Wuppertal, Wuppertal, Germany$^{ a}$ \\
% ...
$ ^{35}$ DESY, Zeuthen, Germany$^{ a}$ \\
% ...
$ ^a$ Supported by the Bundesministerium f\"ur Bildung, Wissenschaft,
      Forschung und Technologie, FRG,
      under contract numbers 7AC17P, 7AC47P, 7DO55P, 7HH17I, 7HH27P,
      7HD17P, 7HD27P, 7KI17I, 6MP17I and 7WT87P \\
% ...
$ ^c$ Supported by FNRS-NFWO, IISN-IIKW \\
% ...
$ ^e$ Supported by the Deutsche Forschungsgemeinschaft \\
% ...
$ ^h$ Supported by Russian Foundation for Basic Researc
      grant no. 96-02-00019 \\
% ...
}
\newpage

%%%%%%%%%%%%%%%%%%%%%%%%%%%%%%%%%%%%%%%%%%%%%%%%%%%%%%%
\section{Introduction}
%%%%%%%%%%%%%%%%%%%%%%%%%%%%%%%%%%%%%%%%%%%%%%%%%%%%%%%

\noindent
Inclusive Deep Inelastic Scattering (DIS) has long
been used as a sensitive probe of proton structure and Quantum
Chromodynamics (QCD).  Since 1992 the experiments H1 and ZEUS have
used the colliding lepton--proton beams of the HERA accelerator to
further extend the phase space of such measurements into new kinematic
regions of large four-momentum transfer squared $Q^2$ and small $x$,
where $x$ is the Bjorken scaling variable. The large integrated
luminosity collected by the experiments has allowed measurements to be
made in the very high $Q^2$ range up to $30\,000$ GeV$^2$. In the
region where $Q^2 \simeq M_Z^2$ or $M_W^2$, the $Z^0$ and $W^{\pm}$
boson masses squared, the effects of the electroweak sector of the
Standard Model can be tested in DIS. In addition any deviation from
the prediction observed at the highest $Q^2$, where the smallest
distance scales of proton structure are probed may indicate new
physics, beyond the Standard Model.

Both contributions to DIS, neutral current (NC) interactions
\mbox{$ep \rightarrow eX$} and charged current (CC) interactions
\mbox{$ep \rightarrow \nu X$}, can be measured at HERA and give
complementary information on the QCD and electroweak parts of the
Standard Model. The cross sections are defined in terms of the
three kinematic variables $Q^2$, $x$, and $y$, where $y$ quantifies the
inelasticity of the interaction. The kinematic variables are related
via $Q^2=sxy$, where $\sqrt{s}$ is the $ep$ centre-of-mass energy.

In this paper we report on NC and CC cross section measurements using
$e^-p$ data taken during 1998 and 1999 with a proton beam energy of
$E_p=920$ GeV and an electron beam energy of $E_e=27.6$ GeV, leading
to a centre-of-mass energy of $\sqrt{s} \approx 320$ GeV.  The
integrated luminosity of this sample is $16.4 \ {\rm pb}^{-1}$ which
represents an increase in integrated luminosity by a factor of
approximately 20 compared with previous HERA measurements of $e^-p$ cross
sections~\cite{h1zeuseminus9394}.

The results are compared with NC and CC measurements of $e^+p$ scattering
from H1~\cite{h1hiq2} and ZEUS~\cite{zeushiq2}
taken at a lower centre-of-mass energy of $\sqrt{s} \approx 300$ GeV.
The $e^+p$ and $e^-p$
NC data from H1 are used to make the first measurements of the
generalised structure function $x\Fz$ in the very high $Q^2$ domain
($Q^2>1\,000$~${\rm GeV}^2$).

%%%%%%%%%%%%%%%%%%%%%%%%%%%%%%%%%%%%%%%%%%%%%%%%%%%%%%%
\section{Neutral and Charged Current Cross Sections}
%%%%%%%%%%%%%%%%%%%%%%%%%%%%%%%%%%%%%%%%%%%%%%%%%%%%%%%

\label{sec:theory}
The DIS cross sections $\sigma_{NC(CC)}$ for NC and CC processes in
$e^{\pm}p$ collisions may be factorised as $\sigma_{NC(CC)} =
\sigma_{NC(CC)}^{Born}(1+\delta^{qed}_{NC(CC)})(1+\delta^{weak}_{NC(CC)})$
where $\sigma_{NC(CC)}^{Born}$ is the Born cross section and
$\delta^{qed}_{NC(CC)}$ and $\delta^{weak}_{NC(CC)}$ are the QED
and weak radiative corrections respectively.

The NC cross section for the process $e^{\pm}p\rightarrow e^{\pm}X$
with unpolarised beams and corrected for QED radiative effects
is given by
\begin{eqnarray}
\label{Snc1}
\frac{{\rm d}^2\sigma_{NC}^{\pm}}{{\rm d}x\;{\rm d}\QQ}
& = & \frac{2\pi \alpha^2}{xQ^4} 
\left[Y_+ \Ftwo \mp Y_{-}x\Fz -y^2 \FL \right]
(1+\delta^{weak}_{NC})\,\,\,,
\end{eqnarray}

where $\alpha$ is the fine structure constant taken to be $\alpha
\equiv \alpha(Q^2=0)$. The $\delta^{weak}_{NC}$ 
corrections are defined in~\cite{workshop} with the Fermi coupling
constant, $G_F$, and $M_Z$ as the other main electroweak inputs. 
The helicity dependences of
the electroweak interactions are contained in \mbox{$Y_{\pm} \equiv 1 \pm
(1-y)^2$}.  The generalised structure functions $\Ftwo$ and $x\Fz$ can
be decomposed as follows~\cite{klein}
\begin{eqnarray}
\label{f2p}
\Ftwo  \equiv & \Fem & - \ v_e  \ \frac{\kappa_w  \QQ}{(\QQ + M_Z^2)}
 \hspace*{0.2cm} \Fint  + (v_e^2+a_e^2)  
\left(\frac{\kappa_w  Q^2}{\QQ + M_Z^2}\right)^2 \Fwk \\
\label{f3p}
x\Fz    \equiv &      & - \ a_e  \ \frac{\kappa_w  \QQ}{(\QQ + M_Z^2)} 
x\Fzint + \hspace*{0.3cm} (2 v_e a_e) \hspace*{0.3cm}
\left(\frac{\kappa_w  Q^2}{\QQ + M_Z^2}\right)^2  x\Fzwk,
\end{eqnarray} 

where $\kappa_w^{-1}=4\frac{M_W^2}{M_Z^2}(1-\frac{M_W^2}{M_Z^2})$ in
the on-shell scheme~\cite{gmu} and $M_W$ is defined in terms of the
electroweak inputs.  The quantities $v_e$ and $a_e$ are the vector and
axial couplings of the electron to the $Z^{0}$~\cite{gmu}.  The
electromagnetic structure function $\Fem$ originates from photon
exchange only. The functions $\Fwk$ and $x \Fzwk$ are the
contributions to $\Ftwo$ and $x\Fz$ from $Z^0$ exchange and the
functions $\Fint$ and $\Fzint$ are the contribution from $\gamma Z^0$
interference. The purely longitudinal structure function $\FL$ may be
decomposed in a manner similar to $\Ftwo$. Its contribution is
significant only at high $y$ and is expected to diminish with
increasing $Q^2$.

Over most of the kinematic domain at HERA the dominant contribution to
the cross section comes from the electromagnetic structure function
$F_2$. Only at large values of $Q^2$ do the contributions due to $Z^0$
exchange become important.  For longitudinally unpolarised lepton
beams $\Ftwo$ is the same for electron and for positron scattering,
while the $x \Fz$ contribution changes sign as can be seen in
eq.~\ref{Snc1}. In $e^-p$ scattering, due to the positive interference
between photon and $Z^0$ exchange, the Standard Model cross section is
larger than that calculated in a model which includes only photon
exchange. Conversely for $e^+p$ scattering, within the HERA kinematic
domain, the negative interference arising from the $x\Fz$ term,
results in a cross section that is smaller than in the photon exchange
only model.

In the quark parton model (QPM) the structure functions $F_2$,
$F_2^{\gamma Z} $ and $F_2^Z$ are related to the sum of the quark and
anti-quark densities
\begin{equation}
\label{eq:f2}
[F_2,F_2^{\gamma Z},F_2^{Z}] = x \sum_q 
[e_q^2, 2 e_q v_q, v_q^2+a_q^2] 
\{q+\bar{q}\} 
\end{equation}
and the structure functions $xF_3^{\gamma Z} $ and $xF_3^Z$ to the
difference between quark and anti-quark densities
\begin{equation}
\label{eq:xf3}
[ x F_3^{\gamma Z},x F_3^{Z} ] = x \sum_q 
[2 e_q a_q, 2 v_q a_q]
\{q -\bar{q} \}.
\end{equation}
The functions $q$ and $\bar{q}$ are the parton distribution functions
(PDFs) for quarks and anti-quarks, $e_q$ is the charge of quark $q$ in
units of the electron charge and $v_q$ and $a_q$ are the vector and
axial-vector couplings of the quarks.

For CC interactions the cross section corrected for QED radiative
effects may  be expressed as
\begin{eqnarray}
\label{eq:cccross}
\frac{{\rm d} ^2 \sigma_{\rm CC}^{\pm}}{{\rm d} x\; {\rm d} Q^2} & = &
\frac{G_F^2 M_W^4}{2 \pi x} \frac{1}{(Q^2+M_W^2)^2} 
\;\phi_{\rm CC} ^\pm \; (1+\delta^{weak}_{CC})\\
\mbox{with } \hspace{1cm} \phi_{\rm CC}^\pm & =& 
\frac{1}{2}(Y_+ \wtwogen^\pm   \mp Y_- \xwthreegen^\pm - y^2 \wlgen^\pm),  
\end{eqnarray}
where $\delta^{weak}_{CC}$ are the CC weak radiative
corrections.  The structure functions for CC interactions $\wlgen$,
$\wtwogen$, and $\xwthreegen$ are defined
in analogy to the NC case.
In the QPM, neglecting contributions from the $t$ and
$b$ quarks, the structure function term for $e^\pm p \rightarrow \nu
X$ can be written as
\begin{equation}
\phi_{\rm CC}^+=
x \left [ (\bar{u}+\bar{c})+(1-y)^2({d}+{s}) \right ],\;\;\;  
\phi_{\rm CC}^-=
x \left [ (u+c)+(1-y)^2(\bar{d}+\bar{s}) \right ]
\label{Scc}
\end{equation}
where $u$, $c$, $d$, $s$ are the quark distributions and $\bar{u}$,
$\bar{c}$, $\bar{d}$, $\bar{s}$ are the anti-quark distributions.

The measured cross sections presented in section~\ref{results}, in
which the effects of QED radiation have been corrected for, correspond
to the differential cross sections ${\rm d}^2 \sigma_{NC(CC)}/{\rm d}x{\rm d}Q^2$
defined in eq.~\ref{Snc1} and~\ref{eq:cccross}.  The corrections
($\delta^{qed}_{NC(CC)}$) are defined in~\cite{h1hiq2} and were
calculated using the program HERACLES~\cite{heracles} as implemented
in DJANGO~\cite{django} and verified with the analytic program
HECTOR~\cite{hector}.  The radiative corrections due to the exchange
of two or more photons between the lepton and the quark lines, which
are not included in DJANGO, vary with the polarity of the lepton beam. 
This variation is small compared to the quoted errors and is
neglected.  The weak corrections ($\delta^{weak}_{NC(CC)}$), are
typically of the order of 1\% and have not been applied to the
measured cross sections, but are applied to determine the
electromagnetic structure function $F_2$ and the CC structure function
term $\phi_{CC}$.

It is convenient to derive the NC and CC ``reduced cross sections'' in
which the dominant part of the $Q^2$ dependence of \mbox{${{\rm d}^2
\sigma}/{{\rm d}x{\rm d}Q^2}$} due to the boson propagators is
removed. The reduced cross sections for NC and CC are defined as

\begin{equation}
\label{Rnc}
\tilde{\sigma}_{NC}(x,Q^2) \equiv  \frac{1}{Y_+} \ 
\frac{ Q^4 \ x  }{2 \pi \alpha^2}
\          \frac{{\rm d}^2 \sigma_{NC}}{{\rm d}x{\rm d}Q^2},\;\;\;
\tilde{\sigma}_{CC}(x,Q^2) \equiv  
\frac{2 \pi  x}{ G_F^2}
\left( \frac {M_W^2+Q^2} {M_W^2} \right)^2
          \frac{{\rm d}^2 \sigma_{CC}}{{\rm d}x{\rm d}Q^2}.\;\;\;
\end{equation}

The expression used to extract the electromagnetic structure
function $F_2$ from the measured NC reduced cross section is:
\begin{equation}
\label{f2corr}
\tilde{\sigma}_{NC}=F_2(1+\Delta_{F_2}+\Delta_{F_3}+\Delta_{F_L})
(1+\delta^{weak}_{NC})=F_2(1+\Delta_{all}),\;\;\;
\end{equation}
where the correction terms\footnote{The explicit definitions are
given in eq.~17 of \cite{h1hiq2}.} $\Delta_{F_2}$ and $\Delta_{F_3}$ account
for the relative contribution of pure $Z^0$ exchange and photon-$Z^0$ interference 
to $\Ftwo$ and $x\Fz$, and $\Delta_{F_L}$ originates from the longitudinal
structure function $\FL$.

%%%%%%%%%%%%%%%%%%%%%%%%%%%%%%%%%%%%%%%%%%%%%%%%%%%%%%%
\section{Experimental Technique}
%%%%%%%%%%%%%%%%%%%%%%%%%%%%%%%%%%%%%%%%%%%%%%%%%%%%%%%

%%%%%%%%%%%%%%%%%%%%%%%%%%%%%%%%%%%%%%%%%%%%%%%%%%%%%%%
\subsection{H1 Apparatus}
%%%%%%%%%%%%%%%%%%%%%%%%%%%%%%%%%%%%%%%%%%%%%%%%%%%%%%%
The co-ordinate system of H1 is defined such that the positive $z$
axis is in the direction of the incident proton beam. The polar
angle $\theta$ is then defined with respect to the positive $z$ axis
which defines the forward direction. The detector
components most relevant to this analysis are the LAr calorimeter,
which measures the angles and energies of particles over the range
$4^\circ<\theta<154^\circ$, a lead-fibre calorimeter (SPACAL) covering
the range $153^\circ<\theta<177^\circ$ and the inner tracking
detectors which measure the angles and momenta of charged particles
over the range $7^\circ<\theta<165^\circ$. In addition the PLUG
calorimeter covers the range $0.7^\circ<\theta<3.3^\circ$.  A full
description of the H1 detector can be found in~\cite{h1detector} and~\cite{spacal}.

%%%%%%%%%%%%%%%%%%%%%%%%%%%%%%%%%%%%%%%%%%%%%%%%%%%%%%%
\subsection{Monte Carlo Generation Programs}
%%%%%%%%%%%%%%%%%%%%%%%%%%%%%%%%%%%%%%%%%%%%%%%%%%%%%%%

In order to determine acceptance corrections and background
contributions for the DIS cross section measurements, the detector
response to events produced by various Monte Carlo (MC) generation programs
is simulated in detail using a program based on GEANT~\cite{GEANT}.
These simulated events are then subjected to the same reconstruction
and analysis chain as the real data.

DIS processes are generated using the DJANGO~\cite{django} program
which is based on HERACLES \cite{heracles} for the electroweak
interaction and on LEPTO~\cite{lepto}, using the colour dipole model
as implemented in ARIADNE \cite{cdm} to generate the QCD dynamics.
The JETSET program is used for the hadron fragmentation~\cite{jetset}.
The simulated events are produced with PDFs from the next to leading
order QCD fit~\cite{h1hiq2} performed on fixed target data from
NMC~\cite{nmc} and BCDMS~\cite{bcdms}, and H1 $e^+p$
data~\cite{h1hiq2}. The fit gives a good description of the data and
is referred to as the ``H1 97 PDF Fit'' in the following.

The dominant $ep$ background contribution to NC and CC processes is
due to photoproduction ($\gamma p$) events.  These are simulated using the
PYTHIA~\cite{pythia} generator with GRV leading order parton
distribution functions for the proton and photon~\cite{ggrv}.

%%%%%%%%%%%%%%%%%%%%%%%%%%%%%%%%%%%%%%%%%%%%%%%%%%%%%%%
\subsection{Kinematic Reconstruction and Calibration}
%%%%%%%%%%%%%%%%%%%%%%%%%%%%%%%%%%%%%%%%%%%%%%%%%%%%%%%

The NC event kinematics are reconstructed using the $e\Sigma$
method~\cite{esigma}, which uses the energy $E_e^{\prime}$ and polar
angle $\theta_e$ of the scattered electron and the quantity
$\Sigma=\sum_i{(E_i-p_{z,i})}$, where the summation is performed over
all objects in the hadronic final state assuming particles of zero rest mass. This
method gives good resolution in $x$ and $Q^2$ throughout the kinematic
range.

The CC event kinematics can only be determined with the hadron method
($h$ method)~\cite{jbmethod}, which uses $\Sigma$ and the hadronic
transverse momentum
\mbox{$P_{T,h}=\sqrt{(\sum_i{p_{x,i}})^2+(\sum_i{p_{y,i}})^2}$}, where
the summation is performed over all hadronic final state particles.

The accessible kinematic range depends on the resolution of the
reconstructed kinematics and is determined by
requiring the purity and stability of any ($x$, $Q^2$) bin to be
larger than $30\%$.  The stability (purity) is defined as the fraction
of events which originate from a bin and which are reconstructed in
it, divided by the number of generated (reconstructed) events in that
bin.

The electromagnetic and hadronic response of the detector is
calibrated using the analysis described in~\cite{h1hiq2}.  The
procedure is found to give an excellent description of the detector
response by the simulation.  The hadronic final state is measured by
combining calorimeter energy deposits (clusters) with low momentum tracks.
Isolated, low energy calorimeter clusters are
classified as noise and are not included in the determination of the
hadronic final state.

%%%%%%%%%%%%%%%%%%%%%%%%%%%%%%%%%%%%%%%%%%%%%%%%%%%%%%%%%%%%
\subsection{Selection of NC Events}
%%%%%%%%%%%%%%%%%%%%%%%%%%%%%%%%%%%%%%%%%%%%%%%%%%%%%%%%%%%%

%----------------------------------------------------------------
\begin{figure}[htb]   
\begin{center} 
\begin{picture}(160,140)(0,0)
\setlength{\unitlength}{1 mm}
\put(0,-10){\epsfig{file=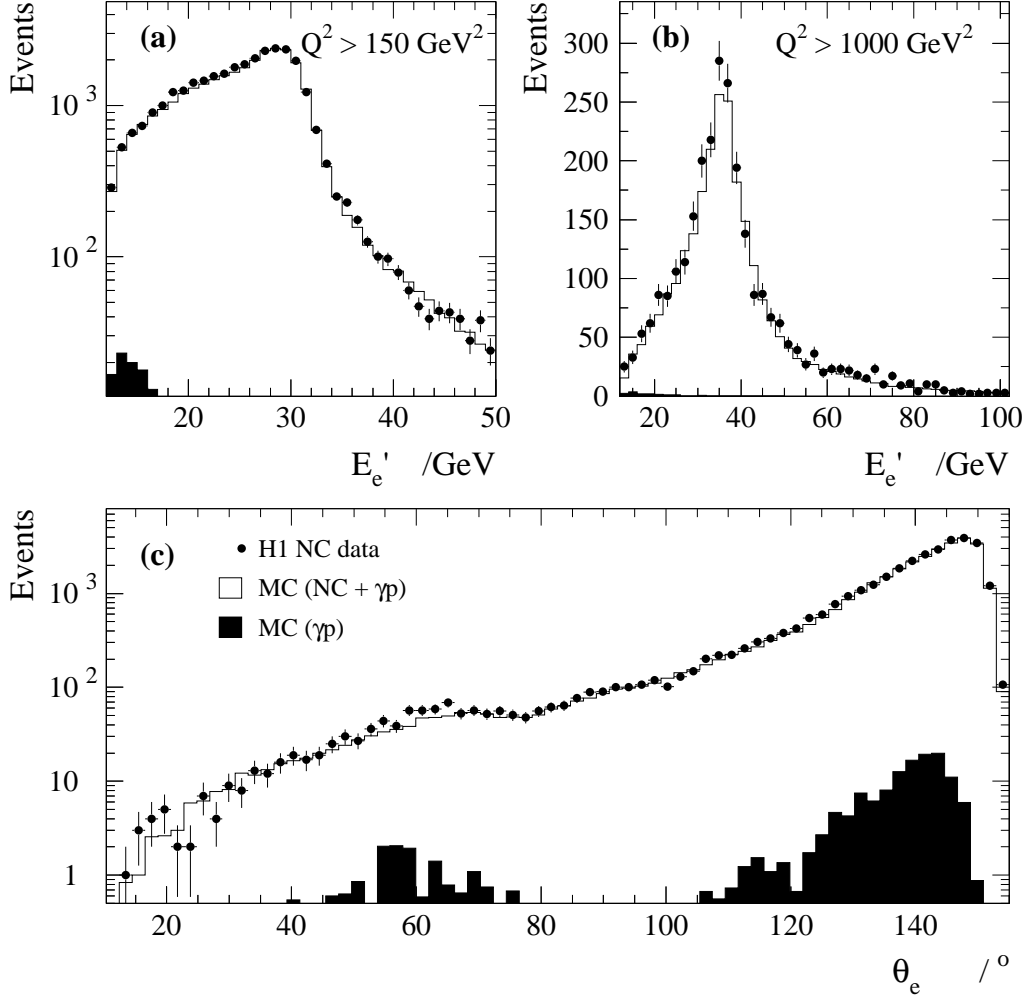,width=15cm}}
\put( 27,119){\bf (a)}
\put( 95,119){\bf (b)}
\put( 27,50) {\bf (c)}
\end{picture}
\end{center}    
\caption{\sl Distributions of (a) $E_e^{\prime} {\rm~for}~Q^2 > 150
\rm~GeV^2$, (b) $E_e^{\prime} {\rm~for}~Q^2 > 1\,000 \rm~GeV^2$ and
(c) $\theta_e$ for NC data (solid points) and simulation (solid
line). The filled histograms show the photoproduction contribution. }
\label{nc_cont}
\end{figure}
%----------------------------------------------------------------

High \qsq NC events are selected by requiring that the event has a
compact electromagnetic cluster in the LAr calorimeter, taken to be
the scattered electron, in addition to an interaction vertex position
within $\pm 35$ cm of its nominal position along the $z$ axis. The
scattered electron energy and polar angle are determined from the
calorimeter cluster. For $Q^2$ less than $890$~GeV$^2$ 
corresponding to a bin boundary, the requirement $y_e<0.63$, where $y$ is determined
from the electron method ($y_e=1-\frac{E_e^{\prime}}{E_e}
\sin^2{\frac{\theta_e}{2}}$), ensures that the measurement is confined to
the region where the trigger efficiency is greater than 99.5\%.
Fiducial cuts are made to remove local regions where the
electromagnetic shower of the scattered electron is not fully
contained in the calorimeter, and where the trigger is not fully
efficient. Further details are given in~\cite{reisert}.

The most significant background in the NC sample arises from
photoproduction interactions where hadronic final state particles are
misidentified as the scattered electron. This background is suppressed
by requiring that $\Sigma+E_e^\prime(1-\cos\theta_e)>35$ GeV,
$y_e<0.9$ and that there is an extrapolated track with a distance of
closest approach to the cluster of less than $12$ cm. This latter
requirement is only applied for $\theta_e\geq 40^{\circ}$, where the
track reconstruction efficiency is greater than 97$\%$.

The final sample of selected NC data consists of about $40\,000$
events. The scattered electron energy spectrum of the data is compared
to simulation in fig.~\ref{nc_cont}(a) for $Q^2>150\rm~GeV^2$.  The
simulation is in good agreement with the data, particularly for the
region $E_e^\prime>26~\rm GeV$ which is sensitive to the details of
the calibration and resolution simulation. In fig.~\ref{nc_cont}(b)
the $E_e^\prime$ spectrum is also shown for $Q^2>1\,000\rm~GeV^2$. The
distribution of the electron polar angle, $\theta_e$, is presented in
fig.~\ref{nc_cont}(c) where the data are well described by the
simulation across the full range of $\theta_e$. The filled histogram
shows the photoproduction contribution which falls rapidly with
decreasing $\theta_e$, in part due to the requirement $y<0.63$. At
higher $Q^2$, and correspondingly lower $\theta_e$, the accessible $y$
range is extended to $y<0.9$ and is responsible for the extra
background contribution visible around $\theta_e\sim 60^\circ$.

%%%%%%%%%%%%%%%%%%%%%%%%%%%%%%%%%%%%%%%%%%%%%%%%%%%%%%%%%%%%
\subsection{Selection of CC Events}
%%%%%%%%%%%%%%%%%%%%%%%%%%%%%%%%%%%%%%%%%%%%%%%%%%%%%%%%%%%%
The selection of CC events is based on the expectation that the unseen
neutrino will result in an observed imbalance in the transverse
momentum, $P_{T,h}$.  A requirement that $P_{T,h}\geq 12$~GeV is
therefore made. In addition the event must have a reconstructed vertex
within $\pm 35$ cm of its nominal position.

The non-$ep$ background in the CC sample is rejected using timing
requirements and a set of topological finders based on track and
calorimeter patterns consistent with cosmic events or particles from
the halo of the proton beam \cite{beate}. The remaining $ep$
background, which is dominantly due to photoproduction events, is
suppressed using the ratio \vap~and the difference in azimuth between
$\vec{P}_{T,h}$ measured in the main detector and the
PLUG calorimeter, $\Delta \phi_{h,{\rm PLUG}}$.  The quantities $V_p$
and $V_{ap}$ are respectively the transverse energy flow parallel and
anti-parallel to $\vec{P}_{T,h}$; they are determined from the
transverse momentum vectors $\vec{P}_{T,i}$ of all the particles $i$
which belong to the hadronic final state according to
\begin{eqnarray}
V_{p}
 = \ \ \sum_{i} \frac{\vec{P}_{T,h}
 \cdot \vec{P}_{T,i}}{P_{T,h}} 
 &
\mbox{for}
&
\vec{P}_{T,h} \cdot \vec{P}_{T,i} >  0 \\
V_{ap}
 = -\sum_{i} \frac{\vec{P}_{T,h}
 \cdot \vec{P}_{T,i}}{P_{T,h}} 
&
\mbox{for}
&
\vec{P}_{T,h} \cdot \vec{P}_{T,i} < 0.
\label{eq:vap} 
\end{eqnarray}

CC events tend to have little energy in the hemisphere around the
direction of the neutrino and consequently have small values of
$V_{ap}/V_{p}$. Conversely the energy is more isotropic in
photoproduction events which generally have higher values of
$V_{ap}/V_{p}$.  One of the main types of photoproduction background
arises from events that contain a jet at low polar angle such that not
all of the energy is recorded in the main detector, resulting in a
measured imbalance in transverse momentum. This missing momentum is,
however, generally tagged in the PLUG calorimeter, with such events
having values of $\Delta \phi_{h,{\rm PLUG}}$ close to $180^\circ$.
The two anti-photoproduction criteria are combined with $P_{T,h}$ so
as to maximise the background rejection whilst still retaining a high
efficiency for CC events.

For $P_{T,h}<25\,{\rm GeV}$ a $P_{T,h}$ dependent cut is applied in
the $\Delta\phi_{h,{\rm PLUG}}$ - $V_{ap}/V_p$ plane,
whereas for $P_{T,h}>25\,{\rm GeV}$, the cut is simplified to $V_{ap}/V_p < 0.2$.
The cut gains a factor of two improvement in background rejection
whilst retaining a similar selection efficiency compared to the cut
\mbox{$V_{ap}/V_p<0.15$}~used in previous
analyses~\cite{h1hiq2}. Further details are available in~\cite{zhang}.

In order to restrict the measurement to a region where the kinematic
reconstruction is optimal the events are required to have
$y_h=\Sigma/2E_e<0.85$.  The CC trigger efficiency is determined using
NC events in which all information associated to the scattered
electron is removed. This method gives a precise measure of the
efficiency which is found to be $63\%$ at $Q^2=300\,{\rm GeV}^2$ and
reaches 98\% at $Q^2=5\,000\,{\rm GeV}^2$.  The measurement is
restricted to the region where the trigger efficiency is everywhere
greater than $40$\% by demanding $y_h>0.03$.

\begin{figure}[htb]   
\begin{center} 
\begin{picture}(160,70)(0,0)
\setlength{\unitlength}{1 mm}
\put(  0,-10){\epsfig{file=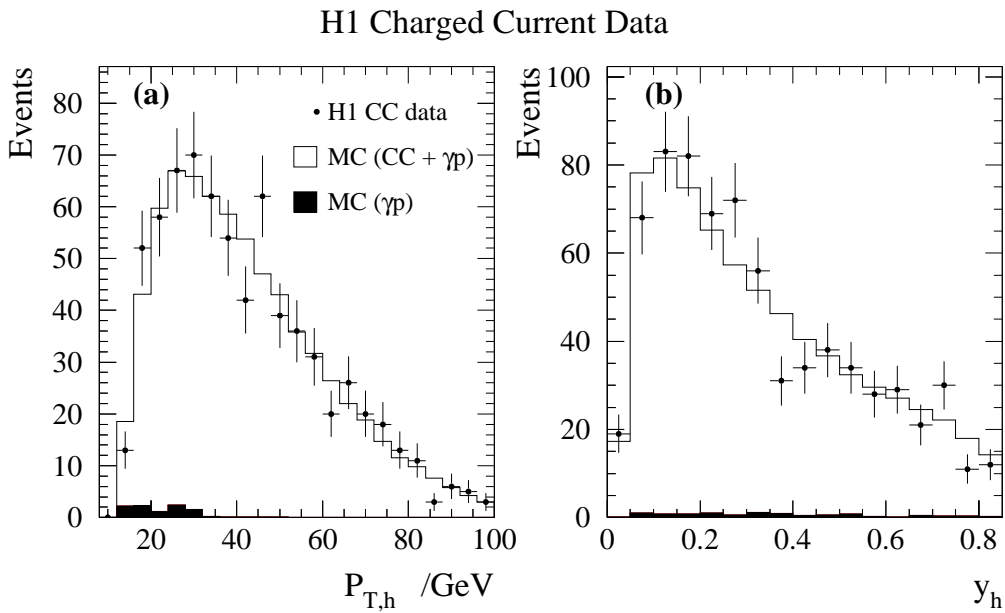,width=15cm}}
\put( 27,60){\bf (a)}
\put( 95,60){\bf (b)}
\end{picture}
\end{center}    
\caption{\sl Distributions of (a) $P_{T,h}$ and (b) $y_h$ for CC data
(solid points) and simulation (solid line). The filled histograms show
the photoproduction contribution.}
\label{cc_cont}
\end{figure}

The final CC data sample contains about 700 events. The data and
simulation are compared in fig.~\ref{cc_cont} for the $P_{T,h}$ and
$y_h$ spectra. In both cases the simulation gives a good description
of the data.

%----------------------------------------------------------------
%----------------------------------------------------------------

%%%%%%%%%%%%%%%%%%%%%%%%%%%%%%%%%%%%%%%%%%%%%%%%%%%%%%%%%%%%
\subsection{Cross Section Determination}
%%%%%%%%%%%%%%%%%%%%%%%%%%%%%%%%%%%%%%%%%%%%%%%%%%%%%%%%%%%%

The photoproduction background is estimated using simulated events
from the PYTHIA generator and checked with a subsample of events which
contain an electron tagged at small scattering angles. The
contribution to the NC cross sections is never more than 5\% at the
highest $y$ and negligible elsewhere.  The background in the CC sample
is at most 13\% at the highest $y$ at low $Q^2$ and below 1\% for
$Q^2>1\,000$~GeV$^2$.  The background is statistically subtracted for
each measurement bin.

The measured distributions are corrected for detector acceptance, migrations,
the effects of QED radiation using the DJANGO simulation, and
converted to cross sections at a specified bin centre by comparison with
the H1 97 PDF Fit~\cite{h1hiq2}.

%%%%%%%%%%%%%%%%%%%%%%%%%%%%%%%%%%%%%%%%%%%%%%%%%%%%%%%%%%%%
\subsection{Systematic Errors}
%%%%%%%%%%%%%%%%%%%%%%%%%%%%%%%%%%%%%%%%%%%%%%%%%%%%%%%%%%%%

The uncertainties on the measurement lead to systematic errors on the
cross sections, which can be split into bin to bin correlated or
uncorrelated parts.  All the correlated systematic errors are found to
be symmetric to a good approximation and are assumed so in the
following. The total systematic error is formed by adding the
individual errors in quadrature. The correlated and the uncorrelated
systematic errors of the NC and CC cross section measurements are
given in tables~\ref{ncfull} and \ref{ccfull} and their origin is
discussed below. 

\begin{itemize}

\item An uncertainty on the electron energy of $1.5\%$ if the $z$
impact position of the electron at the calorimeter surface ($z_{imp}$)
is in the backward part ($z_{imp}<-150$~cm), of $1\%$ in the region $-150
< z_{imp}<20$~cm, of $ 2\%$ for $20< z_{imp}<110$~cm and of $3\%$ in the
forward part ($z_{imp} >110$~cm).  These uncertainties are obtained by
the quadratic sum of an uncorrelated uncertainty and a bin to bin
correlated uncertainty.  The correlated uncertainty comes mainly from
the potential bias of the calibration method and is estimated to be
$0.5\%$ in the whole LAr calorimeter.  It results in a correlated
systematic error on the NC cross section which is $\approx \ 3\%$ at
low $y$ and $Q^2 \ \lapprox \ 1\,000{\rm GeV}^2$.  \item A correlated
uncertainty of $3 \ {\rm {mrad}}$ on the determination of the electron
polar angle. This leads to an uncertainty on the NC reduced cross
section which does not exceed 5\%.

\item A $2\%$ uncertainty on the hadronic energy in the LAr
calorimeter which is obtained from the quadratic sum of an
uncorrelated systematic uncertainty of $1.7\%$ and a correlated
uncertainty of $1\%$ originating from the calibration method and from
the uncertainty of the reference scale ($P_{T,e}$).  The resulting
correlated systematic error increases at low $y$, and is typically
$\lapprox \ 4\%$.

\item
A correlated $25\%$ uncertainty on the amount of noise energy
subtracted in the LAr calor\-i\-meter, which gives rise to a
correlated systematic error at low $y$, e.g.  $\simeq 5\%$ at $x=0.65$
and $Q^2 \le 2\,000$~${\rm GeV}^2$ in the NC measurements.

\item
A $7\%$ ($3\%$) uncertainty on the energy of the hadronic final state
measured in the SPACAL (tracking system).  The influence on the
cross section is small compared to the uncorrelated uncertainty of the
LAr calorimeter energy, and so the three contributions (LAr, SPACAL,
tracks) have been added quadratically, giving rise to the uncorrelated
hadronic error which is given in table~\ref{ncfull} for the NC data
and in table \ref{ccfull} for the CC data.

\item The correlated error due to the uncertainty of the efficiency of
  the anti-photoproduction cut is estimated by varying the quantity
  $V_{ap}/V_p$ by $\pm 0.02$.  This leads to a maximum error at low
  $P_{T,h}$ in the CC analysis of $12\%$.

\item
The  $30\%$ uncertainty on  the subtracted  
photoproduction background 
results in a correlated systematic error  
always smaller than $5\%$  in any bin, both for the NC and CC measurements.

\end{itemize}
The following uncertainties, which lead to 
equivalent uncorrelated  systematic errors on the cross sections,
 have also  been taken into account as listed below.
\begin{itemize}
\item
A $2\%$ error originating from the electron identification efficiency
in the NC analysis.

\item A  1\% error on the efficiency of the track-cluster link requirement
in the NC
analysis.

\item
{A $0.5\%$ error on the trigger efficiency in the NC analysis,
 and from $2$ to $6\%$  in the CC analysis.}
\item
{An error of $1\%$  (NC), $3\%$  (CC) on the cross section 
originating from  the QED radiative corrections.}
\item
{A $3\%$ error on the efficiency of the non-$ep$ background finders in the CC
analysis.}
\item
{A $2\%$ error  ($5\%$  for
  $y < 0.1$)  on the vertex finding efficiency for CC events.}
\end{itemize}
Further details can be found in~\cite{reisert,beate}.
Overall the typical total systematic error for the NC (CC) double
differential cross section is about $5\%$ ($12\%$). For the ${\rm d}
\sigma_{NC(CC)} / {\rm d} Q^2$ measurements, the equivalent error is
typically $3\%$ ($8\%$).  In addition a $1.8\%$ normalisation error,
due to the luminosity uncertainty, must be considered, but is not
included in the systematic error of the measurements given in the
tables, or shown in the figures.

%%%%%%%%%%%%%%%%%%%%%%%%%%%%%%%%%%%%%%%%%%%%%%%%%%%%%%%%%%%%
\section{Results}
%%%%%%%%%%%%%%%%%%%%%%%%%%%%%%%%%%%%%%%%%%%%%%%%%%%%%%%%%%%%
\label{results}

%%%%%%%%%%%%%%%%%%%%%%%%%%%%%%%%%%%%%%%%%%%%%%%%%%%%%%%%%%
\subsection{{\boldmath NC and CC Cross Sections ${\rm d}\sigma/\rm{ d}x$}}
%%%%%%%%%%%%%%%%%%%%%%%%%%%%%%%%%%%%%%%%%%%%%%%%%%%%%%%%%%

\begin{figure}[ht]
 \setlength{\unitlength}{1 mm}
\begin{center} 
\begin{picture}(160,140)(0,0)
\put(  0,-10){\epsfig{file=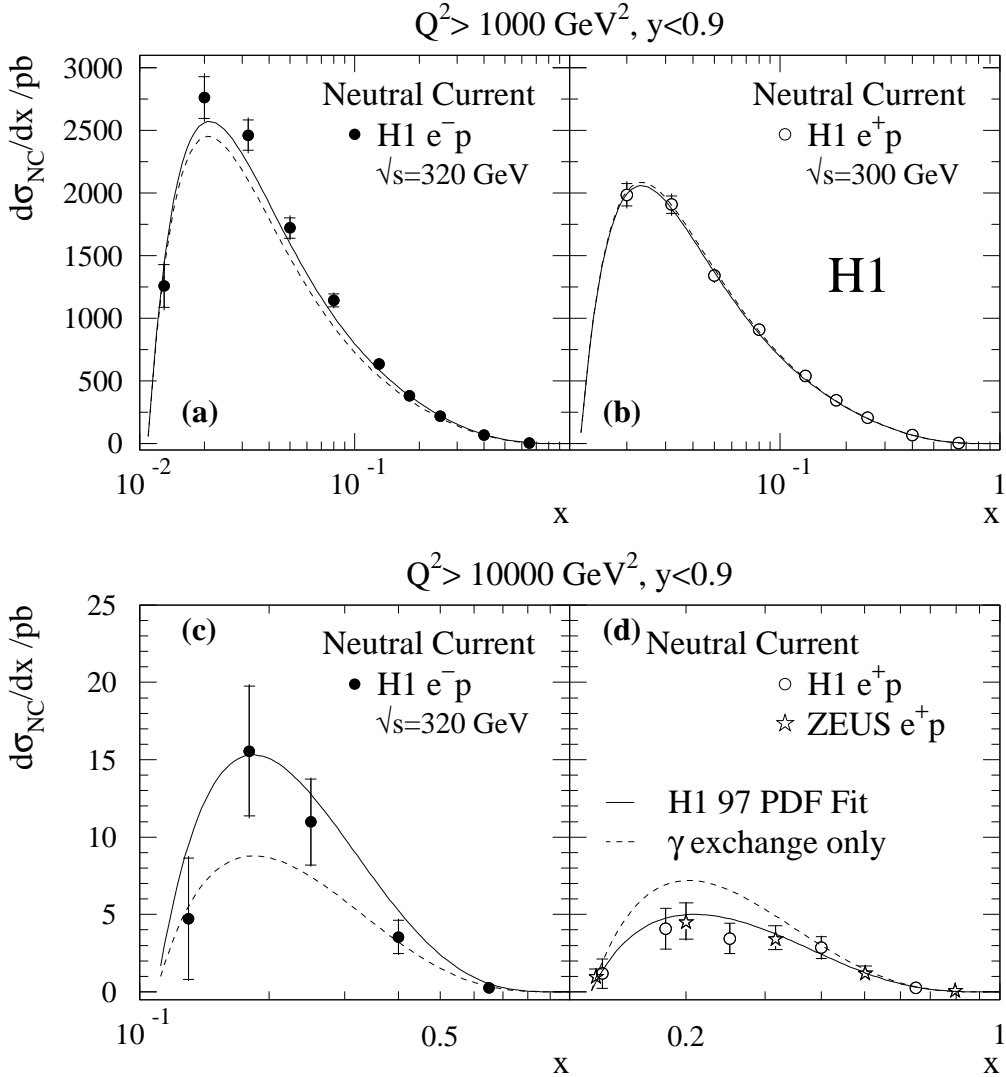,width=15cm}}
\put( 30,80){\bf (a)}
\put( 86,80){\bf (b)}
\put( 30,51){\bf (c)}
\put( 86,51){\bf (d)}
\end{picture}
\end{center}
\caption{\sl The NC cross sections ${\rm d} \sigma_{\rm NC}/{\rm d} x$ for the
$e^-p$ data are shown in (a) for $Q^2>1\,000$~${\rm GeV}^2$ and in (c)
for $Q^2>10\,000$~${\rm GeV}^2$. The H1 $e^+p$ cross sections~\cite{h1hiq2} are
shown in (b) and (d) for $Q^2>1\,000$~${\rm GeV}^2$ and
$10\,000$~${\rm GeV}^2$ respectively.  In addition the ZEUS $e^+p$
data~\cite{zeushiq2} are shown for $Q^2>10\,000$~${\rm GeV}^2$.  The
solid curves show the Standard Model expectation based on the H1 97
PDF Fit. The dashed curves show the contribution of photon exchange
only. All cross sections are shown for $y<0.9$. The inner error bars
represent the statistical error, and the outer error bars show the
total error.  The normalisation uncertainties are not included in
the error bars.}
\label{nc_dsdx} 
\end{figure}

The dependence of the NC cross sections as a function of $x$ is shown
in fig.~\ref{nc_dsdx} for both $e^-p$ and $e^+p$ scattering. The data
are shown for $y<0.9$ and $Q^2>1\,000$~GeV$^2$ in
fig.~\ref{nc_dsdx}(a,b) and listed in table~\ref{ncdx1}.  The data for
$y<0.9$ and $Q^2>10\,000$~GeV$^2$ are shown in fig.~\ref{nc_dsdx}(c,d)
and listed in table~\ref{ncdx2}. Fig.~\ref{nc_dsdx} also shows the
expectation from the Standard Model, derived from the H1 97
PDF Fit to the H1 $e^+p$ data~\cite{h1hiq2}. The data are also compared to a
model of pure photon exchange, where the effects of $Z^0$ exchange are
neglected.
\label{sec:dsigdx}
%----------------------------------------------------------------
%----------------------------------------------------------------

The cross sections for $Q^2>1\,000$~GeV$^2$ are in agreement with the
Standard Model, although for $e^-p$ they have a tendency to be larger
at low $x$ and are also found to be larger compared to the pure photon
exchange model. For $Q^2>10\,000$~GeV$^2$ the $e^-p$ cross section is
found to be approximately a factor of four larger than for $e^+p$
scattering at $x \simeq 0.2$ of which only 12\% is due to the
different centre-of-mass energies. The $e^-p$ cross sections are seen
to lie above the pure photon exchange model, whereas $e^+p$ cross
sections lie below it. Both measurements agree with the Standard
Model.

The observed difference at high $Q^2$ between the $e^+p$ and $e^-p$
data and between both sets of data and the pure photon exchange model
is understood in the Standard Model by the effects of $Z^0$ exchange.
It can be deduced from eq.~\ref{f2p} that the generalised structure
function $\Ftwo$ is always larger than the electromagnetic structure
function $\Fem$.  Furthermore since $|v| \ll a^2\kappa_w$ the increase
at large $Q^2$ is expected to be mainly due to the $Z^0$ exchange contribution
$\Fwk$.  Conversely the main contribution to $x\Fz$ is from the
photon-$Z^0$ interference term.  For $e^-p$ scattering the cross
section is enhanced relative to pure photon exchange, since the
contributions from $\Fwk$ and $x\Fz$ are both positive (see
eq.~\ref{Snc1}). For $e^+p$ scattering $\Fwk$ gives a positive
contribution and $x\Fz$ gives a negative contribution. In the HERA
kinematic range the contribution from $x\Fz$ is larger than that from
$\Fwk$ and so the resulting cross section is smaller than that from
pure photon exchange.

\begin{figure}[ht]
\setlength{\unitlength}{1 mm}
\begin{center} 
{\epsfig{file=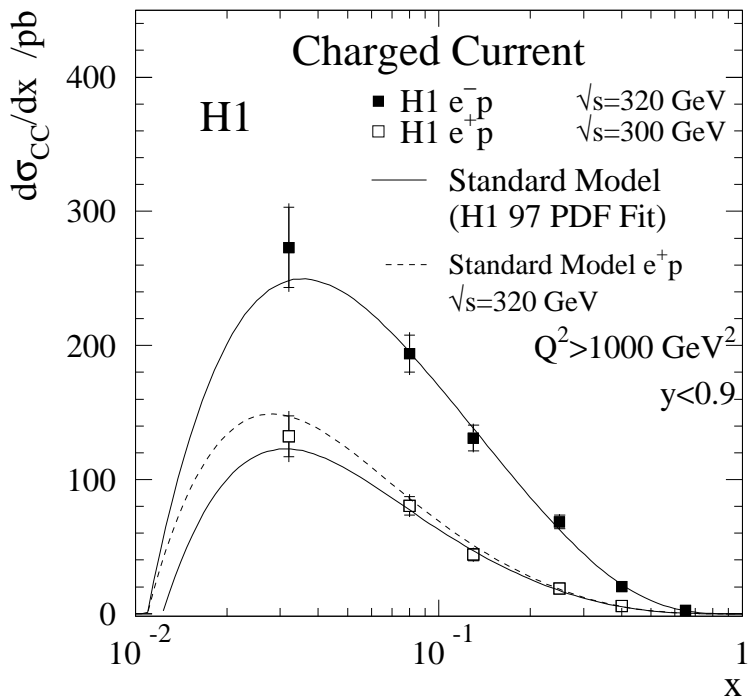,width=11cm}}
\end{center}
\caption{\sl The CC cross section ${\rm d} \sigma_{\rm CC}/{\rm d} x$ for
$Q^2>1\,000$~${\rm GeV}^2$ and $y<0.9$ is shown for the H1 $e^-p$ data
(solid points) and the H1 $e^+p$ data~\cite{h1hiq2}(open points). The solid curves
show the Standard Model expectation based on the H1 97 PDF Fit. The
dashed curve shows the $e^+p$ cross section for an increased centre-of-mass
energy. The inner error bars represent the statistical error, and the
outer error bars show the total error.  The normalisation
uncertainties are not included in the error bars.}
\label{cc_dsdx} 
\end{figure}

The CC cross section ${\rm d}\sigma/{\rm d}x$, is measured for
$Q^2>1\,000$~GeV$^2$ and $y<0.9$ and is shown in fig.~\ref{cc_dsdx}
and listed in table~\ref{ccdx}.  A correction has been made for the
cuts $0.03<y<0.85$ using the H1 97 PDF Fit and is given in
table~\ref{ccdx}.  Fig.~\ref{cc_dsdx} also shows the corresponding
measurement in $e^+p$ scattering and the Standard Model expectation.

%----------------------------------------------------------------
%----------------------------------------------------------------

The CC $e^-p$ cross section is larger than that for $e^+p$ scattering
by a factor of two at low $x$ and a factor of four at high $x$. The
difference between the cross sections is understood to be due to the different
contributions of quark flavours probed by the exchanged $W^\pm$ bosons
and the helicity structure of the CC interaction (see eq.~\ref{Scc}). 
In the valence region at high $x$ the dominant CC
process is the scattering off
$u$ quarks for $e^-p$ interactions and off $d$ quarks for $e^+p$
interactions. The $e^-p$ cross section is expected to be larger  (see
eq.~\ref{Scc}) since
there are two $u$ valence quarks and only one $d$ valence quark in the
proton. Furthermore, scattering off $d$ quarks is suppressed by a
factor of $(1-y)^2$ compared to $u$ quarks. The effect of the increased
centre-of-mass energy accounts only for a small part of the
difference, and is shown by the dashed curve in fig.~\ref{cc_dsdx}.

%%%%%%%%%%%%%%%%%%%%%%%%%%%%%%%%%%%%%%%%%%%%%%%%%%%%%%%%%%%
\subsection{{\boldmath NC and CC Cross Sections ${\rm d}\sigma/\rm{ d}Q^2$}}
%%%%%%%%%%%%%%%%%%%%%%%%%%%%%%%%%%%%%%%%%%%%%%%%%%%%%%%%%%%

\begin{figure}[t]
  \setlength{\unitlength}{1 mm}
\begin{center}
\begin{picture}(160,160)(0,0)
\put( 20, 45){\epsfig{file=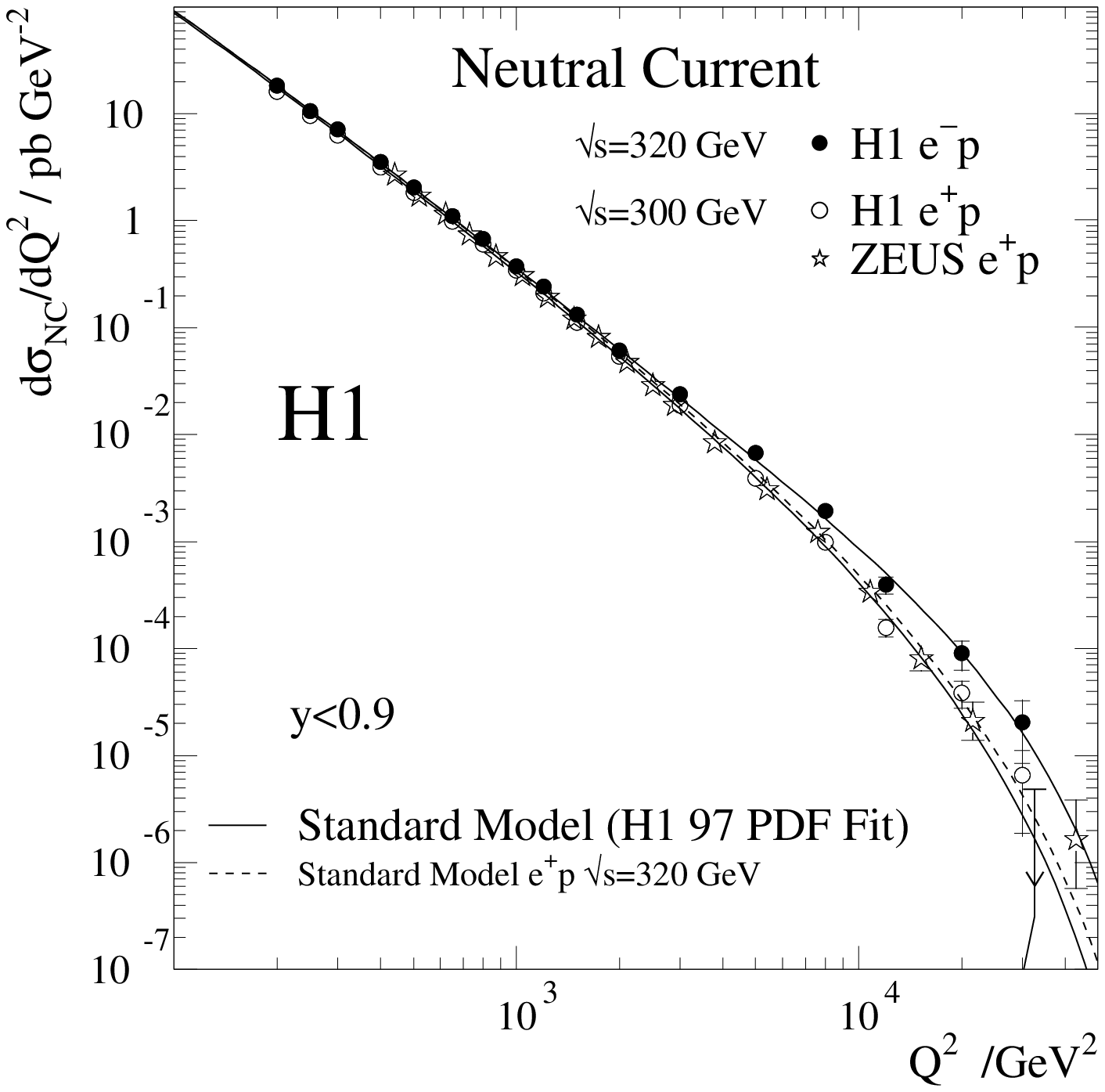,width=11.0cm}}
\put( 20, -5){\epsfig{file=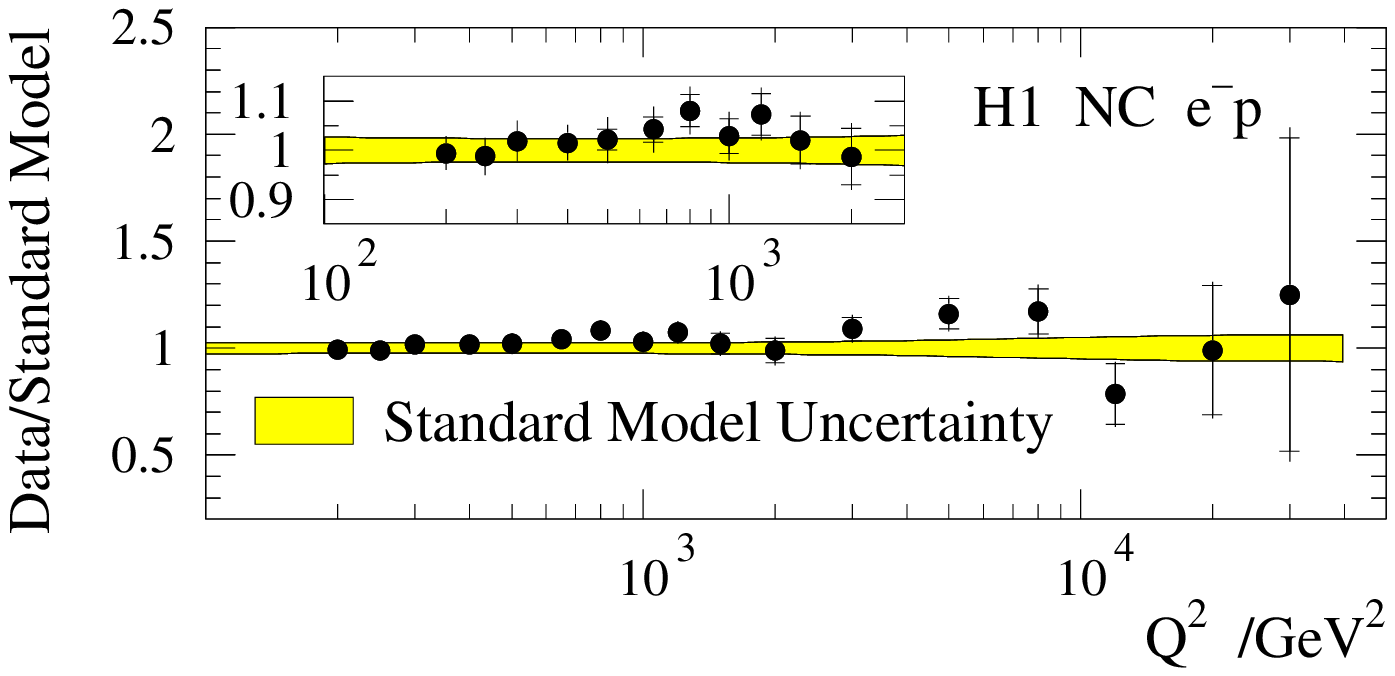,width=11.0cm}}
\end{picture}
\end{center}
\caption{\sl The $Q^2$ dependence of the NC cross section ${\rm d}
\sigma_{\rm NC}/{\rm d} Q^2$ is shown for the H1 $e^-p$ data (solid points) and
$e^+p$ measurements (open points) from H1~\cite{h1hiq2} and
ZEUS~\cite{zeushiq2}. The data are compared with the Standard Model
expectation determined from the H1 97 PDF Fit including the H1 $e^+p$
data.  The dashed curve shows the influence of an increased
centre-of-mass energy on the $e^+p$ cross section.  The ratio of the $e^-p$
data to the Standard Model expectation is shown in the lower figure.
The Standard Model uncertainty is shown as the shaded band.  The inner
error bars represent the statistical error, and the outer error bars
show the total error.  The normalisation uncertainties are not
included in the error bars.}
\label{dsdq2nc}
\end{figure}

The NC cross section ${\rm d}\sigma/{\rm d}Q^2$ for $e^-p$ data is
shown in fig.~\ref{dsdq2nc} for $y<0.9$ and is listed in
table~\ref{ncdq2}.  The cross section is corrected for the effect of
the cut $y<0.63$ for $Q^2<890$~${\rm GeV}^2$, the correction is also
given in table~\ref{ncdq2}. Also included in fig.~\ref{dsdq2nc} are
the measurements of the $e^+p$ data~\cite{h1hiq2,zeushiq2} and the
expectations from the Standard Model. The lower plot shows the ratio
of the measurement to the Standard Model expectation.  The Standard
Model uncertainty represents the uncertainty of the expectation due to
assumptions made in the H1 97 PDF Fit, as well as the uncertainties of
the experimental data entering the fit~\cite{h1hiq2}.

%----------------------------------------------------------------
%----------------------------------------------------------------

The NC data span a range of more than two orders of magnitude in $Q^2$
and are everywhere in good agreement with the Standard Model.  At
$Q^2<1\,000$~GeV$^2$ there is only a small difference of $\approx 7\%$ between the
$e^-p$ and $e^+p$ measurements due to the increased
centre-of-mass energy of the $e^-$ data.  For $Q^2>2\,000$ GeV$^2$ the
$e^-p$ cross section is observed to be systematically larger than the
$e^+p$ cross section.  This difference cannot be explained by the
increased centre-of-mass energy, for which the expected effect on the
$e^+p$ cross section is
indicated by the dashed line.  The observed asymmetry between NC
$e^-p$ and $e^+p$ scattering is well described by the Standard Model,
where the effects of $Z^0$ exchange result in an enhancement of the
$e^-p$ cross section compared to the $e^+p$ cross section.

%----------------------------------------------------------------
\begin{figure}[htb]
\setlength{\unitlength}{1 mm}
\begin{center}
\begin{picture}(160,160)(0,0)
\put(20, 45){\epsfig{file=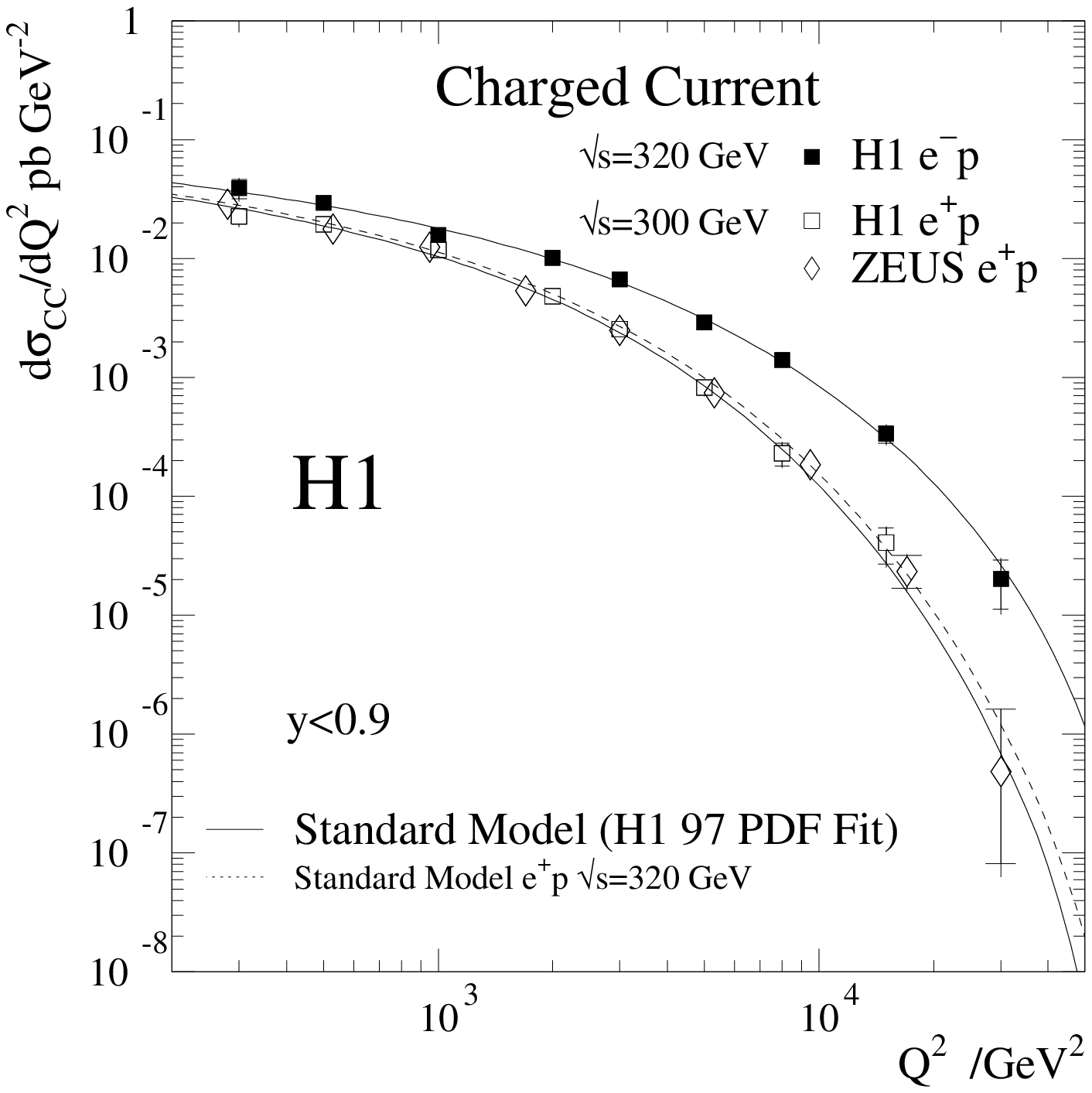,width=11.0cm}}
\put(20,-5){\epsfig{file=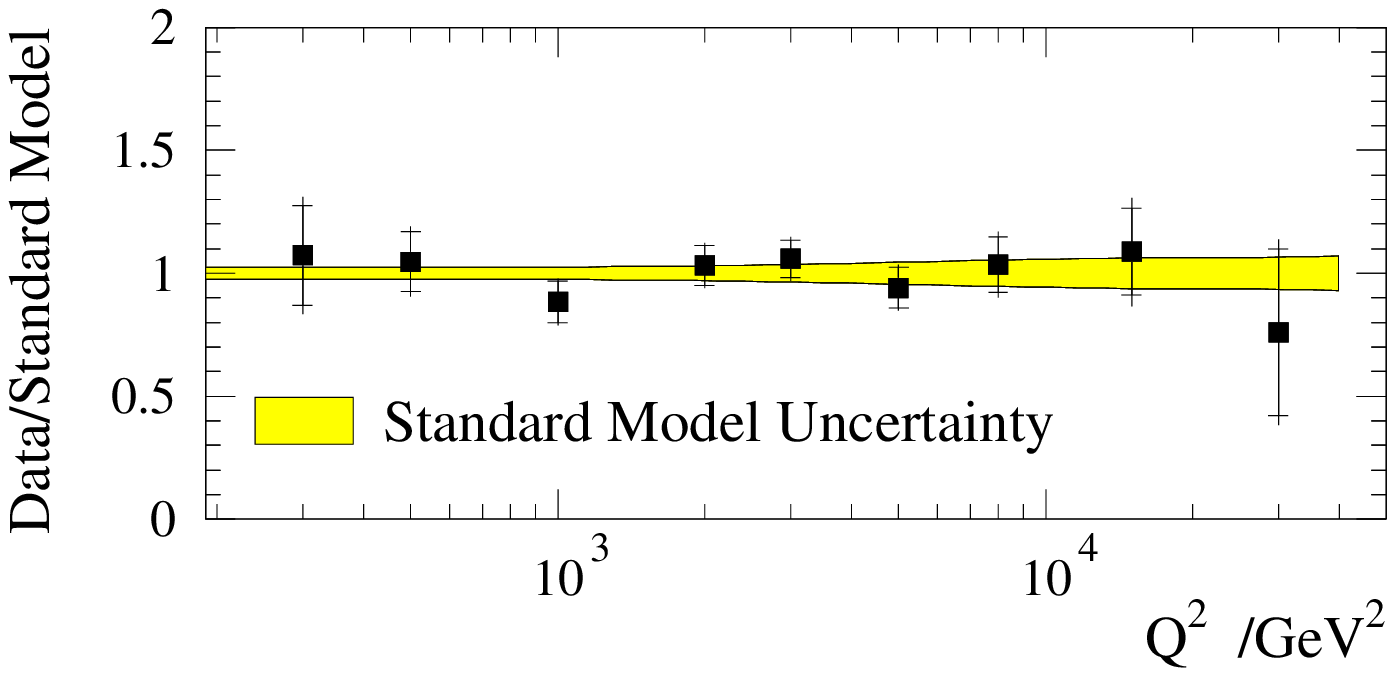,width=11.0cm}}
\end{picture}
\end{center}
\caption{\sl The $Q^2$ dependence of the CC cross section ${\rm d}
\sigma_{\rm CC}/{\rm d} Q^2$ is shown for the $e^-p$ data (solid points) and
$e^+p$ measurements (open points) from H1 and
ZEUS~\cite{zeushiq2}. The data are compared with the Standard Model
expectation determined from the H1 97 PDF Fit including the H1 $e^+p$
data.  The dashed curve shows the influence of an increased
centre-of-mass energy on the $e^+p$ data.  The ratio of the $e^-p$
data to the Standard Model expectation is shown in the lower figure
. The Standard Model uncertainty is shown as the shaded band.  The
inner error bars represent the statistical error, and the outer error
bars show the total error. The normalisation uncertainties are not
included in the error bars.}
\label{dsdq2cc}
\end{figure}
%----------------------------------------------------------------

The CC cross section ${\rm d}\sigma/{\rm d}Q^2$ for $e^-p$ data is
shown in fig.~\ref{dsdq2cc} for $y<0.9$ and is listed in
table~\ref{ccdq2}.  The cross section is corrected for the cuts
$0.03<y<0.85$ and $P_{T,h}>12$~GeV.  The lower plot shows the ratio of
the measurements to the Standard Model expectation. The $e^-p$ cross
section is found to be larger than the $e^+p$ cross section.  The
difference increases with $Q^2$ reaching a factor 10 at $Q^2=15\,000~
\rm GeV^2$ since at higher $Q^2$ the average $x$ and $y$ values are
larger due to the kinematic constraint $Q^2=sxy$.  At lower $Q^2$, and
hence lower $x$, the most important contribution arises from the sea
quarks, which are approximately flavour symmetric~\cite{nusea} and
contribute roughly equally for $e^-p$ and $e^+p$ scattering.  As
$Q^2$, $x$ and $y$ increase the contribution of the valence quarks
becomes more important. 
The CC electron scattering cross section is in good agreement with the
Standard Model expectation throughout the $Q^2$ range.

%%%%%%%%%%%%%%%%%%%%%%%%%%%%%%%%%%%%%%%%%%%%%%%%%%%%%%%%%%
\subsection{\boldmath NC and CC Reduced Cross Sections}
%%%%%%%%%%%%%%%%%%%%%%%%%%%%%%%%%%%%%%%%%%%%%%%%%%%%%%%%%%%

%----------------------------------------------------------------
\begin{figure}[th]
\center
\epsfig{file=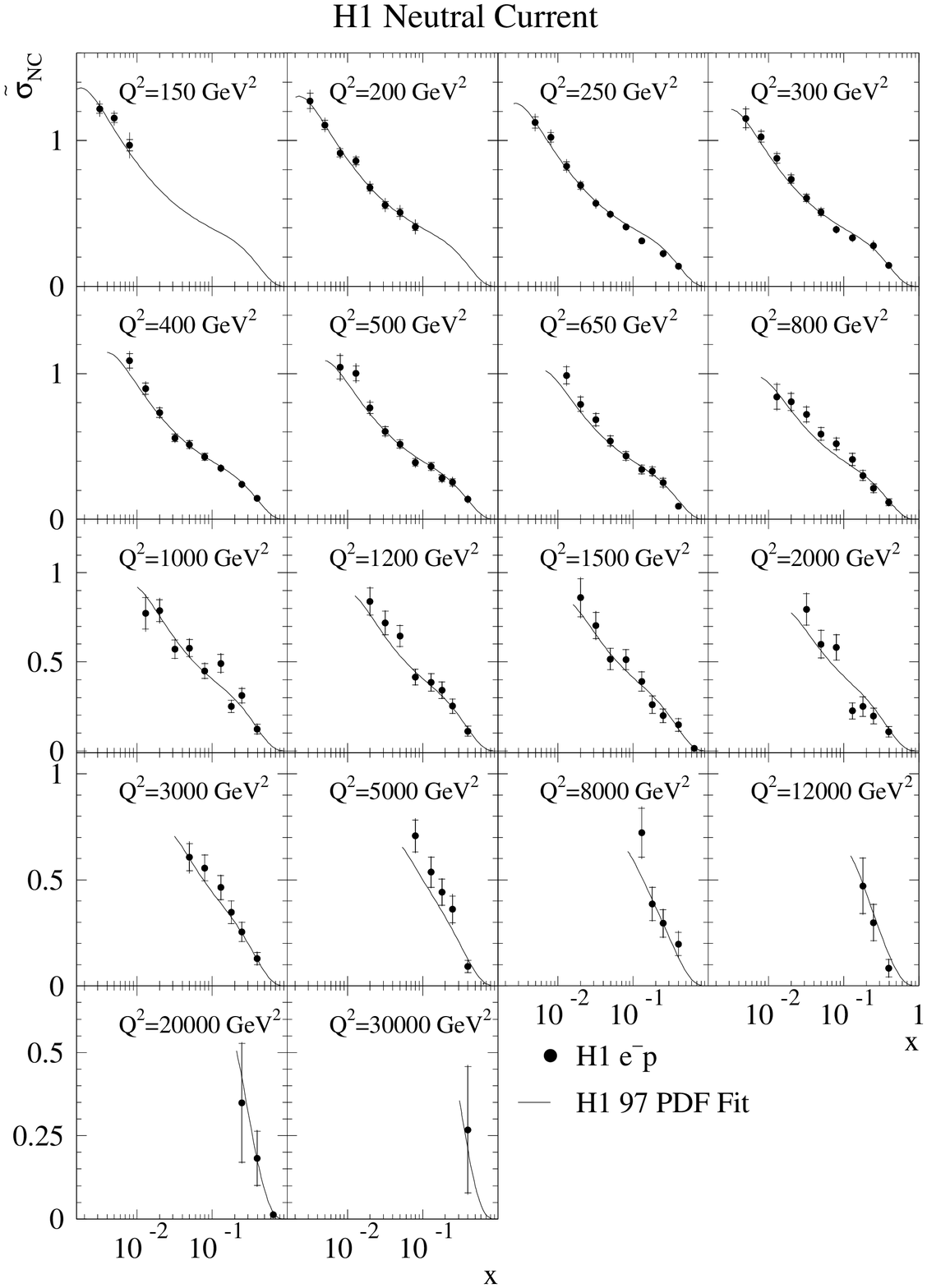,width=15cm,bbllx=30pt,bblly=40pt,bburx=580pt,bbury=800pt}
\caption{\sl The NC $e^-p$ reduced cross section
$\tilde{\sigma}_{NC}(x,Q^2)$ is compared with the H1 97 PDF Fit. The
inner error bars represent the statistical error, and the outer error
bars show the total error.  The normalisation uncertainty is not
included in the error bars.}
\label{nc_stamp} 
\end{figure}
%----------------------------------------------------------------

%----------------------------------------------------------------
\begin{figure}[th]
\center \epsfig{file=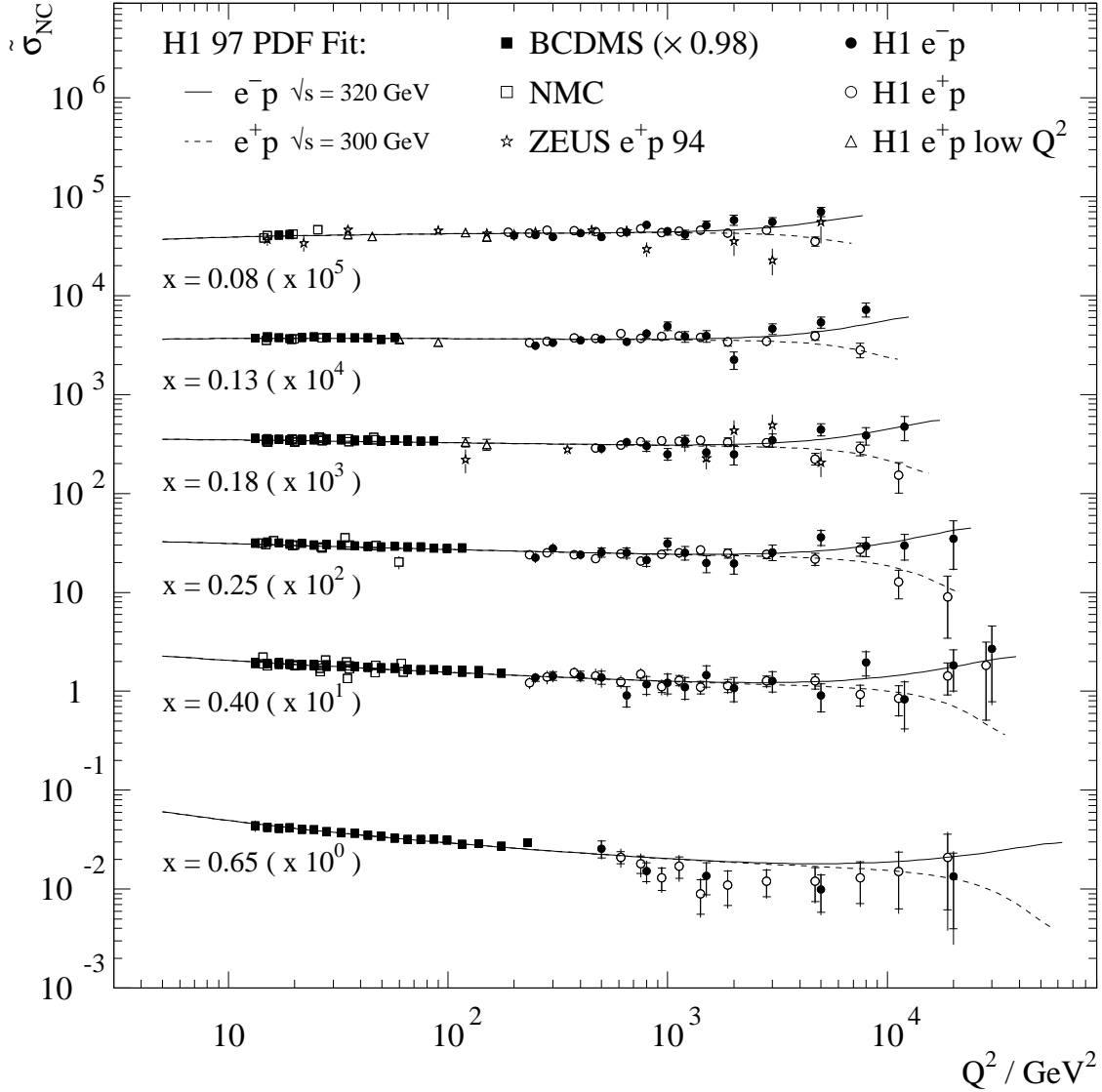,width=15cm}
\caption{\sl The NC reduced cross section $\tilde{\sigma}_{NC}(x,Q^2)$
is shown at high $x$ compared with the H1 97 PDF Fit. The $e^-p$ data
with $\sqrt{s} \approx 320$~${\rm GeV}$ (solid circles) are compared
with the H1 $e^+p$ data at $\sqrt{s} \approx 300$~${\rm GeV}$ (open
circles), ZEUS $e^+p$ data~\cite{zeushiq2}, and fixed target data from
BCDMS~\cite{bcdms} and NMC~\cite{nmc}.  The solid curves represent the
Standard Model expectation based on the H1 97 PDF Fit. The inner error
bars represent the statistical error, and the outer error bars show
the total error.  The normalisation uncertainties are not included
in the error bars.}
\label{nc_hix} 
\end{figure}
%----------------------------------------------------------------

%----------------------------------------------------------------
\begin{figure}[ht]
\center \epsfig{file=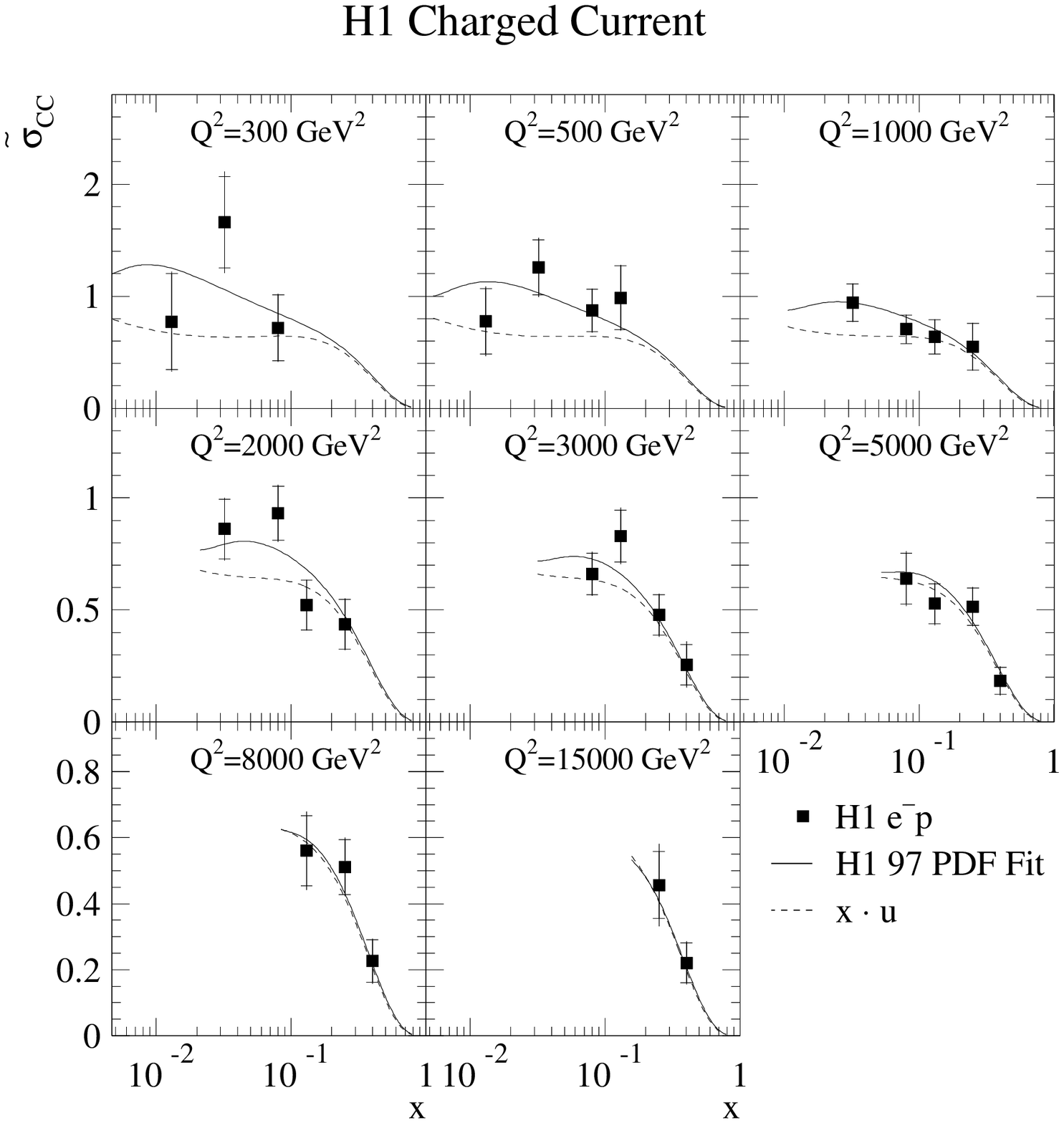,width=15cm}
\caption{\sl The CC reduced cross section $\tilde{\sigma}_{CC}(x,Q^2)$
is shown for $e^-p$ scattering at $\sqrt{s} \approx 320$~${\rm GeV}$
(solid points). The data are compared with the H1 97 PDF Fit.  The
inner error bars represent the statistical error, and the outer error
bars show the total error. The normalisation uncertainty is not
included in the error bars.}
\label{cc_stamp}
\end{figure}
%----------------------------------------------------------------

The double differential NC reduced cross section measurements are
listed in table~\ref{ncdxdq2} and are also given in table~\ref{ncfull}
where the contribution of each of the main systematic uncertainties is
listed separately.  In addition table~\ref{ncdxdq2} gives the
extracted value of the electromagnetic structure function $F_2$, where
the correction factor, $(1+\Delta_{all})$, was determined from the H1
97 PDF Fit. The NC reduced cross section is shown in
fig.~\ref{nc_stamp} over the full $x$ and $Q^2$ range of the
measurement which reaches $x=0.65$ and $Q^2=30\,000$~GeV$^2$. The data
exhibit a strong rise with decreasing $x$ which can be interpreted
(see eq.~\ref{eq:f2}) as due to the high density of low $x$ quarks in
the proton. The H1 97 PDF Fit is found to give a good description of
the $x$, $Q^2$ behaviour of the data, though at low $x$ the
expectation has a slight tendency to be lower than the measured cross
sections.

In fig.~\ref{nc_hix} the reduced cross section at high $x$ is compared
with HERA measurements of $e^+p$ scattering and fixed target data as a
function of $Q^2$. The $e^-p$ data are found to be in agreement with
the $e^+p$ measurements for $Q^2<1\,000$~GeV$^2$. The reduced cross
section exhibits approximate scaling with $Q^2$ as is expected from
the scattering of virtual photons off point-like constituents.  At
larger values of $Q^2$ the $e^-p$ data are generally higher than the
$e^+p$ data, as is expected from the effects of $Z^0$ exchange. The
data are compared with the H1 97 PDF Fit, which can be seen to give a
good description of all the data up to $x=0.4$. At $x=0.65$ the fit,
which in this kinematic region is mainly constrained by the BCDMS
data~\cite{bcdms}, lies slightly above the H1 data. The level of
agreement may be quantified by comparing the measurement of
${\rm d}\sigma/{\rm d}x$ at $x=0.65$ for $Q^2>1\,000~\rm GeV^2$ (see
table~\ref{ncdx1}) with the expectation from the H1 97 PDF Fit which
yields a cross section of $7.05~\rm pb$. The difference is found to
be less than $2.5$ standard deviations. The H1 data are not yet
precise enough to clearly distinguish whether the effect is due to a
statistical fluctuation in the H1 data or to systematic differences
between the experiments at high $x$.

The double differential $e^-p$ CC reduced cross section is shown in
fig.~\ref{cc_stamp} and compared with the H1 97 PDF Fit. The data
agree well with the expectation for all $x$ and $Q^2$. Also shown in
fig.~\ref{cc_stamp} is the expected contribution from the $u$ quark
which dominates the $e^-p$ CC cross section for all $x$ and $Q^2$.
The good agreement observed between the $e^-p$ CC reduced cross
section and the H1 97 PDF Fit indicates that the $u$ quark density of
the proton is well understood across the kinematic range in the
measurements.

%%%%%%%%%%%%%%%%%%%%%%%%%%%%%%%%%%%%%%%%%%%%%%%%%%%%%%%%%%%
\subsection{\boldmath Measurement of $x\Fz$}
%%%%%%%%%%%%%%%%%%%%%%%%%%%%%%%%%%%%%%%%%%%%%%%%%%%%%%%%%%%
At large $Q^2$ the NC $e^+p$ and $e^-p$ cross sections differ, as
expected from the effects of $Z^0$ exchange. The data are thus
sensitive to the generalised structure function $x\Fz$.  Since at HERA
the dominant contribution to $x\Fz$ is from photon-$Z^0$ interference
we also evaluate the structure function $x\Fzint$, which is more
closely related to the quark structure of the proton.

\begin{figure}[p]
\begin{picture}(160,170)(0,0)
\setlength{\unitlength}{1 mm}
\put(25,60){\epsfig{file=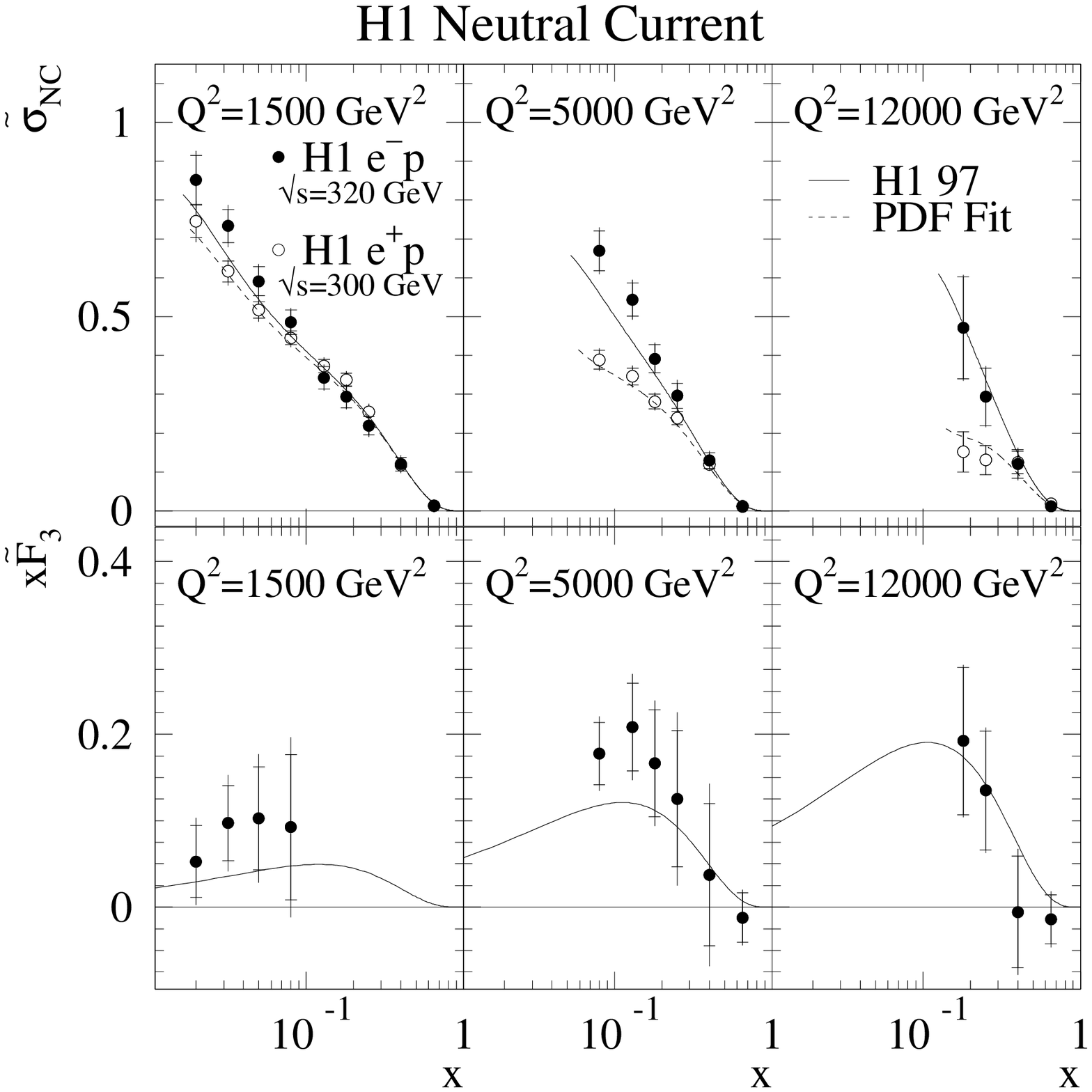,width=12cm}}
%\put(25, 0){\epsfig{file=xf3gz.eps,width=12cm}}
\put(30, 2){\epsfig{file=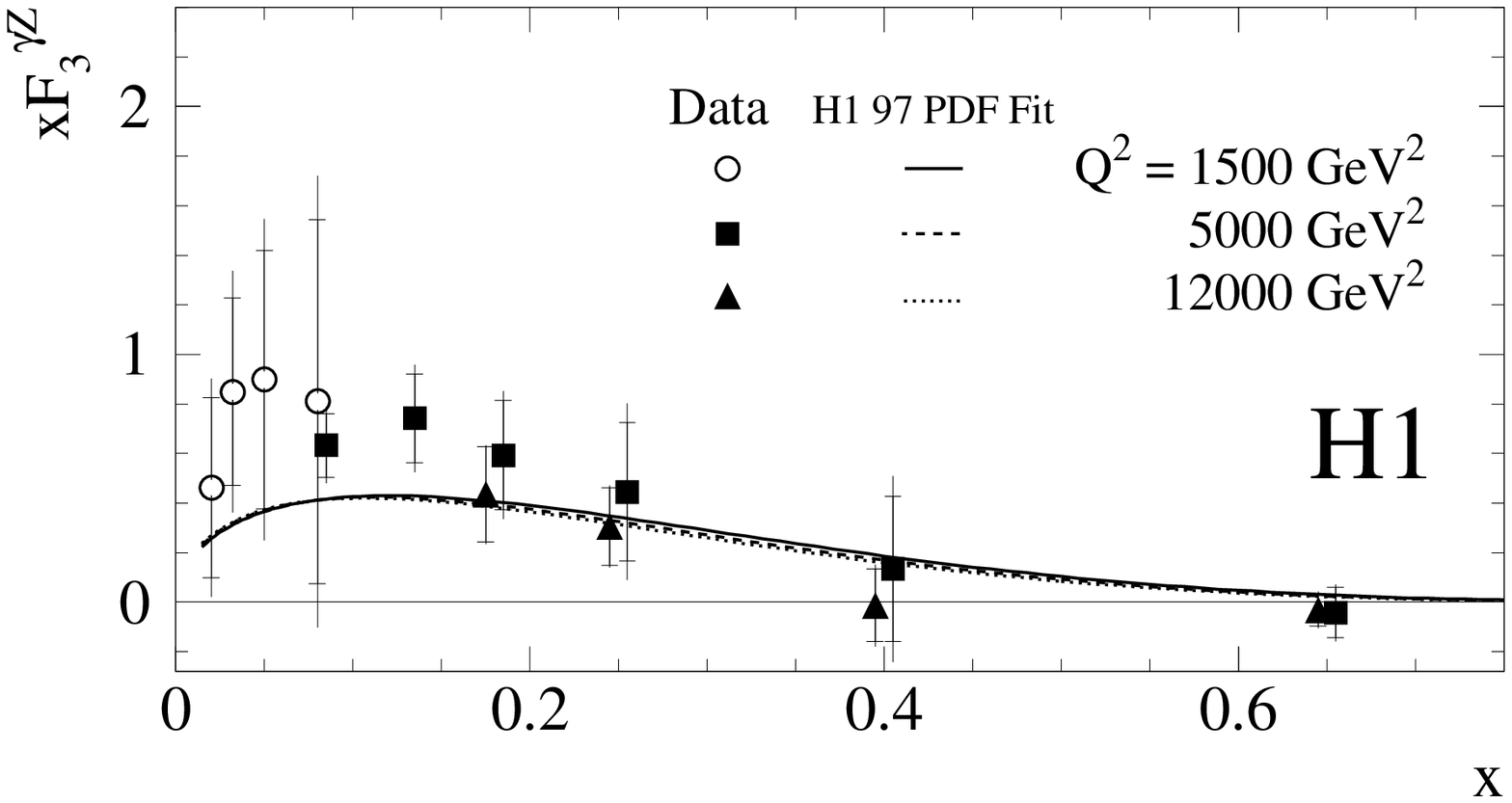,width=11cm}}
\put(45,140){(a)}
\put(45,105){(b)}
\put(45,55){(c)}
\end{picture}
\caption{\sl The NC reduced cross section $\tilde{\sigma}_{NC}(x,Q^2)$
is compared with the H1 97 PDF Fit for three different $Q^2$
values (a).  The $e^-p$ data with $\sqrt{s} \approx 320$~{\rm GeV}
(solid points) are compared with the H1 $e^+p$ data at $\sqrt{s}
\approx 300$~{\rm GeV} (open points).  The structure function $x\Fz$
is compared with the H1 97 PDF Fit (b).  The structure function
$x\Fzint$ is compared with the H1 97 PDF Fit (c). In all figures the
inner error bars represent the statistical error, and the outer error
bars show the total error. The normalisation uncertainties of the
$e^-p$ and $e^+p$ data sets are included in the systematic errors.}
\label{xf3}
\end{figure}

In order to optimise the sensitivity to $x\Fz$, both the $e^+p$ and
the $e^-p$ data samples with $Q^2$ greater than the bin boundary at
$Q^2=1\,125$~GeV$^2$ are rebinned into three $Q^2$ bins with centres
$Q^2=1\,500$~GeV$^2$, $5\,000$~GeV$^2$ and $12\,000$~GeV$^2$. The
reduced cross section measured in these bins is shown in
fig.~\ref{xf3}(a), where the $e^-p$ reduced cross section is seen to
be significantly higher than the $e^+p$ reduced cross section in many
of the ($x,Q^2$) bins shown. The structure function $x\Fz$ is then
evaluated using the equation
\begin{eqnarray}
\tilde{\sigma}^{-}_{NC} - \tilde{\sigma}^{+}_{NC} & = &
x\Fz \left[ 
\frac{ Y_{-~920}} {Y_{+~920}} + \frac{Y_{-~820}} {Y_{+~820}}
\right]
- \FL \left[ \frac{y^2_{920}}{Y_{+~920}} - \frac{y^2_{820}}{Y_{+~820}} 
\right]\;\;\;, 
\end{eqnarray}
where $y_{920}$ and $y_{820}$ are the inelasticities, and $Y_{\pm~920}$
and $Y_{\pm~820}$ are the helicity functions (see section
\ref{sec:theory}) evaluated for fixed $x$ and $Q^2$ for the given
proton beam energy $920\,{\rm GeV}$ and $820\,{\rm GeV}$.
The contribution of $\FL$ was estimated from the QCD fit and is found
to be $\simeq 10\%$ at the lowest $x$ and negligible elsewhere.  The
resulting generalised structure function $x\Fz$, shown in
fig.~\ref{xf3}(b) as a function of $x$ is expected to rise with $Q^2$
for fixed $x$ due to the $Z^0$ propagator factor (see eq.~\ref{f3p}). 
At high $x$ and low $Q^2$ the data are insensitive to $x\Fz$, and
therefore the corresponding points are removed.

%----------------------------------------------------------------
%----------------------------------------------------------------

The structure function $x\Fzint$ is evaluated by dividing $x\Fz$ by
the factor \mbox{$-a_e\kappa_wQ^2/(Q^2+M_Z^2)$}. The contribution of
$x\Fzwk$ is estimated to be less than $3\%$ at the highest $Q^2$ and
so is neglected. The measurement of $x\Fzint$ is shown in
fig.~\ref{xf3}(c) as a function of $x$ for three values of $Q^2$. The
change of $x\Fzint$ at fixed $x$ over the measured $Q^2$ range is
expected to be very small, because it arises only from QCD scaling
violations for a non-singlet structure function. It is thus reasonable
to directly compare $x\Fzint$ at the different $Q^2$ values.

The measurement of $x\Fzint$ is the first at high $Q^2$ and also
extends to lower $x$ than previous measurements. It has the advantage
compared to fixed target determinations~\cite{ccfr} in that the target
is a proton, and therefore there are no corrections for nuclear
effects. It should be noted that due to the quark charges and
couplings (see eq.~\ref{eq:xf3}) $x\Fzint$ measured in $ep$
interactions is not the same quantity as $xF_3^\nu$ as measured in
$\nu N$ interactions.

The results in fig.~\ref{xf3}(c) are consistent with zero at large $x$,
rising to $\sim 0.7$ at $x \simeq 0.1$. At lower $x$ the 
data are consistent with expectation albeit with large errors.
These observations are in agreement with the expectations from QCD in
which $x\Fzint$ is dependent on the difference between quark and
anti-quark densities (see eq.~\ref{eq:xf3}) and is therefore primarily
sensitive to the valence quark distributions. The QCD expectation for
$x\Fzint$, which is also shown in fig.~\ref{xf3}(c), is seen to rise to
a maximum at $x \approx 0.1$. The data are found to be in agreement
with the expectation at $x \gapprox 0.2$, but lie above at lower $x$
values. In order to quantify the level of agreement of the data and
the expectation the sum rule
\begin{equation} 
\int_0^1 \Fzint {\rm d}x = 2 e_u a_u N_u + 2 e_d a_d N_d=\frac{5}{3} \cdot 
{\cal O}(1-\alpha_s/\pi)
\label{sumrule}
\end{equation}
is determined~\cite{H1sumrule} by analogy with the Gross
Llewellyn-Smith sum rule~\cite{gls} for neutrino scattering which has
been found to be valid \cite{CCFRsumrule}. The sum rule in
eq.~\ref{sumrule} follows from eq.~\ref{eq:xf3} by replacing the
differences between the quark and anti-quark distributions by the
valence distributions which, when integrated yield $N_u$ and $N_d$,
the numbers of $u$ and $d$ valence quarks, respectively. The term
${\cal O}(1-\alpha_s/\pi)$ represents the QCD radiative
corrections~\cite{QCDrc}. The H1 data at fixed $x$ are averaged by
taking weighted means, and integrated yielding
$$\int_{0.02}^{0.65} \Fzint {\rm d}x = 1.88 \pm 0.35 (\rm stat.) \pm 0.27 (\rm syst.).$$ \\ 
The integral obtained from the H1 97 PDF Fit gives
$\int_{0.02}^{0.65}~\Fzint~dx~=~1.11$ and when integrated over the full
range in $x$ is found\footnote{Note that the conditions $N_u=2$ and $N_d=1$ were
constraints of the H1 97 PDF Fit.} to be $\int_{0}^1~\Fzint~dx~=~1.59$.
 The data and expectation are
found to agree within less than two standard deviations.

%%%%%%%%%%%%%%%%%%%%%%%%%%%%%%%%%%%%%%%%%%%%%%%%%%%%%%%%%%%%%%%%%%%%%%%%%%
\subsection{\boldmath Measurement of the Total CC Cross Section and $M_W$}
%%%%%%%%%%%%%%%%%%%%%%%%%%%%%%%%%%%%%%%%%%%%%%%%%%%%%%%%%%%%%%%%%%%%%%%%%%
The total CC cross section has been measured in the region \mbox{$Q^2>1\,000
\rm~GeV^2$} and $0.03<y<0.85$. A small correction factor for the $y$ cuts is
applied\footnote{The factor is determined to be 1.04 from the H1 97
PDF Fit.} and the cross section for the range \mbox{$Q^2>1\,000 \rm~GeV^2$} and
\mbox{$y<0.9$} is
$$\sigma^{tot}_{CC}(e^-p)~=~43.08 \pm 1.84 (\rm stat.) \pm 1.74(\rm
syst.) \rm ~pb,$$
where the $1.8\%$ normalisation uncertainty is included in the
systematic error. This is consistent with the expectation from the H1 97
PDF Fit where \mbox{$\sigma^{tot}_{CC}(e^-p)~=~42.70~\pm~1.65~\rm~pb.$}

Within the Standard Model CC interactions are mediated by the
$t$-channel exchange of a $W$ boson, and therefore, are sensitive to
the $W$ mass in the space-like regime. Recent determinations of the
virtual $W$ mass from high $Q^2$ HERA data based on $e^+p$ scattering
have been published~\cite{h1hiq2,zeushiq2}. Due to the enhanced CC
cross section for $e^-p$ compared to $e^+p$ scattering the new data
presented here allow an improved measurement of the $W$ mass. The fit
procedure is defined in~\cite{h1hiq2} and uses PDFs from the H1 Low
$Q^2$ QCD Fit~\cite{h1hiq2} performed on data from NMC and BCDMS, and
low $Q^2$ $F_2$ data from H1~\cite{h1f294}. No data for
$Q^2>100$~${\rm GeV}^2$ are used in this QCD fit. The propagator mass 
$M_{W}$ of eq.~\ref{eq:cccross} is fitted to the
double differential CC cross section data, yields a $\chi^2$ per
degree of freedom of $30.4/(29-1)=1.09$ and a $W$ mass of
$$M_{W}~=~79.9 \pm 2.2 (\rm stat.) \pm 0.9 (\rm syst.)  \pm 2.1 (\rm
theo.)~\rm~GeV.$$
The Standard Model uncertainty (theory) is determined by varying the
assumptions for the Low $Q^2$ QCD Fit and is detailed
in~\cite{h1hiq2}. The weak corrections are taken into
account using the HECTOR program, but are found to have negligible
influence on the results. Despite a smaller luminosity than for $e^+p$, the
larger cross section of the $e^-p$ data yield a more precise
measurement of the $W$ mass.

%%%%%%%%%%%%%%%%%%%%%%%%%%%%%%%%%%%%%%%%%%%%%%%%%%%%%%%%%%%%%%%%%%%%%%%%%%
\section{Summary}
%%%%%%%%%%%%%%%%%%%%%%%%%%%%%%%%%%%%%%%%%%%%%%%%%%%%%%%%%%%%%%%%%%%%%%%%%%
Cross sections in NC and CC interactions have been measured
for $e^-p$ scattering at a centre-of-mass energy of $\sqrt{s} \approx
320$~GeV which complement earlier $e^+p$ measurements~\cite{h1hiq2}. 
Standard Model expectations for deep inelastic scattering derived from
the H1 97 PDF Fit to NMC, BCDMS and H1 $e^+p$ data provide a good
description of the measured $e^- p$ cross sections thus corroborating
the universality of the underlying theory.

The double differential CC reduced cross section is presented for the
range $300 \leq Q^2 \leq 15\,000$ GeV$^2$, and $0.013 \leq x \leq
0.4$.  The $Q^2$ dependence of the CC cross section is used
to measure the $W$ propagator mass in the space-like regime.  The
value obtained of $M_W=79.9 \pm 3.2$~GeV is in good agreement with
direct measurements in the time-like domain from LEP~\cite{lepmw} and
the TEVATRON~\cite{tevatronmw} thus supporting the overall consistency
of the Standard Model description.

Detailed comparisons of the $e^-p$ cross sections with the H1
measurements of NC and CC $e^+p$ cross sections are made.  The NC
$e^-p$ measurement of ${\rm d}\sigma/{\rm d}Q^2$ shows a clear
increase with respect to $e^+p$ scattering cross sections at high
$Q^2$, consistent with the expectation of the contribution of $Z^0$
exchange.  At \qsq\ $>10\,000$ GeV$^2$ the cross
section is found to be approximately four times larger than for $e^+p$
scattering.  The CC cross section at high $Q^2$ is observed to be
larger for electron scattering than for positron scattering by up to a
factor of ten.  The major part is due to different quark flavours and
helicities entering the $e^-p$ and $e^+p$ cross sections.  The
influence of the larger centre-of-mass energy is responsible for only
a small part of the increase in the NC and CC cross sections.

The double differential NC reduced cross sections are measured in the
range $150 \leq Q^2 \leq 30\,000$ GeV$^2$, and $0.002 \leq x \leq
0.65$.  The data agree well with measurements for $e^+p$ scattering
for $Q^2<1\,000$ GeV$^2$.  At higher $Q^2$ the expected difference of
the cross sections due to $Z^0$ exchange is observed and the
generalised structure function $x\Fz$~ is measured in the range
$0.02<x<0.65$ and $1\,500 \leq Q^2 \leq 12\,000$ GeV$^2$.  This
structure function is dominated by the interference of $\gamma$ and
$Z^0$ which constitutes a probe of the valence quark structure of the
proton.  The structure function, $\Fzint$, is explicitly derived from
the measurement.  The integral $\int_{0.02}^{0.65} \Fzint {\rm d}x$ is
evaluated and found to be consistent with expectation within
experimental errors.

%%%%%%%%%%%%%%%%%%%%%%%%%%%%%%%%%%%%%%%%%%%%%%%%%%%%%%%%%%%
\section*{Acknowledgements}
%%%%%%%%%%%%%%%%%%%%%%%%%%%%%%%%%%%%%%%%%%%%%%%%%%%%%%%%%%%
We are grateful to the HERA machine group whose outstanding efforts
have made and continue to make this experiment possible.  We thank the
engineers and technicians for their work in constructing and now
maintaining the H1 detector, our funding agencies for financial
support, the DESY technical staff for continual assistance, and the
DESY directorate for the hospitality which they extend to the non-DESY
members of the collaboration.

%%%%%%%%%%%%%%%%%%%%%%%%%%%%%%%%%%%%%%%%%%%%%%%%%%%%%%%%%%%

\clearpage
 \begin{table}[htb]
 \begin{center}
 \begin{tabular}{|c|c||r|r|r|r|}
 \hline
  $x$  &
 ${\rm d}\sigma_{NC}/{\rm d}x~(\rm pb )$ &
 $\delta_{sta}$ & $\delta_{unc}$ & $\delta_{cor}$ &
 $\delta_{tot}$ \\
  &$Q^2>1\,000~\rm GeV^2, y<0.9$ &
 $(\%)$ &$(\%)$ &$(\%)$ &$(\%)$ \\
 \hline
$0.013$&$0.126 \cdot 10^{ 4}$&$ 13.6$&$  5.7$&$  4.6$&$ 15.5$\\
$0.020$&$0.276 \cdot 10^{ 4}$&$  6.0$&$  3.1$&$  1.9$&$  7.0$\\
$0.032$&$0.246 \cdot 10^{ 4}$&$  4.9$&$  2.9$&$  1.0$&$  5.8$\\
$0.050$&$0.172 \cdot 10^{ 4}$&$  4.8$&$  3.0$&$  1.1$&$  5.8$\\
$0.080$&$0.115 \cdot 10^{ 4}$&$  4.5$&$  3.1$&$  1.1$&$  5.6$\\
$0.130$&$0.636 \cdot 10^{ 3}$&$  5.4$&$  3.5$&$  1.0$&$  6.5$\\
$0.180$&$0.383 \cdot 10^{ 3}$&$  6.2$&$  3.3$&$  1.3$&$  7.1$\\
$0.250$&$0.221 \cdot 10^{ 3}$&$  6.9$&$  6.0$&$  2.8$&$  9.6$\\
$0.400$&$0.693 \cdot 10^{ 2}$&$  9.8$&$  9.1$&$  4.2$&$ 14.0$\\
$0.650$&$0.421 \cdot 10^{ 1}$&$ 24.3$&$ 14.8$&$  6.3$&$ 29.2$\\
 \hline
 \end{tabular}
 \end{center}
 \caption[RESULT]
 {\sl \label{ncdx1}
 The NC cross section
 ${\rm d}\sigma_{NC}/{\rm d}x$ measured for $y<0.9$
 and $Q^2>1\,000$~GeV$^2$.
 The statistical $(\delta_{sta})$,
 uncorrelated systematic $(\delta_{unc})$,
 correlated systematic $(\delta_{cor})$,
 and total $(\delta_{tot})$ errors are also given.
 The normalisation uncertainty of 1.8\% is
 not included in the errors.}
 \end{table}

 \begin{table}[htb]
 \begin{center}
 \begin{tabular}{|c|c||r|r|r|r|}
 \hline
  $x$  &
 ${\rm d}\sigma_{NC}/{\rm d}x~(\rm pb )$ &
 $\delta_{sta}$ & $\delta_{unc}$ & $\delta_{cor}$ &
 $\delta_{tot}$ \\
  &$Q^2>10\,000~\rm GeV^2, y<0.9$ &
 $(\%)$ &$(\%)$ &$(\%)$ &$(\%)$ \\
 \hline
$0.130$&$0.473 \cdot 10^{ 1}$&$ 82.9$&$ 19.3$&$  5.2$&$ 85.3$\\
$0.180$&$0.156 \cdot 10^{ 2}$&$ 26.9$&$  6.5$&$  3.1$&$ 27.9$\\
$0.250$&$0.110 \cdot 10^{ 2}$&$ 25.2$&$  7.2$&$  2.1$&$ 26.3$\\
$0.400$&$0.354 \cdot 10^{ 1}$&$ 30.2$&$ 13.7$&$  3.5$&$ 33.3$\\
$0.650$&$0.024 \cdot 10^{ 1}$&$ 70.7$&$ 33.9$&$ 13.4$&$ 79.6$\\
 \hline
 \end{tabular}
 \end{center}
 \caption[RESULT]
 {\sl \label{ncdx2}
 The NC cross section
 ${\rm d}\sigma_{NC}/{\rm d}x$ measured for $y<0.9$
 and $Q^2>10\,000$~GeV$^2$.
 The statistical $(\delta_{sta})$,
 uncorrelated systematic $(\delta_{unc})$,
 correlated systematic $(\delta_{cor})$,
 and total $(\delta_{tot})$ errors are also given.
 The normalisation uncertainty of 1.8\% is
 not included in the errors.}
 \end{table}

\begin{table}[htb]
  \begin{center}
    \begin{tabular}{|c|c||c|c|c|c|c|}
      \hline
      $x$  &$ {\rm d} \sigma_{CC}/{\rm d} x~(\rm pb)$  & $k_{\rm cor}$ &
      $\delta_{sta}$ & $\delta_{unc}$ &
      $\delta_{cor}$ &$\delta_{tot}$ \htab \\
      & $Q^2>1\,000\,{\rm GeV}^2, y<0.9$ &
      &$(\%)$ & $(\%)$ &$(\%)$
      & $(\%)$ \\
\hline

$0.032$ & $0.273\cdot 10^{ 3}
$ & $1.064$ & $ 11.0$ & $  4.6$ & $  2.4$ & $ 12.1$ \\
$0.080$ & $0.194\cdot 10^{ 3}
$ & $1.033$ & $  7.1$ & $  3.9$ & $  1.2$ & $  8.2$ \\
$0.130$ & $0.131\cdot 10^{ 3}
$ & $1.023$ & $  7.3$ & $  3.4$ & $  1.0$ & $  8.1$ \\
$0.250$ & $0.686\cdot 10^{ 2}
$ & $1.014$ & $  7.4$ & $  3.9$ & $  1.1$ & $  8.4$ \\
$0.400$ & $0.201\cdot 10^{ 2}
$ & $1.051$ & $ 14.1$ & $ 11.0$ & $  4.8$ & $ 18.5$ \\
$0.650$ & $0.264\cdot 10^{ 1}
$ & $1.197$ & $ 49.8$ & $ 21.1$ & $ 16.9$ & $ 56.7$ \\

\hline
    \end{tabular}
  \end{center}
\caption[RESULT]
{\sl \label{ccdx} The CC cross-section ${\rm d} \sigma_{CC}/{\rm d} x$
measured for $Q^2>1\,000\,{\rm GeV}^2$ and $0.03 < y<0.85 $
after correction ($k_{\rm cor}$) to $y < 0.9$.
The  statistical ($\delta_{sta}$),
uncorrelated systematic ($\delta_{unc}$),  
correlated  systematic ($\delta_{cor}$),
and total ($\delta_{tot}$) errors are also given.
The normalisation uncertainty of 1.8\% is not included in the errors.}
\end{table}                                                

 \begin{table}[htb]
 \begin{center}
 \begin{tabular}{|r|l|l||r|r|r|r|}
 \hline
 $Q^2$  &
 ${\rm d}\sigma_{NC}/{\rm d}Q^2$ &
 $k_{\rm cor}$ &
 $\delta_{sta}$ & $\delta_{unc}$ & $\delta_{cor}$ &
 $\delta_{tot}$ \\
 $(\rm GeV^2)$ &$(\rm pb / \rm GeV^2)$ &
      & 
 $(\%)$ &$(\%)$ &$(\%)$ &$(\%)$ \\
  & $y < 0.9$ & $   $ & & & & \\
 \hline
$  200$&$1.822 \cdot 10^{ 1}$&$1.078$&$  1.3$&$  3.0$&$  1.1$&$  3.4$\\
$  250$&$1.067 \cdot 10^{ 1}$&$1.075$&$  1.3$&$  3.2$&$  1.7$&$  3.9$\\
$  300$&$7.109 \cdot 10^{ 0}$&$1.070$&$  1.5$&$  3.4$&$  1.7$&$  4.1$\\
$  400$&$3.536 \cdot 10^{ 0}$&$1.060$&$  1.7$&$  2.8$&$  1.3$&$  3.6$\\
$  500$&$2.062 \cdot 10^{ 0}$&$1.048$&$  2.1$&$  3.5$&$  2.2$&$  4.6$\\
$  650$&$1.103 \cdot 10^{ 0}$&$1.031$&$  2.5$&$  3.3$&$  1.7$&$  4.5$\\
$  800$&$0.683 \cdot 10^{ 0}$&$1.013$&$  3.0$&$  3.2$&$  1.1$&$  4.5$\\
$ 1000$&$0.372 \cdot 10^{ 0}$&$1.000$&$  3.4$&$  3.1$&$  1.3$&$  4.8$\\
$ 1200$&$0.245 \cdot 10^{ 0}$&$1.000$&$  3.9$&$  3.0$&$  0.9$&$  5.0$\\
$ 1500$&$0.132 \cdot 10^{ 0}$&$1.000$&$  4.8$&$  3.0$&$  1.0$&$  5.7$\\
$ 2000$&$0.615 \cdot 10^{-1}$&$1.000$&$  5.8$&$  3.3$&$  1.7$&$  6.8$\\
$ 3000$&$0.239 \cdot 10^{-1}$&$1.000$&$  5.0$&$  3.1$&$  0.9$&$  6.0$\\
$ 5000$&$0.675 \cdot 10^{-2}$&$1.000$&$  6.1$&$  3.5$&$  0.8$&$  7.1$\\
$ 8000$&$0.191 \cdot 10^{-2}$&$1.000$&$  9.1$&$  5.7$&$  1.0$&$ 10.8$\\
$12000$&$0.395 \cdot 10^{-3}$&$1.000$&$ 18.2$&$  6.7$&$  1.3$&$ 19.4$\\
$20000$&$0.901 \cdot 10^{-4}$&$1.000$&$ 30.6$&$ 10.2$&$  1.4$&$ 32.2$\\
$30000$&$0.204 \cdot 10^{-4}$&$1.000$&$ 58.5$&$ 21.8$&$  3.0$&$ 62.5$\\
 \hline
 \end{tabular}
 \end{center}
 \caption[RESULT]
 {\sl \label{ncdq2} The NC cross section 
 ${\rm d}\sigma_{NC}/{\rm d}Q^2$ for $y<0.9$ after
 correction ($k_{\rm cor}$) according to the 
 Standard Model expectation for
 the kinematic cuts $y<0.63$
 for $Q^2<890$~GeV$^2$. The
 statistical $(\delta_{sta})$,
 uncorrelated systematic $(\delta_{unc})$,
 correlated systematic $(\delta_{cor})$,
 and total $(\delta_{tot})$ errors are also given.
 The normalisation uncertainty of 1.8\% is
 not included in the errors.}
 \end{table}

\begin{table}[htb]
  \begin{center}
    \begin{tabular}{|c|c||*{5}{c|}|c|}
      \hline
      $Q^2$  &$ {\rm d} \sigma_{CC}/{\rm d} Q^2$  &
      $k_{\rm cor}$ &
      $\delta_{sta}$ & $\delta_{unc}$ &
      $\delta_{cor}$ &$\delta_{tot}$ & $\delta_{qed}$ \htab \\
      $( \rm GeV^2)$ &$(\rm pb/GeV^2)$    
      & 
      &\footnotesize $(\%)$ &\footnotesize  $(\%)$ &\footnotesize  $(\%)$ 
      &\footnotesize $(\%)$ &\footnotesize $(\%)$ \\
       & \footnotesize $y<0.9$  &  &  &  &  &  & \\
\hline

$  300$ & $0.389\cdot 10^{-1}$ & $1.495
$ & $ 18.8$ & $  7.2$ & $  9.5$ & $ 22.3$ & $  4.2$ \\
$  500$ & $0.292\cdot 10^{-1}$ & $1.245
$ & $ 11.7$ & $  5.3$ & $  4.3$ & $ 13.6$ & $  2.1$ \\
$ 1000$ & $0.158\cdot 10^{-1}$ & $1.070
$ & $  9.7$ & $  4.2$ & $  2.0$ & $ 10.7$ & $ -0.6$ \\
$ 2000$ & $0.102\cdot 10^{-1}$ & $1.024
$ & $  7.8$ & $  4.0$ & $  1.4$ & $  8.8$ & $ -1.5$ \\
$ 3000$ & $0.667\cdot 10^{-2}$ & $1.026
$ & $  7.3$ & $  4.2$ & $  0.6$ & $  8.4$ & $ -0.8$ \\
$ 5000$ & $0.292\cdot 10^{-2}$ & $1.034
$ & $  8.8$ & $  4.3$ & $  0.9$ & $  9.9$ & $ -4.2$ \\
$ 8000$ & $0.140\cdot 10^{-2}$ & $1.046
$ & $ 10.8$ & $  6.6$ & $  2.5$ & $ 12.9$ & $ -8.0$ \\
$15000$ & $0.336\cdot 10^{-3}$ & $1.080
$ & $ 16.3$ & $ 11.2$ & $  4.8$ & $ 20.3$ & $-13.6$ \\
$30000$ & $0.202\cdot 10^{-4}$ & $1.183
$ & $ 44.6$ & $ 18.7$ & $ 10.9$ & $ 49.6$ & $-20.7$ \\

\hline
    \end{tabular}
  \end{center}
\caption{\sl The CC cross-section ${\rm d} \sigma_{CC}/{\rm d} Q^2$ for $y<0.9$ after 
correction $(k_{\rm cor})$ according to Standard Model expectations for
kinematic cuts $0.03<y<0.85 $ and $P_{T,h}>12\,{\rm{GeV}}$.
The statistical ($\delta_{sta}$), uncorrelated systematic ($\delta_{unc}$),
correlated systematic ($\delta_{cor}$), and total ($\delta_{tot}$) errors
are also given. The final column gives the correction for QED radiative effets
$\delta^{qed}_{CC}$. 
The normalisation uncertainty of 1.8\% is not included in the errors.}
\label{ccdq2}
\end{table}                                                

 \begin{table}[htb]
 \begin{center}
 \begin{tabular}{|r|c|c||c|r|r|r||c|r|r|r|r|}
 \hline
 $Q^2$  &$x$ &$y$ & $\tilde{\sigma}_{NC}$ &
 $\delta_{sta}$ & $\delta_{sys}$ & $\delta_{tot}$
 & $F_2$ & $\Delta_{all}$&
 $\Delta_{F_2}$& $\Delta_{F_3}$& $\Delta_{F_L}$ \\
 $(\rm GeV^2)$ & & & &
 $(\%)$ &$(\%)$ &$(\%)$ & &$(\%)$ &$(\%)$ &$(\%)$ &
 $(\%)$ \\ \hline \hline
$  150$&$0.0032$&$0.462$&$1.218$&$  2.7$&$  3.8$&$  4.7$&$1.253$&$ -2.8$&$  0.1$&$  0.1$&$ -3.0  $\\
$  150$&$0.0050$&$0.295$&$1.154$&$  2.8$&$  3.4$&$  4.4$&$1.164$&$ -0.8$&$  0.1$&$  0.1$&$ -1.0  $\\
$  150$&$0.0080$&$0.185$&$0.968$&$  4.1$&$  8.2$&$  9.1$&$0.969$&$ -0.2$&$  0.1$&$  0.0$&$ -0.3  $\\ \hline
$  200$&$0.0032$&$0.615$&$1.271$&$  4.1$&$  4.5$&$  6.1$&$1.344$&$ -5.4$&$  0.2$&$  0.1$&$ -5.7  $\\
$  200$&$0.0050$&$0.394$&$1.107$&$  2.8$&$  3.6$&$  4.6$&$1.125$&$ -1.6$&$  0.2$&$  0.1$&$ -1.8  $\\
$  200$&$0.0080$&$0.246$&$0.915$&$  3.0$&$  3.3$&$  4.5$&$0.918$&$ -0.4$&$  0.2$&$  0.1$&$ -0.6  $\\
$  200$&$0.0130$&$0.152$&$0.860$&$  3.2$&$  3.5$&$  4.7$&$0.859$&$  0.0$&$  0.2$&$  0.1$&$ -0.2  $\\
$  200$&$0.0200$&$0.099$&$0.677$&$  3.8$&$  5.3$&$  6.5$&$0.676$&$  0.2$&$  0.2$&$  0.1$&$ -0.1  $\\
$  200$&$0.0320$&$0.062$&$0.558$&$  4.5$&$  7.4$&$  8.6$&$0.556$&$  0.2$&$  0.2$&$  0.1$&$  0.0  $\\
$  200$&$0.0500$&$0.039$&$0.506$&$  5.2$&$  8.4$&$  9.9$&$0.505$&$  0.2$&$  0.2$&$  0.1$&$  0.0  $\\
$  200$&$0.0800$&$0.025$&$0.407$&$  5.9$&$ 10.9$&$ 12.4$&$0.406$&$  0.2$&$  0.1$&$  0.0$&$  0.0  $\\ \hline
$  250$&$0.0050$&$0.492$&$1.123$&$  3.5$&$  4.0$&$  5.3$&$1.154$&$ -2.7$&$  0.2$&$  0.1$&$ -3.0  $\\
$  250$&$0.0080$&$0.308$&$1.021$&$  3.2$&$  4.2$&$  5.3$&$1.027$&$ -0.6$&$  0.2$&$  0.1$&$ -0.9  $\\
$  250$&$0.0130$&$0.189$&$0.825$&$  3.4$&$  4.5$&$  5.7$&$0.825$&$  0.0$&$  0.2$&$  0.1$&$ -0.3  $\\
$  250$&$0.0200$&$0.123$&$0.691$&$  3.5$&$  4.0$&$  5.4$&$0.689$&$  0.2$&$  0.2$&$  0.1$&$ -0.1  $\\
$  250$&$0.0320$&$0.077$&$0.569$&$  3.8$&$  4.7$&$  6.1$&$0.567$&$  0.3$&$  0.2$&$  0.1$&$  0.0  $\\
$  250$&$0.0500$&$0.049$&$0.493$&$  4.3$&$  3.7$&$  5.7$&$0.492$&$  0.3$&$  0.2$&$  0.1$&$  0.0  $\\
$  250$&$0.0800$&$0.031$&$0.407$&$  4.7$&$  3.9$&$  6.1$&$0.406$&$  0.3$&$  0.2$&$  0.1$&$  0.0  $\\
$  250$&$0.1300$&$0.019$&$0.311$&$  5.3$&$  5.8$&$  7.8$&$0.310$&$  0.2$&$  0.2$&$  0.1$&$  0.0  $\\
$  250$&$0.2500$&$0.010$&$0.225$&$  7.5$&$  9.5$&$ 12.1$&$0.224$&$  0.2$&$  0.2$&$  0.0$&$  0.0  $\\
$  250$&$0.4000$&$0.006$&$0.138$&$  9.3$&$  7.2$&$ 11.8$&$0.138$&$  0.2$&$  0.2$&$  0.0$&$  0.0  $\\ \hline
$  300$&$0.0050$&$0.591$&$1.152$&$  5.6$&$  4.6$&$  7.2$&$1.202$&$ -4.1$&$  0.3$&$  0.2$&$ -4.6  $\\
$  300$&$0.0080$&$0.369$&$1.026$&$  3.6$&$  3.6$&$  5.1$&$1.036$&$ -0.9$&$  0.3$&$  0.2$&$ -1.4  $\\
$  300$&$0.0130$&$0.227$&$0.878$&$  3.8$&$  3.7$&$  5.3$&$0.878$&$  0.0$&$  0.3$&$  0.2$&$ -0.4  $\\
$  300$&$0.0200$&$0.148$&$0.735$&$  4.0$&$  4.3$&$  5.9$&$0.733$&$  0.3$&$  0.3$&$  0.1$&$ -0.1  $\\
$  300$&$0.0320$&$0.092$&$0.605$&$  4.2$&$  4.1$&$  5.8$&$0.603$&$  0.3$&$  0.3$&$  0.1$&$  0.0  $\\
$  300$&$0.0500$&$0.059$&$0.509$&$  4.5$&$  5.1$&$  6.8$&$0.507$&$  0.4$&$  0.3$&$  0.1$&$  0.0  $\\
$  300$&$0.0800$&$0.037$&$0.390$&$  5.2$&$  4.6$&$  6.9$&$0.389$&$  0.3$&$  0.2$&$  0.1$&$  0.0  $\\
$  300$&$0.1300$&$0.023$&$0.332$&$  5.4$&$  7.0$&$  8.8$&$0.331$&$  0.3$&$  0.2$&$  0.1$&$  0.0  $\\
$  300$&$0.2500$&$0.012$&$0.277$&$  6.9$&$ 10.8$&$ 12.8$&$0.277$&$  0.3$&$  0.2$&$  0.0$&$  0.0  $\\
$  300$&$0.4000$&$0.007$&$0.143$&$ 10.3$&$  9.8$&$ 14.2$&$0.142$&$  0.2$&$  0.2$&$  0.0$&$  0.0  $\\ \hline
$  400$&$0.0080$&$0.492$&$1.088$&$  4.5$&$  4.1$&$  6.1$&$1.109$&$ -1.9$&$  0.4$&$  0.3$&$ -2.6  $\\
$  400$&$0.0130$&$0.303$&$0.897$&$  4.3$&$  3.6$&$  5.6$&$0.898$&$ -0.1$&$  0.4$&$  0.3$&$ -0.8  $\\
$  400$&$0.0200$&$0.197$&$0.732$&$  4.5$&$  3.6$&$  5.8$&$0.729$&$  0.4$&$  0.4$&$  0.3$&$ -0.3  $\\
$  400$&$0.0320$&$0.123$&$0.560$&$  4.8$&$  3.8$&$  6.1$&$0.557$&$  0.5$&$  0.4$&$  0.2$&$ -0.1  $\\
$  400$&$0.0500$&$0.079$&$0.514$&$  5.0$&$  3.7$&$  6.3$&$0.511$&$  0.5$&$  0.4$&$  0.2$&$  0.0  $\\
$  400$&$0.0800$&$0.049$&$0.429$&$  5.5$&$  4.3$&$  7.0$&$0.427$&$  0.5$&$  0.4$&$  0.2$&$  0.0  $\\
$  400$&$0.1300$&$0.030$&$0.352$&$  5.6$&$  5.0$&$  7.5$&$0.351$&$  0.5$&$  0.3$&$  0.1$&$  0.0  $\\
 \hline
 \end{tabular}
 \end{center}
 \caption[RESULT]
 {\sl \label{ncdxdq2} The NC reduced cross section
 $\tilde{\sigma}_{NC}(x,Q^2)$ with statistical
 $(\delta_{sta})$, systematic $(\delta_{sys})$,
 and total $(\delta_{tot})$ errors. The
 electromagnetic structure function $F_2$ is
 also given with the corrections $\Delta_{all}$,
 $\Delta_{F_2}$,$\Delta_{F_3}$,$\Delta_{F_L}$
 as defined in eq.~\ref{f2corr}. 
 The normalisation uncertainty of 1.8\% is
 not included in the errors.
 The table continues on the next 3 pages.}
 \end{table}
 \begin{table}[htb]
 \begin{center}
 \begin{tabular}{|r|c|c||c|r|r|r||c|r|r|r|r|}
 \hline
 $Q^2$  &$x$ &$y$ & $\tilde{\sigma}_{NC}$ &
 $\delta_{sta}$ & $\delta_{sys}$ & $\delta_{tot}$
 & $F_2$ & $\Delta_{all}$&
 $\Delta_{F_2}$& $\Delta_{F_3}$& $\Delta_{F_L}$ \\
 $(\rm GeV^2)$ & & & &
 $(\%)$ &$(\%)$ &$(\%)$ & &$(\%)$ &$(\%)$ &$(\%)$ &
 $(\%)$ \\ \hline \hline
$  400$&$0.2500$&$0.016$&$0.240$&$  7.6$&$  7.4$&$ 10.6$&$0.239$&$  0.4$&$  0.3$&$  0.1$&$  0.0  $\\
$  400$&$0.4000$&$0.010$&$0.143$&$ 10.8$&$  8.4$&$ 13.7$&$0.143$&$  0.3$&$  0.3$&$  0.1$&$  0.0  $\\ \hline
$  500$&$0.0080$&$0.615$&$1.044$&$  7.8$&$  5.1$&$  9.3$&$1.080$&$ -3.3$&$  0.5$&$  0.5$&$ -4.3  $\\
$  500$&$0.0130$&$0.379$&$1.003$&$  5.1$&$  4.5$&$  6.8$&$1.006$&$ -0.3$&$  0.5$&$  0.4$&$ -1.2  $\\
$  500$&$0.0200$&$0.246$&$0.765$&$  5.1$&$  4.8$&$  7.0$&$0.761$&$  0.5$&$  0.5$&$  0.4$&$ -0.4  $\\
$  500$&$0.0320$&$0.154$&$0.604$&$  5.3$&$  4.5$&$  7.0$&$0.600$&$  0.7$&$  0.5$&$  0.4$&$ -0.1  $\\
$  500$&$0.0500$&$0.099$&$0.517$&$  5.6$&$  4.0$&$  6.9$&$0.513$&$  0.8$&$  0.5$&$  0.3$&$  0.0  $\\
$  500$&$0.0800$&$0.062$&$0.392$&$  6.4$&$  6.5$&$  9.2$&$0.389$&$  0.7$&$  0.5$&$  0.3$&$  0.0  $\\
$  500$&$0.1300$&$0.038$&$0.363$&$  7.2$&$  4.9$&$  8.7$&$0.361$&$  0.6$&$  0.4$&$  0.2$&$  0.0  $\\
$  500$&$0.1800$&$0.027$&$0.283$&$  8.2$&$  8.1$&$ 11.5$&$0.281$&$  0.6$&$  0.4$&$  0.2$&$  0.0  $\\
$  500$&$0.2500$&$0.020$&$0.254$&$ 10.5$&$  9.5$&$ 14.2$&$0.253$&$  0.5$&$  0.4$&$  0.1$&$  0.0  $\\
$  500$&$0.4000$&$0.012$&$0.139$&$ 15.4$&$ 15.1$&$ 21.6$&$0.138$&$  0.5$&$  0.4$&$  0.1$&$  0.0  $\\
$  500$&$0.6500$&$0.008$&$0.026$&$ 19.6$&$ 10.9$&$ 22.4$&$0.026$&$  0.4$&$  0.4$&$  0.1$&$  0.0  $\\ \hline
$  650$&$0.0130$&$0.492$&$0.988$&$  6.0$&$  4.1$&$  7.3$&$0.995$&$ -0.7$&$  0.7$&$  0.8$&$ -2.2  $\\
$  650$&$0.0200$&$0.320$&$0.791$&$  6.3$&$  4.4$&$  7.7$&$0.785$&$  0.7$&$  0.7$&$  0.7$&$ -0.7  $\\
$  650$&$0.0320$&$0.200$&$0.684$&$  6.1$&$  4.3$&$  7.4$&$0.677$&$  1.1$&$  0.7$&$  0.6$&$ -0.2  $\\
$  650$&$0.0500$&$0.128$&$0.538$&$  6.5$&$  5.2$&$  8.3$&$0.532$&$  1.2$&$  0.7$&$  0.5$&$ -0.1  $\\
$  650$&$0.0800$&$0.080$&$0.436$&$  7.1$&$  5.8$&$  9.2$&$0.431$&$  1.1$&$  0.7$&$  0.4$&$  0.0  $\\
$  650$&$0.1300$&$0.049$&$0.343$&$  8.8$&$  5.8$&$ 10.5$&$0.339$&$  1.0$&$  0.6$&$  0.3$&$  0.0  $\\
$  650$&$0.1800$&$0.036$&$0.330$&$  9.1$&$  7.5$&$ 11.8$&$0.327$&$  0.9$&$  0.6$&$  0.3$&$  0.0  $\\
$  650$&$0.2500$&$0.026$&$0.251$&$ 11.9$&$ 10.6$&$ 15.9$&$0.249$&$  0.8$&$  0.6$&$  0.2$&$  0.0  $\\
$  650$&$0.4000$&$0.016$&$0.090$&$ 22.9$&$  9.6$&$ 24.9$&$0.090$&$  0.7$&$  0.5$&$  0.1$&$  0.0  $\\ \hline
$  800$&$0.0130$&$0.606$&$0.842$&$ 10.2$&$  5.8$&$ 11.7$&$0.854$&$ -1.4$&$  1.0$&$  1.1$&$ -3.5  $\\
$  800$&$0.0200$&$0.394$&$0.806$&$  7.2$&$  4.9$&$  8.8$&$0.799$&$  0.9$&$  1.0$&$  1.0$&$ -1.1  $\\
$  800$&$0.0320$&$0.246$&$0.721$&$  7.1$&$  5.0$&$  8.7$&$0.709$&$  1.6$&$  1.0$&$  0.9$&$ -0.3  $\\
$  800$&$0.0500$&$0.158$&$0.587$&$  7.4$&$  4.4$&$  8.6$&$0.577$&$  1.6$&$  0.9$&$  0.8$&$ -0.1  $\\
$  800$&$0.0800$&$0.099$&$0.518$&$  7.8$&$  5.2$&$  9.4$&$0.510$&$  1.5$&$  0.9$&$  0.7$&$  0.0  $\\
$  800$&$0.1300$&$0.061$&$0.411$&$ 10.0$&$  6.2$&$ 11.8$&$0.406$&$  1.3$&$  0.9$&$  0.5$&$  0.0  $\\
$  800$&$0.1800$&$0.044$&$0.302$&$ 11.6$&$  6.7$&$ 13.4$&$0.298$&$  1.2$&$  0.8$&$  0.4$&$  0.0  $\\
$  800$&$0.2500$&$0.032$&$0.212$&$ 14.1$&$  8.2$&$ 16.4$&$0.210$&$  1.1$&$  0.8$&$  0.3$&$  0.0  $\\
$  800$&$0.4000$&$0.020$&$0.117$&$ 20.9$&$ 12.6$&$ 24.4$&$0.116$&$  0.9$&$  0.7$&$  0.2$&$  0.0  $\\
$  800$&$0.6500$&$0.012$&$0.015$&$ 21.8$&$ 14.9$&$ 26.5$&$0.015$&$  0.8$&$  0.7$&$  0.1$&$  0.0  $\\ \hline
$ 1000$&$0.0130$&$0.757$&$0.773$&$ 11.5$&$  6.9$&$ 13.5$&$0.795$&$ -2.8$&$  1.4$&$  1.7$&$ -5.8  $\\
$ 1000$&$0.0200$&$0.492$&$0.787$&$  7.9$&$  4.7$&$  9.2$&$0.778$&$  1.2$&$  1.4$&$  1.6$&$ -1.8  $\\
$ 1000$&$0.0320$&$0.308$&$0.572$&$  9.0$&$  4.4$&$ 10.0$&$0.560$&$  2.3$&$  1.3$&$  1.4$&$ -0.5  $\\
$ 1000$&$0.0500$&$0.197$&$0.577$&$  8.4$&$  4.5$&$  9.5$&$0.564$&$  2.4$&$  1.3$&$  1.2$&$ -0.2  $\\
$ 1000$&$0.0800$&$0.123$&$0.450$&$  9.3$&$  5.6$&$ 10.8$&$0.440$&$  2.2$&$  1.2$&$  1.0$&$  0.0  $\\
$ 1000$&$0.1300$&$0.076$&$0.491$&$ 10.3$&$  5.3$&$ 11.6$&$0.482$&$  1.9$&$  1.2$&$  0.8$&$  0.0  $\\
 \hline
 \end{tabular}
 \end{center}
 \end{table}
 \begin{table}[htb]
 \begin{center}
 \begin{tabular}{|r|c|c||c|r|r|r||c|r|r|r|r|}
 \hline
 $Q^2$  &$x$ &$y$ & $\tilde{\sigma}_{NC}$ &
 $\delta_{sta}$ & $\delta_{sys}$ & $\delta_{tot}$
 & $F_2$ & $\Delta_{all}$&
 $\Delta_{F_2}$& $\Delta_{F_3}$& $\Delta_{F_L}$ \\
 $(\rm GeV^2)$ & & & &
 $(\%)$ &$(\%)$ &$(\%)$ & &$(\%)$ &$(\%)$ &$(\%)$ &
 $(\%)$ \\ \hline \hline
$ 1000$&$0.1800$&$0.055$&$0.249$&$ 13.5$&$  5.7$&$ 14.6$&$0.245$&$  1.7$&$  1.1$&$  0.6$&$  0.0  $\\
$ 1000$&$0.2500$&$0.039$&$0.311$&$ 13.0$&$  9.2$&$ 15.9$&$0.306$&$  1.5$&$  1.1$&$  0.5$&$  0.0  $\\
$ 1000$&$0.4000$&$0.025$&$0.122$&$ 22.9$&$ 14.0$&$ 26.9$&$0.120$&$  1.3$&$  1.0$&$  0.3$&$  0.0  $\\ \hline
$ 1200$&$0.0200$&$0.591$&$0.839$&$  9.1$&$  4.0$&$ 10.0$&$0.829$&$  1.3$&$  1.8$&$  2.3$&$ -2.7  $\\
$ 1200$&$0.0320$&$0.369$&$0.719$&$  9.2$&$  3.7$&$  9.9$&$0.698$&$  3.0$&$  1.7$&$  2.1$&$ -0.7  $\\
$ 1200$&$0.0500$&$0.236$&$0.645$&$  9.3$&$  3.6$&$  9.9$&$0.624$&$  3.2$&$  1.7$&$  1.8$&$ -0.2  $\\
$ 1200$&$0.0800$&$0.148$&$0.415$&$ 10.7$&$  3.4$&$ 11.2$&$0.403$&$  3.0$&$  1.6$&$  1.4$&$ -0.1  $\\
$ 1200$&$0.1300$&$0.091$&$0.384$&$ 12.6$&$  4.5$&$ 13.4$&$0.375$&$  2.6$&$  1.5$&$  1.1$&$  0.0  $\\
$ 1200$&$0.1800$&$0.066$&$0.341$&$ 13.6$&$  5.3$&$ 14.6$&$0.333$&$  2.3$&$  1.4$&$  0.9$&$  0.0  $\\
$ 1200$&$0.2500$&$0.047$&$0.251$&$ 15.8$&$  7.0$&$ 17.3$&$0.246$&$  2.0$&$  1.4$&$  0.7$&$  0.0  $\\
$ 1200$&$0.4000$&$0.030$&$0.110$&$ 25.0$&$ 12.0$&$ 27.7$&$0.109$&$  1.7$&$  1.3$&$  0.4$&$  0.0  $\\ \hline
$ 1500$&$0.0200$&$0.738$&$0.860$&$ 12.4$&$  5.5$&$ 13.5$&$0.850$&$  1.2$&$  2.4$&$  3.4$&$ -4.6  $\\
$ 1500$&$0.0320$&$0.462$&$0.704$&$ 10.4$&$  4.7$&$ 11.4$&$0.675$&$  4.3$&$  2.4$&$  3.2$&$ -1.2  $\\
$ 1500$&$0.0500$&$0.295$&$0.515$&$ 11.7$&$  3.6$&$ 12.2$&$0.492$&$  4.7$&$  2.3$&$  2.7$&$ -0.4  $\\
$ 1500$&$0.0800$&$0.185$&$0.512$&$ 11.0$&$  4.0$&$ 11.7$&$0.490$&$  4.3$&$  2.2$&$  2.2$&$ -0.1  $\\
$ 1500$&$0.1300$&$0.114$&$0.390$&$ 13.9$&$  5.0$&$ 14.8$&$0.376$&$  3.7$&$  2.1$&$  1.7$&$  0.0  $\\
$ 1500$&$0.1800$&$0.082$&$0.260$&$ 18.6$&$  4.3$&$ 19.1$&$0.251$&$  3.3$&$  2.0$&$  1.3$&$  0.0  $\\
$ 1500$&$0.2500$&$0.059$&$0.197$&$ 19.6$&$  7.7$&$ 21.1$&$0.191$&$  2.9$&$  1.9$&$  1.0$&$  0.0  $\\
$ 1500$&$0.4000$&$0.037$&$0.145$&$ 24.3$&$ 12.8$&$ 27.4$&$0.142$&$  2.4$&$  1.7$&$  0.7$&$  0.0  $\\
$ 1500$&$0.6500$&$0.023$&$0.014$&$ 35.4$&$ 16.1$&$ 38.9$&$0.013$&$  2.0$&$  1.6$&$  0.4$&$  0.0  $\\ \hline
$ 2000$&$0.0320$&$0.615$&$0.796$&$ 11.1$&$  4.4$&$ 11.9$&$0.747$&$  6.6$&$  3.6$&$  5.4$&$ -2.4  $\\
$ 2000$&$0.0500$&$0.394$&$0.599$&$ 13.0$&$  5.0$&$ 13.9$&$0.557$&$  7.6$&$  3.5$&$  4.8$&$ -0.7  $\\
$ 2000$&$0.0800$&$0.246$&$0.582$&$ 12.3$&$  4.3$&$ 13.0$&$0.544$&$  7.0$&$  3.3$&$  3.9$&$ -0.2  $\\
$ 2000$&$0.1300$&$0.152$&$0.224$&$ 20.0$&$  4.6$&$ 20.6$&$0.212$&$  6.0$&$  3.1$&$  2.9$&$  0.0  $\\
$ 2000$&$0.1800$&$0.109$&$0.249$&$ 21.9$&$  6.3$&$ 22.7$&$0.236$&$  5.2$&$  3.0$&$  2.3$&$  0.0  $\\
$ 2000$&$0.2500$&$0.079$&$0.197$&$ 22.4$&$  6.8$&$ 23.4$&$0.188$&$  4.5$&$  2.8$&$  1.8$&$  0.0  $\\
$ 2000$&$0.4000$&$0.049$&$0.108$&$ 27.7$&$ 10.1$&$ 29.5$&$0.104$&$  3.7$&$  2.6$&$  1.1$&$  0.0  $\\ \hline
$ 3000$&$0.0500$&$0.591$&$0.606$&$ 10.6$&$  6.4$&$ 12.4$&$0.530$&$ 14.4$&$  6.0$&$ 10.0$&$ -1.7  $\\
$ 3000$&$0.0800$&$0.369$&$0.556$&$ 10.9$&$  4.5$&$ 11.8$&$0.489$&$ 13.6$&$  5.7$&$  8.3$&$ -0.4  $\\
$ 3000$&$0.1300$&$0.227$&$0.464$&$ 12.4$&$  4.0$&$ 13.0$&$0.416$&$ 11.4$&$  5.4$&$  6.2$&$ -0.1  $\\
$ 3000$&$0.1800$&$0.164$&$0.347$&$ 15.3$&$  5.1$&$ 16.1$&$0.315$&$  9.9$&$  5.1$&$  4.8$&$  0.0  $\\
$ 3000$&$0.2500$&$0.118$&$0.255$&$ 17.8$&$  7.0$&$ 19.1$&$0.235$&$  8.5$&$  4.8$&$  3.7$&$  0.0  $\\
$ 3000$&$0.4000$&$0.074$&$0.128$&$ 23.0$&$ 10.9$&$ 25.5$&$0.120$&$  6.8$&$  4.4$&$  2.4$&$  0.0  $\\ \hline
$ 5000$&$0.0800$&$0.615$&$0.707$&$ 10.6$&$  4.8$&$ 11.7$&$0.545$&$ 29.7$&$ 10.8$&$ 20.3$&$ -1.4  $\\
$ 5000$&$0.1300$&$0.379$&$0.536$&$ 13.1$&$  5.3$&$ 14.2$&$0.428$&$ 25.2$&$ 10.1$&$ 15.4$&$ -0.3  $\\
$ 5000$&$0.1800$&$0.274$&$0.442$&$ 14.0$&$  5.2$&$ 14.9$&$0.364$&$ 21.5$&$  9.6$&$ 12.0$&$ -0.1  $\\
$ 5000$&$0.2500$&$0.197$&$0.361$&$ 17.4$&$ 10.5$&$ 20.3$&$0.306$&$ 18.0$&$  9.0$&$  9.0$&$  0.0  $\\
$ 5000$&$0.4000$&$0.123$&$0.091$&$ 31.6$&$ 11.1$&$ 33.5$&$0.080$&$ 14.0$&$  8.3$&$  5.7$&$  0.0  $\\
 \hline
 \end{tabular}
 \end{center}
 \end{table}
 \begin{table}[htb]
 \begin{center}
 \begin{tabular}{|r|c|c||c|r|r|r||c|r|r|r|r|}
 \hline
 $Q^2$  &$x$ &$y$ & $\tilde{\sigma}_{NC}$ &
 $\delta_{sta}$ & $\delta_{sys}$ & $\delta_{tot}$
 & $F_2$ & $\Delta_{all}$&
 $\Delta_{F_2}$& $\Delta_{F_3}$& $\Delta_{F_L}$ \\
 $(\rm GeV^2)$ & & & &
 $(\%)$ &$(\%)$ &$(\%)$ & &$(\%)$ &$(\%)$ &$(\%)$ &
 $(\%)$ \\ \hline \hline
$ 5000$&$0.6500$&$0.076$&$0.010$&$ 41.0$&$ 18.8$&$ 45.1$&$0.009$&$ 11.1$&$  7.6$&$  3.5$&$  0.0  $\\ \hline
$ 8000$&$0.1300$&$0.606$&$0.722$&$ 16.0$&$  6.5$&$ 17.2$&$0.485$&$ 49.0$&$ 16.6$&$ 33.3$&$ -0.9  $\\
$ 8000$&$0.1800$&$0.438$&$0.386$&$ 20.4$&$  5.8$&$ 21.2$&$0.272$&$ 41.8$&$ 15.7$&$ 26.4$&$ -0.3  $\\
$ 8000$&$0.2500$&$0.315$&$0.295$&$ 21.8$&$  8.2$&$ 23.3$&$0.219$&$ 34.5$&$ 14.8$&$ 19.8$&$ -0.1  $\\
$ 8000$&$0.4000$&$0.197$&$0.197$&$ 27.7$&$ 16.8$&$ 32.4$&$0.156$&$ 26.0$&$ 13.6$&$ 12.5$&$  0.0  $\\ \hline
$12000$&$0.1800$&$0.656$&$0.471$&$ 27.8$&$  7.6$&$ 28.8$&$0.277$&$ 70.0$&$ 22.3$&$ 48.6$&$ -0.9  $\\
$12000$&$0.2500$&$0.473$&$0.298$&$ 28.9$&$  8.6$&$ 30.2$&$0.189$&$ 58.1$&$ 21.0$&$ 37.4$&$ -0.3  $\\
$12000$&$0.4000$&$0.295$&$0.083$&$ 50.0$&$ 19.6$&$ 53.7$&$0.058$&$ 42.7$&$ 19.3$&$ 23.5$&$ -0.1  $\\ \hline
$20000$&$0.2500$&$0.788$&$0.349$&$ 51.1$&$ 10.8$&$ 52.2$&$0.174$&$101.1$&$ 29.4$&$ 72.8$&$ -1.1  $\\
$20000$&$0.4000$&$0.492$&$0.182$&$ 44.7$&$ 13.3$&$ 46.7$&$0.103$&$ 76.3$&$ 27.0$&$ 49.5$&$ -0.2  $\\
$20000$&$0.6500$&$0.303$&$0.014$&$ 70.7$&$ 36.9$&$ 79.8$&$0.009$&$ 54.2$&$ 24.8$&$ 29.4$&$  0.0  $\\ \hline
$30000$&$0.4000$&$0.738$&$0.268$&$ 70.7$&$ 17.5$&$ 72.9$&$0.125$&$113.7$&$ 32.8$&$ 81.4$&$ -0.6  $\\ \hline
 \end{tabular}
 \end{center}
 \end{table}

\begin{table}[htb]
  \begin{center}
    \begin{tabular}{|c|c|c|l|c|c|c|c|r|}
      \hline
      $Q^2$  & $x$ & $y$ & ${\rm d} ^2 \sigma_{CC}/{\rm d} x{\rm d} Q^2$ & $\phi_{CC}$ &
      $\delta_{sta}$ & $\delta_{sys}$ & $\delta_{tot}$ & $\delta^{qed}_{CC}$ \\
      $(\rm GeV^2)$ & & & $(\rm{pb/GeV^2})$ & & 
      $(\%)$ & $(\%)$ & $(\%)$ & $(\%)$ \\
 \hline

$  300$ & $0.013$ & $0.227$ & \phantom{C} $0.458\cdot 10^{ 0}
$ & $0.773$ & $ 55.4$ & $ 15.7$ & $ 57.6$ & $  3.5$ \\
$  300$ & $0.032$ & $0.092$ & \phantom{C} $0.399\cdot 10^{ 0}
$ & $1.658$ & $ 24.5$ & $ 12.0$ & $ 27.3$ & $  5.6$ \\
$  300$ & $0.080$ & $0.037$ & \phantom{C} $0.690\cdot 10^{-1}
$ & $0.717$ & $ 40.7$ & $ 11.6$ & $ 42.3$ & $  7.0$ \\
\hline
$  500$ & $0.013$ & $0.379$ & \phantom{C} $0.433\cdot 10^{ 0}
$ & $0.775$ & $ 37.6$ & $ 13.3$ & $ 39.9$ & $ -0.3$ \\
$  500$ & $0.032$ & $0.154$ & \phantom{C} $0.285\cdot 10^{ 0}
$ & $1.257$ & $ 19.6$ & $  7.8$ & $ 21.0$ & $  5.0$ \\
$  500$ & $0.080$ & $0.061$ & \phantom{C} $0.790\cdot 10^{-1}
$ & $0.870$ & $ 21.8$ & $  5.1$ & $ 22.4$ & $  1.4$ \\
$  500$ & $0.130$ & $0.038$ & \phantom{C} $0.551\cdot 10^{-1}
$ & $0.986$ & $ 29.0$ & $  7.0$ & $ 29.9$ & $  3.3$ \\
\hline
$ 1000$ & $0.032$ & $0.308$ & \phantom{C} $0.186\cdot 10^{ 0}
$ & $0.941$ & $ 17.5$ & $  4.9$ & $ 18.2$ & $ -3.8$ \\
$ 1000$ & $0.080$ & $0.123$ & \phantom{C} $0.556\cdot 10^{-1}
$ & $0.703$ & $ 17.9$ & $  4.3$ & $ 18.4$ & $  2.1$ \\
$ 1000$ & $0.130$ & $0.076$ & \phantom{C} $0.310\cdot 10^{-1}
$ & $0.637$ & $ 24.0$ & $  4.6$ & $ 24.5$ & $  3.4$ \\
$ 1000$ & $0.250$ & $0.039$ & \phantom{C} $0.139\cdot 10^{-1}
$ & $0.548$ & $ 37.6$ & $ 10.6$ & $ 39.1$ & $ -1.7$ \\
\hline
$ 2000$ & $0.032$ & $0.615$ & \phantom{C} $0.132\cdot 10^{ 0}
$ & $0.859$ & $ 15.5$ & $  4.9$ & $ 16.2$ & $ -2.0$ \\
$ 2000$ & $0.080$ & $0.246$ & \phantom{C} $0.571\cdot 10^{-1}
$ & $0.929$ & $ 13.0$ & $  3.9$ & $ 13.6$ & $ -2.5$ \\
$ 2000$ & $0.130$ & $0.152$ & \phantom{C} $0.197\cdot 10^{-1}
$ & $0.521$ & $ 21.2$ & $  4.5$ & $ 21.7$ & $ -0.2$ \\
$ 2000$ & $0.250$ & $0.079$ & \phantom{C} $0.855\cdot 10^{-2}
$ & $0.435$ & $ 25.6$ & $  6.5$ & $ 26.4$ & $ -0.1$ \\
\hline
$ 3000$ & $0.080$ & $0.369$ & \phantom{C} $0.324\cdot 10^{-1}
$ & $0.659$ & $ 14.0$ & $  4.8$ & $ 14.8$ & $  0.1$ \\
$ 3000$ & $0.130$ & $0.227$ & \phantom{C} $0.250\cdot 10^{-1}
$ & $0.827$ & $ 14.0$ & $  6.1$ & $ 15.2$ & $ -0.7$ \\
$ 3000$ & $0.250$ & $0.118$ & \phantom{C} $0.749\cdot 10^{-2}
$ & $0.476$ & $ 18.9$ & $  7.0$ & $ 20.1$ & $ -1.5$ \\
$ 3000$ & $0.400$ & $0.074$ & \phantom{C} $0.251\cdot 10^{-2}
$ & $0.255$ & $ 35.2$ & $ 19.6$ & $ 40.3$ & $ -1.1$ \\
\hline
$ 5000$ & $0.080$ & $0.615$ & \phantom{C} $0.213\cdot 10^{-1}
$ & $0.637$ & $ 17.9$ & $  6.7$ & $ 19.2$ & $ -5.3$ \\
$ 5000$ & $0.130$ & $0.379$ & \phantom{C} $0.108\cdot 10^{-1}
$ & $0.525$ & $ 16.8$ & $  7.0$ & $ 18.2$ & $ -3.8$ \\
$ 5000$ & $0.250$ & $0.197$ & \phantom{C} $0.550\cdot 10^{-2}
$ & $0.512$ & $ 16.3$ & $  4.4$ & $ 16.9$ & $ -2.9$ \\
$ 5000$ & $0.400$ & $0.123$ & \phantom{C} $0.123\cdot 10^{-2}
$ & $0.183$ & $ 33.1$ & $ 13.1$ & $ 35.6$ & $ -4.9$ \\
\hline
$ 8000$ & $0.130$ & $0.606$ & \phantom{C} $0.722\cdot 10^{-2}
$ & $0.557$ & $ 18.9$ & $  9.3$ & $ 21.1$ & $ -9.9$ \\
$ 8000$ & $0.250$ & $0.315$ & \phantom{C} $0.342\cdot 10^{-2}
$ & $0.508$ & $ 16.3$ & $  6.2$ & $ 17.4$ & $ -5.0$ \\
$ 8000$ & $0.400$ & $0.197$ & \phantom{C} $0.946\cdot 10^{-3}
$ & $0.225$ & $ 28.6$ & $ 10.3$ & $ 30.4$ & $ -7.9$ \\
\hline
$15000$ & $0.250$ & $0.591$ & \phantom{C} $0.139\cdot 10^{-2}
$ & $0.453$ & $ 22.1$ & $ 16.0$ & $ 27.3$ & $-10.9$ \\
$15000$ & $0.400$ & $0.369$ & \phantom{C} $0.419\cdot 10^{-3}
$ & $0.219$ & $ 27.5$ & $ 10.7$ & $ 29.5$ & $-17.7$ \\

\hline
\end{tabular}
\caption{\sl
The CC double differential cross section ${\rm d}^2\sigma_{CC}/{\rm d}x {\rm d} Q^2$
and the structure function term $\phi_{CC}(x,Q^2)$
computed assuming $M_W=80.41\,{\rm GeV}$. Also given are the statistical
$(\delta_{sta})$, systematic $(\delta_{sys})$, and total $(\delta_{tot})$
errors. The last column gives the correction for QED radiative effects
$\delta^{qed}_{CC}$.
The normalisation uncertainty of 1.8\% is not included in the errors.}
\label{cctab}
\end{center}
\end{table}

 \begin{table}[htb]
 \begin{center}
 \tiny
 \begin{tabular}{|r|c||c|r|r||r|r|r||r|r|r|r|r|r|}
 \hline
 $Q^2$  &$x$ & $\tilde{\sigma}_{NC}$ &
 $\delta_{tot}$ & $\delta_{sta}$ & $\delta_{unc}$ &
 $\delta_{unc}^{E}$ &
 $\delta_{unc}^{h}$&
 $\delta_{cor}$ &
 $\delta_{cor}^{E^+}$ &
 $\delta_{cor}^{\theta^+}$&
 $\delta_{cor}^{h^+}$&
 $\delta_{cor}^{N^+}$&
 $\delta_{cor}^{B^+}$ \\
 $(\rm GeV^2)$ & & &
 $(\%)$ &$(\%)$ &$(\%)$ &$(\%)$ &$(\%)$ &$(\%)$ &
 $(\%)$ &$(\%)$ &$(\%)$ &$(\%)$ &$(\%)$ 
 \\\hline\hline
$  150$&$0.0032$&$1.218$&$  4.7$&$  2.7$&$  3.0$&$  0.5$&$  0.6$&$  2.3$&$  0.1$&$ -1.8$&$  1.3$&$  0.1$&$ -0.7$\\
$  150$&$0.0050$&$1.154$&$  4.4$&$  2.8$&$  3.0$&$  0.2$&$  0.5$&$  1.5$&$  0.4$&$  1.4$&$  0.3$&$  0.2$&$ -0.1$\\
$  150$&$0.0080$&$0.968$&$  9.1$&$  4.1$&$  5.4$&$  4.0$&$  1.0$&$  6.1$&$ -1.6$&$  5.9$&$ -0.7$&$  0.1$&$  0.0$\\ \hline
$  200$&$0.0032$&$1.271$&$  6.1$&$  4.1$&$  3.5$&$  0.8$&$  0.2$&$  2.9$&$ -1.0$&$ -2.5$&$  0.5$&$  0.2$&$ -0.9$\\
$  200$&$0.0050$&$1.107$&$  4.6$&$  2.8$&$  3.1$&$  0.7$&$  0.9$&$  1.8$&$ -0.5$&$ -1.0$&$  1.4$&$  0.2$&$ -0.3$\\
$  200$&$0.0080$&$0.915$&$  4.5$&$  3.0$&$  3.2$&$  1.1$&$  0.4$&$  0.9$&$  0.4$&$ -0.6$&$  0.4$&$ -0.2$&$ -0.1$\\
$  200$&$0.0130$&$0.860$&$  4.7$&$  3.2$&$  3.3$&$  1.2$&$  0.1$&$  1.1$&$ -0.3$&$  0.8$&$ -0.6$&$  0.4$&$  0.0$\\
$  200$&$0.0200$&$0.677$&$  6.5$&$  3.8$&$  4.5$&$  3.1$&$  0.1$&$  2.9$&$ -1.2$&$  2.6$&$ -0.4$&$  0.1$&$  0.0$\\
$  200$&$0.0320$&$0.558$&$  8.6$&$  4.5$&$  6.1$&$  5.0$&$  0.5$&$  4.2$&$ -2.0$&$  3.5$&$ -1.1$&$  0.3$&$  0.0$\\
$  200$&$0.0500$&$0.506$&$  9.9$&$  5.2$&$  6.8$&$  5.7$&$  0.7$&$  4.9$&$ -1.6$&$  4.5$&$ -0.6$&$ -1.1$&$  0.0$\\
$  200$&$0.0800$&$0.407$&$ 12.4$&$  5.9$&$  8.4$&$  7.4$&$  0.8$&$  6.9$&$ -3.0$&$  6.2$&$  0.4$&$  1.0$&$  0.0$\\ \hline
$  250$&$0.0050$&$1.123$&$  5.3$&$  3.5$&$  3.3$&$  1.1$&$  0.7$&$  2.3$&$  0.6$&$ -1.7$&$  1.3$&$  0.2$&$ -0.5$\\
$  250$&$0.0080$&$1.021$&$  5.3$&$  3.2$&$  3.7$&$  2.0$&$  0.7$&$  2.0$&$ -0.8$&$ -1.6$&$  0.8$&$  0.1$&$  0.0$\\
$  250$&$0.0130$&$0.825$&$  5.7$&$  3.4$&$  4.0$&$  2.6$&$  0.5$&$  2.1$&$  0.6$&$ -2.0$&$  0.3$&$  0.3$&$  0.0$\\
$  250$&$0.0200$&$0.691$&$  5.4$&$  3.5$&$  3.6$&$  1.8$&$  0.5$&$  1.8$&$  0.6$&$ -1.7$&$ -0.2$&$  0.5$&$  0.0$\\
$  250$&$0.0320$&$0.569$&$  6.1$&$  3.8$&$  3.9$&$  2.3$&$  0.7$&$  2.6$&$  1.0$&$ -2.2$&$ -0.9$&$ -0.3$&$  0.0$\\
$  250$&$0.0500$&$0.493$&$  5.7$&$  4.3$&$  3.4$&$  0.8$&$  0.5$&$  1.4$&$ -0.2$&$ -1.1$&$ -0.6$&$  0.6$&$  0.0$\\
$  250$&$0.0800$&$0.407$&$  6.1$&$  4.7$&$  3.5$&$  1.0$&$  0.6$&$  1.6$&$ -0.6$&$ -0.4$&$  0.3$&$  1.4$&$  0.0$\\
$  250$&$0.1300$&$0.311$&$  7.8$&$  5.3$&$  5.2$&$  3.1$&$  2.2$&$  2.6$&$  1.6$&$ -0.9$&$ -1.4$&$ -1.3$&$  0.0$\\
$  250$&$0.2500$&$0.225$&$ 12.1$&$  7.5$&$  8.6$&$  6.9$&$  2.8$&$  4.1$&$  2.4$&$ -1.5$&$ -1.8$&$ -2.3$&$  0.0$\\
$  250$&$0.4000$&$0.138$&$ 11.8$&$  9.3$&$  6.8$&$  4.2$&$  1.4$&$  2.3$&$  0.4$&$ -1.0$&$ -0.4$&$ -2.0$&$  0.0$\\ \hline
$  300$&$0.0050$&$1.152$&$  7.2$&$  5.6$&$  3.9$&$  1.2$&$  0.3$&$  2.3$&$  0.4$&$ -2.2$&$  0.2$&$  0.1$&$ -0.6$\\
$  300$&$0.0080$&$1.026$&$  5.1$&$  3.6$&$  3.3$&$  0.2$&$  1.0$&$  1.5$&$ -0.6$&$ -0.8$&$  1.1$&$  0.1$&$ -0.1$\\
$  300$&$0.0130$&$0.878$&$  5.3$&$  3.8$&$  3.3$&$  0.7$&$  0.8$&$  1.7$&$ -0.3$&$ -1.5$&$  0.7$&$  0.2$&$  0.0$\\
$  300$&$0.0200$&$0.735$&$  5.9$&$  4.0$&$  3.9$&$  2.1$&$  0.3$&$  1.8$&$  0.9$&$ -1.5$&$ -0.1$&$  0.3$&$ -0.1$\\
$  300$&$0.0320$&$0.605$&$  5.8$&$  4.2$&$  3.8$&$  2.0$&$  0.3$&$  1.5$&$  0.7$&$ -1.3$&$ -0.2$&$  0.3$&$  0.0$\\
$  300$&$0.0500$&$0.509$&$  6.8$&$  4.5$&$  4.5$&$  3.1$&$  0.1$&$  2.3$&$  1.0$&$ -2.0$&$  0.2$&$  0.6$&$  0.0$\\
$  300$&$0.0800$&$0.390$&$  6.9$&$  5.2$&$  4.1$&$  2.3$&$  0.3$&$  2.0$&$  0.9$&$ -1.7$&$ -0.3$&$  0.5$&$  0.0$\\
$  300$&$0.1300$&$0.332$&$  8.8$&$  5.4$&$  6.7$&$  5.6$&$  1.1$&$  2.1$&$  1.1$&$ -1.6$&$ -0.7$&$ -0.4$&$  0.0$\\
$  300$&$0.2500$&$0.277$&$ 12.8$&$  6.9$&$  9.4$&$  7.7$&$  3.4$&$  5.3$&$  3.1$&$ -2.6$&$ -2.1$&$ -2.7$&$  0.0$\\
$  300$&$0.4000$&$0.143$&$ 14.2$&$ 10.3$&$  8.9$&$  6.8$&$  1.9$&$  4.2$&$  2.8$&$ -2.6$&$ -0.8$&$ -1.6$&$  0.0$\\ \hline
$  400$&$0.0080$&$1.088$&$  6.1$&$  4.5$&$  3.5$&$  0.2$&$  0.9$&$  2.2$&$  0.3$&$ -1.8$&$  1.3$&$  0.1$&$ -0.2$\\
$  400$&$0.0130$&$0.897$&$  5.6$&$  4.3$&$  3.4$&$  0.3$&$  0.8$&$  1.2$&$  0.7$&$ -0.8$&$  0.6$&$  0.2$&$  0.0$\\
$  400$&$0.0200$&$0.732$&$  5.8$&$  4.5$&$  3.5$&$  1.0$&$  0.4$&$  1.1$&$  0.2$&$ -0.9$&$ -0.5$&$  0.2$&$  0.0$\\
$  400$&$0.0320$&$0.560$&$  6.1$&$  4.8$&$  3.6$&$  1.2$&$  0.4$&$  1.3$&$  0.6$&$ -1.1$&$  0.5$&$  0.1$&$ -0.1$\\
$  400$&$0.0500$&$0.514$&$  6.3$&$  5.0$&$  3.6$&$  1.0$&$  0.3$&$  1.0$&$  0.4$&$  0.6$&$ -0.6$&$  0.4$&$  0.0$\\
$  400$&$0.0800$&$0.429$&$  7.0$&$  5.5$&$  3.8$&$  1.4$&$  0.5$&$  2.1$&$  1.5$&$ -1.2$&$ -0.1$&$  0.8$&$  0.0$\\
$  400$&$0.1300$&$0.352$&$  7.5$&$  5.6$&$  4.4$&$  2.4$&$  1.1$&$  2.2$&$  1.5$&$ -1.4$&$ -0.7$&$ -0.1$&$  0.0$\\
$  400$&$0.2500$&$0.240$&$ 10.6$&$  7.6$&$  6.7$&$  4.5$&$  2.7$&$  3.2$&$  2.1$&$ -0.8$&$ -1.6$&$ -1.6$&$  0.0$\\
$  400$&$0.4000$&$0.143$&$ 13.7$&$ 10.8$&$  7.3$&$  4.1$&$  2.4$&$  4.1$&$  1.9$&$ -1.8$&$ -1.3$&$ -2.8$&$  0.0$\\ \hline
$  500$&$0.0080$&$1.044$&$  9.3$&$  7.8$&$  4.6$&$  1.3$&$  0.5$&$  2.0$&$ -0.8$&$ -1.6$&$  0.9$&$ -0.2$&$ -0.5$\\
$  500$&$0.0130$&$1.003$&$  6.8$&$  5.1$&$  3.9$&$  1.2$&$  1.1$&$  2.3$&$  0.6$&$ -2.1$&$  1.0$&$  0.2$&$ -0.1$\\
$  500$&$0.0200$&$0.765$&$  7.0$&$  5.1$&$  4.2$&$  2.1$&$  1.2$&$  2.3$&$  0.9$&$ -1.9$&$  0.8$&$  0.2$&$  0.0$\\
$  500$&$0.0320$&$0.604$&$  7.0$&$  5.3$&$  3.8$&$  1.5$&$  0.5$&$  2.5$&$  1.3$&$ -2.0$&$  0.2$&$  0.5$&$  0.0$\\
$  500$&$0.0500$&$0.517$&$  6.9$&$  5.6$&$  3.8$&$  1.3$&$  0.6$&$  1.4$&$  0.5$&$ -1.2$&$ -0.2$&$  0.5$&$  0.0$\\
$  500$&$0.0800$&$0.392$&$  9.2$&$  6.4$&$  5.8$&$  4.3$&$  1.3$&$  3.0$&$  2.0$&$ -2.0$&$ -0.8$&$ -0.3$&$  0.0$\\
$  500$&$0.1300$&$0.363$&$  8.7$&$  7.2$&$  4.1$&$  0.4$&$  0.8$&$  2.7$&$  0.5$&$ -1.7$&$ -0.5$&$  2.0$&$  0.0$\\
$  500$&$0.1800$&$0.283$&$ 11.5$&$  8.2$&$  6.6$&$  4.5$&$  2.4$&$  4.7$&$  3.5$&$ -2.1$&$ -1.6$&$ -1.5$&$  0.0$\\
$  500$&$0.2500$&$0.254$&$ 14.2$&$ 10.5$&$  8.4$&$  6.0$&$  2.9$&$  4.5$&$  3.3$&$ -2.4$&$ -1.6$&$ -1.1$&$  0.0$\\
$  500$&$0.4000$&$0.139$&$ 21.6$&$ 15.4$&$ 12.0$&$  8.5$&$  3.9$&$  9.2$&$  7.2$&$ -3.5$&$ -2.3$&$ -3.9$&$  0.0$\\
$  500$&$0.6500$&$0.026$&$ 22.4$&$ 19.6$&$ 10.5$&$  2.8$&$  2.4$&$  3.0$&$ -1.5$&$ -0.8$&$ -2.1$&$ -1.3$&$  0.0$\\ \hline
$  650$&$0.0130$&$0.988$&$  7.3$&$  6.0$&$  4.0$&$  0.5$&$  0.9$&$  1.1$&$ -0.4$&$ -0.3$&$  0.9$&$  0.1$&$ -0.2$\\
$  650$&$0.0200$&$0.791$&$  7.7$&$  6.3$&$  4.2$&$  1.2$&$  0.9$&$  1.5$&$  0.6$&$ -0.9$&$  1.0$&$ -0.3$&$  0.0$\\
$  650$&$0.0320$&$0.684$&$  7.4$&$  6.1$&$  4.0$&$  0.7$&$  1.0$&$  1.6$&$  0.5$&$ -1.4$&$  0.6$&$  0.3$&$  0.0$\\
$  650$&$0.0500$&$0.538$&$  8.3$&$  6.5$&$  4.5$&$  2.2$&$  0.3$&$  2.7$&$  1.9$&$ -1.8$&$  0.2$&$  0.5$&$  0.0$\\
$  650$&$0.0800$&$0.436$&$  9.2$&$  7.1$&$  5.0$&$  2.7$&$  1.2$&$  3.1$&$  2.2$&$ -1.6$&$ -1.2$&$ -0.3$&$  0.0$\\
$  650$&$0.1300$&$0.343$&$ 10.5$&$  8.8$&$  5.4$&$  2.1$&$  2.2$&$  2.3$&$  1.5$&$ -0.6$&$  1.0$&$  1.2$&$  0.0$\\
$  650$&$0.1800$&$0.330$&$ 11.8$&$  9.1$&$  6.8$&$  4.4$&$  2.0$&$  3.2$&$  2.1$&$ -1.6$&$ -1.7$&$ -0.5$&$  0.0$\\
$  650$&$0.2500$&$0.251$&$ 15.9$&$ 11.9$&$  9.3$&$  6.3$&$  3.9$&$  5.0$&$  3.6$&$ -1.1$&$ -2.1$&$ -2.7$&$  0.0$\\
$  650$&$0.4000$&$0.090$&$ 24.9$&$ 22.9$&$  9.3$&$  3.2$&$  1.6$&$  2.4$&$  1.9$&$  0.9$&$  1.2$&$  0.2$&$  0.0$\\ \hline
$  800$&$0.0130$&$0.842$&$ 11.7$&$ 10.2$&$  5.2$&$  0.6$&$  0.2$&$  2.5$&$  1.3$&$ -2.2$&$ -0.3$&$  0.0$&$ -0.2$\\
$  800$&$0.0200$&$0.806$&$  8.8$&$  7.2$&$  4.6$&$  1.3$&$  0.9$&$  1.7$&$  1.1$&$  1.1$&$  0.5$&$  0.0$&$ -0.5$\\
$  800$&$0.0320$&$0.721$&$  8.7$&$  7.1$&$  4.8$&$  1.9$&$  1.2$&$  1.3$&$  0.8$&$ -0.2$&$  1.1$&$  0.2$&$  0.0$\\
$  800$&$0.0500$&$0.587$&$  8.6$&$  7.4$&$  4.4$&$  0.9$&$  0.7$&$  0.8$&$ -0.3$&$  0.1$&$ -0.6$&$  0.3$&$  0.0$\\
$  800$&$0.0800$&$0.518$&$  9.4$&$  7.8$&$  4.9$&$  1.9$&$  0.7$&$  1.6$&$  1.1$&$  0.6$&$ -1.0$&$  0.4$&$  0.0$\\
$  800$&$0.1300$&$0.411$&$ 11.8$&$ 10.0$&$  5.8$&$  2.5$&$  0.6$&$  2.2$&$  1.5$&$ -1.4$&$ -0.5$&$  0.5$&$  0.0$\\
$  800$&$0.1800$&$0.302$&$ 13.4$&$ 11.6$&$  6.2$&$  2.5$&$  1.6$&$  2.7$&$  1.9$&$ -1.4$&$ -0.9$&$  0.8$&$  0.0$\\
$  800$&$0.2500$&$0.212$&$ 16.4$&$ 14.1$&$  7.7$&$  4.3$&$  2.0$&$  2.9$&$  2.0$&$ -0.9$&$ -1.0$&$ -1.5$&$  0.0$\\
$  800$&$0.4000$&$0.117$&$ 24.4$&$ 20.9$&$ 11.6$&$  6.6$&$  3.3$&$  5.1$&$  3.7$&$ -2.2$&$ -2.1$&$ -1.6$&$  0.0$\\
$  800$&$0.6500$&$0.015$&$ 26.5$&$ 21.8$&$ 12.5$&$  9.0$&$  3.9$&$  8.1$&$  6.3$&$ -3.9$&$ -1.6$&$ -3.1$&$  0.0$\\ \hline
 \end{tabular}
 \end{center}
 \caption[RESULT]
 {\sl \label{ncfull} The NC reduced cross section
 $\tilde{\sigma}_{NC}(x,Q^2)$ with statistical
 $(\delta_{sta})$,
 total $(\delta_{tot})$, and
 uncorrelated systematic $(\delta_{unc})$ errors,
 and its contributions from the electron
 energy error ($\delta_{unc}^{E}$), 
 and the hadronic energy error 
 ($\delta_{unc}^{h}$)
 . The effect of the other uncorrelated
 systematic errors is included in $(\delta_{unc})$
 . In addition the correlated systematic 
 $(\delta_{cor})$, and its contributions from a
 positive variation of one standard deviation of the
 electron energy error ($\delta_{cor}^{E^+}$), of
 the polar electron angle error
 ($\delta_{cor}^{\theta^+}$), of the hadronic
 energy error ($\delta_{cor}^{h^+}$), of the error
 due to noise subtraction ($\delta_{cor}^{N^+}$),
 and of the error due to background subtraction
 ($\delta_{cor}^{B^+}$).
 The normalisation uncertainty of 1.8\% is
 not included in the errors.
 The table continues on the next page.}
 \end{table}
 \begin{table}[htb]
 \begin{center}
 \tiny
 \begin{tabular}{|r|c||c|r|r||r|r|r||r|r|r|r|r|r|}
 \hline
 $Q^2$  &$x$ & $\tilde{\sigma}_{NC}$ &
 $\delta_{tot}$ & $\delta_{sta}$ & $\delta_{unc}$ &
 $\delta_{unc}^{E}$ &
 $\delta_{unc}^{h}$&
 $\delta_{cor}$ &
 $\delta_{cor}^{E^+}$ &
 $\delta_{cor}^{\theta^+}$&
 $\delta_{cor}^{h^+}$&
 $\delta_{cor}^{N^+}$&
 $\delta_{cor}^{B^+}$ \\
 $(\rm GeV^2)$ & & &
 $(\%)$ &$(\%)$ &$(\%)$ &$(\%)$ &$(\%)$ &$(\%)$ &
 $(\%)$ &$(\%)$ &$(\%)$ &$(\%)$ &$(\%)$ 
 \\\hline\hline
$ 1000$&$0.0130$&$0.773$&$ 13.5$&$ 11.5$&$  6.5$&$  2.6$&$  1.8$&$  2.3$&$ -0.7$&$ -0.8$&$ -1.5$&$ -0.2$&$ -1.5$\\
$ 1000$&$0.0200$&$0.787$&$  9.2$&$  7.9$&$  4.3$&$  1.5$&$  1.3$&$  1.9$&$  1.0$&$ -0.9$&$  1.3$&$  0.1$&$ -0.2$\\
$ 1000$&$0.0320$&$0.572$&$ 10.0$&$  9.0$&$  4.2$&$  0.8$&$  1.2$&$  1.3$&$  0.4$&$ -0.8$&$  0.9$&$  0.2$&$ -0.1$\\
$ 1000$&$0.0500$&$0.577$&$  9.5$&$  8.4$&$  4.3$&$  0.5$&$  1.4$&$  1.3$&$  0.1$&$ -1.0$&$  0.5$&$  0.5$&$  0.0$\\
$ 1000$&$0.0800$&$0.450$&$ 10.8$&$  9.3$&$  4.9$&$  2.4$&$  1.2$&$  2.8$&$  2.3$&$ -1.1$&$ -1.0$&$ -0.3$&$  0.0$\\
$ 1000$&$0.1300$&$0.491$&$ 11.6$&$ 10.3$&$  5.0$&$  1.1$&$  0.2$&$  1.9$&$ -1.3$&$ -1.4$&$  0.4$&$  0.3$&$  0.0$\\
$ 1000$&$0.1800$&$0.249$&$ 14.6$&$ 13.5$&$  5.4$&$  1.9$&$  1.1$&$  1.6$&$  1.2$&$ -0.8$&$ -0.3$&$ -0.4$&$  0.0$\\
$ 1000$&$0.2500$&$0.311$&$ 15.9$&$ 13.0$&$  8.2$&$  5.3$&$  2.8$&$  4.2$&$  2.7$&$ -2.2$&$ -2.0$&$  1.0$&$  0.0$\\
$ 1000$&$0.4000$&$0.122$&$ 26.9$&$ 22.9$&$ 13.3$&$  9.6$&$  4.2$&$  4.4$&$  2.9$&$  0.8$&$ -1.8$&$ -2.7$&$  0.0$\\ \hline
$ 1200$&$0.0200$&$0.839$&$ 10.0$&$  9.1$&$  3.6$&$  0.7$&$  0.8$&$  1.8$&$ -0.6$&$ -0.9$&$  1.3$&$  0.0$&$ -0.6$\\
$ 1200$&$0.0320$&$0.719$&$  9.9$&$  9.2$&$  3.5$&$  0.4$&$  1.4$&$  1.1$&$  0.3$&$ -0.6$&$  0.9$&$  0.1$&$  0.0$\\
$ 1200$&$0.0500$&$0.645$&$  9.9$&$  9.3$&$  3.5$&$  1.1$&$  0.7$&$  0.7$&$ -0.1$&$ -0.3$&$  0.6$&$  0.2$&$  0.0$\\
$ 1200$&$0.0800$&$0.415$&$ 11.2$&$ 10.7$&$  3.4$&$  0.8$&$  0.6$&$  0.6$&$  0.4$&$ -0.2$&$  0.2$&$  0.3$&$  0.0$\\
$ 1200$&$0.1300$&$0.384$&$ 13.4$&$ 12.6$&$  4.3$&$  2.2$&$  0.6$&$  1.2$&$  1.0$&$  0.3$&$ -0.3$&$ -0.5$&$  0.0$\\
$ 1200$&$0.1800$&$0.341$&$ 14.6$&$ 13.6$&$  4.7$&$  2.5$&$  0.6$&$  2.3$&$  1.9$&$ -0.7$&$ -0.9$&$  0.6$&$  0.0$\\
$ 1200$&$0.2500$&$0.251$&$ 17.3$&$ 15.8$&$  6.4$&$  4.5$&$  1.8$&$  2.9$&$  2.6$&$ -1.1$&$ -0.5$&$  0.2$&$  0.0$\\
$ 1200$&$0.4000$&$0.110$&$ 27.7$&$ 25.0$&$ 10.2$&$  7.7$&$  4.0$&$  6.2$&$  4.2$&$ -1.1$&$ -3.2$&$ -3.2$&$  0.0$\\ \hline
$ 1500$&$0.0200$&$0.860$&$ 13.5$&$ 12.4$&$  5.2$&$  0.1$&$  1.3$&$  1.4$&$ -0.8$&$ -0.4$&$ -0.4$&$ -0.2$&$ -1.0$\\
$ 1500$&$0.0320$&$0.704$&$ 11.4$&$ 10.4$&$  4.3$&$  1.5$&$  2.2$&$  1.8$&$ -0.6$&$ -0.6$&$  1.6$&$  0.2$&$ -0.1$\\
$ 1500$&$0.0500$&$0.515$&$ 12.2$&$ 11.7$&$  3.5$&$  0.5$&$  0.5$&$  0.8$&$  0.4$&$ -0.6$&$  0.4$&$  0.1$&$  0.0$\\
$ 1500$&$0.0800$&$0.512$&$ 11.7$&$ 11.0$&$  3.8$&$  1.4$&$  0.9$&$  1.4$&$  0.7$&$ -1.1$&$  0.4$&$  0.2$&$  0.0$\\
$ 1500$&$0.1300$&$0.390$&$ 14.8$&$ 13.9$&$  4.7$&$  2.8$&$  0.7$&$  1.8$&$  1.6$&$ -0.7$&$  0.4$&$ -0.2$&$  0.0$\\
$ 1500$&$0.1800$&$0.260$&$ 19.1$&$ 18.6$&$  4.2$&$  0.3$&$  1.0$&$  1.1$&$ -0.9$&$ -0.4$&$  0.3$&$  0.5$&$  0.0$\\
$ 1500$&$0.2500$&$0.197$&$ 21.1$&$ 19.6$&$  6.9$&$  4.5$&$  3.1$&$  3.4$&$  2.9$&$ -0.4$&$ -1.7$&$ -0.3$&$  0.0$\\
$ 1500$&$0.4000$&$0.145$&$ 27.4$&$ 24.3$&$ 10.6$&$  7.8$&$  4.1$&$  7.3$&$  6.0$&$ -1.7$&$ -3.2$&$ -2.1$&$  0.0$\\
$ 1500$&$0.6500$&$0.014$&$ 38.9$&$ 35.4$&$ 14.5$&$ 10.7$&$  6.4$&$  7.2$&$  5.8$&$ -0.3$&$ -2.7$&$ -3.3$&$  0.0$\\ \hline
$ 2000$&$0.0320$&$0.796$&$ 11.9$&$ 11.1$&$  3.9$&$  0.9$&$  0.9$&$  2.0$&$  0.3$&$ -1.3$&$  1.5$&$  0.0$&$ -0.5$\\
$ 2000$&$0.0500$&$0.599$&$ 13.9$&$ 13.0$&$  4.4$&$  1.7$&$  1.8$&$  2.3$&$  1.3$&$ -1.4$&$  1.2$&$  0.2$&$  0.0$\\
$ 2000$&$0.0800$&$0.582$&$ 13.0$&$ 12.3$&$  4.0$&$  1.5$&$  0.4$&$  1.5$&$  1.3$&$ -0.8$&$  0.3$&$  0.1$&$  0.0$\\
$ 2000$&$0.1300$&$0.224$&$ 20.6$&$ 20.0$&$  4.4$&$  1.7$&$  0.9$&$  1.5$&$ -1.0$&$ -0.9$&$ -0.6$&$ -0.4$&$  0.0$\\
$ 2000$&$0.1800$&$0.249$&$ 22.7$&$ 21.9$&$  5.7$&$  3.5$&$  1.2$&$  2.6$&$  2.2$&$  0.8$&$  0.9$&$  0.8$&$  0.0$\\
$ 2000$&$0.2500$&$0.197$&$ 23.4$&$ 22.4$&$  6.1$&$  3.8$&$  1.2$&$  2.9$&$  2.6$&$ -1.2$&$ -0.2$&$ -0.6$&$  0.0$\\
$ 2000$&$0.4000$&$0.108$&$ 29.5$&$ 27.7$&$  9.2$&$  6.6$&$  3.1$&$  4.2$&$  3.2$&$ -0.7$&$ -2.0$&$ -1.8$&$  0.0$\\ \hline
$ 3000$&$0.0500$&$0.606$&$ 12.4$&$ 10.6$&$  5.8$&$  2.0$&$  2.2$&$  2.7$&$  0.6$&$  0.2$&$  1.9$&$  0.2$&$ -1.7$\\
$ 3000$&$0.0800$&$0.556$&$ 11.8$&$ 10.9$&$  4.4$&$  2.0$&$  1.6$&$  0.9$&$ -0.3$&$ -0.2$&$  0.8$&$  0.1$&$  0.0$\\
$ 3000$&$0.1300$&$0.464$&$ 13.0$&$ 12.4$&$  3.9$&$  0.9$&$  0.6$&$  1.0$&$ -0.8$&$ -0.4$&$ -0.5$&$  0.2$&$  0.0$\\
$ 3000$&$0.1800$&$0.347$&$ 16.1$&$ 15.3$&$  4.6$&$  2.3$&$  0.1$&$  2.2$&$  1.9$&$ -1.0$&$  0.3$&$  0.2$&$  0.0$\\
$ 3000$&$0.2500$&$0.255$&$ 19.1$&$ 17.8$&$  6.4$&$  4.3$&$  1.7$&$  2.9$&$  2.7$&$  0.3$&$ -0.6$&$  0.6$&$  0.0$\\
$ 3000$&$0.4000$&$0.128$&$ 25.5$&$ 23.0$&$  9.7$&$  7.4$&$  3.4$&$  5.0$&$  3.7$&$  0.2$&$ -2.8$&$ -1.8$&$  0.0$\\ \hline
$ 5000$&$0.0800$&$0.707$&$ 11.7$&$ 10.6$&$  4.4$&$  1.4$&$  2.1$&$  2.0$&$  0.4$&$ -0.7$&$  1.8$&$  0.1$&$ -0.3$\\
$ 5000$&$0.1300$&$0.536$&$ 14.2$&$ 13.1$&$  5.2$&$  2.9$&$  1.6$&$  1.1$&$  0.3$&$  0.5$&$  0.9$&$  0.1$&$  0.0$\\
$ 5000$&$0.1800$&$0.442$&$ 14.9$&$ 14.0$&$  5.1$&$  3.0$&$  0.3$&$  0.9$&$ -0.7$&$ -0.5$&$  0.3$&$ -0.3$&$  0.0$\\
$ 5000$&$0.2500$&$0.361$&$ 20.3$&$ 17.4$&$  9.6$&$  8.0$&$  2.2$&$  4.3$&$  3.7$&$  1.5$&$ -1.7$&$  0.1$&$  0.0$\\
$ 5000$&$0.4000$&$0.091$&$ 33.5$&$ 31.6$&$ 10.9$&$  8.9$&$  1.2$&$  1.9$&$  1.5$&$  0.8$&$  0.8$&$ -0.3$&$  0.0$\\
$ 5000$&$0.6500$&$0.010$&$ 45.1$&$ 41.0$&$ 17.7$&$ 15.7$&$  5.0$&$  6.5$&$  5.7$&$  1.3$&$ -2.6$&$ -1.2$&$  0.0$\\ \hline
$ 8000$&$0.1300$&$0.722$&$ 17.2$&$ 16.0$&$  6.2$&$  3.4$&$  2.1$&$  1.9$&$  1.4$&$ -0.2$&$  1.3$&$  0.3$&$ -0.3$\\
$ 8000$&$0.1800$&$0.386$&$ 21.2$&$ 20.4$&$  5.3$&$  1.5$&$  1.7$&$  2.3$&$ -1.0$&$ -1.6$&$  1.4$&$  0.1$&$  0.0$\\
$ 8000$&$0.2500$&$0.295$&$ 23.3$&$ 21.8$&$  7.4$&$  5.2$&$  0.8$&$  3.5$&$  2.8$&$  2.0$&$ -0.7$&$ -0.2$&$  0.0$\\
$ 8000$&$0.4000$&$0.197$&$ 32.4$&$ 27.7$&$ 16.7$&$ 14.9$&$  1.0$&$  2.1$&$  1.3$&$  1.5$&$  0.7$&$  0.3$&$  0.0$\\ \hline
$12000$&$0.1800$&$0.471$&$ 28.8$&$ 27.8$&$  7.1$&$  1.9$&$  0.7$&$  2.7$&$  1.6$&$ -1.8$&$ -1.2$&$ -0.3$&$ -0.1$\\
$12000$&$0.2500$&$0.298$&$ 30.2$&$ 28.9$&$  8.2$&$  5.2$&$  2.3$&$  2.7$&$ -0.7$&$ -1.2$&$  2.3$&$  0.3$&$  0.0$\\
$12000$&$0.4000$&$0.083$&$ 53.7$&$ 50.0$&$ 18.1$&$ 16.2$&$  0.2$&$  7.6$&$  6.3$&$  4.3$&$ -0.7$&$  0.0$&$  0.0$\\ \hline
$20000$&$0.2500$&$0.349$&$ 52.2$&$ 51.1$&$ 10.6$&$  2.5$&$  0.5$&$  2.2$&$  0.7$&$  1.8$&$  0.7$&$  0.0$&$ -0.6$\\
$20000$&$0.4000$&$0.182$&$ 46.7$&$ 44.7$&$ 13.1$&$  8.8$&$  1.9$&$  2.3$&$  1.4$&$ -1.6$&$  0.9$&$  0.0$&$  0.0$\\
$20000$&$0.6500$&$0.014$&$ 79.8$&$ 70.7$&$ 34.4$&$ 31.7$&$  2.7$&$ 13.4$&$  9.9$&$  8.9$&$ -1.0$&$  0.0$&$  0.0$\\ \hline
$30000$&$0.4000$&$0.268$&$ 72.9$&$ 70.7$&$ 16.9$&$  4.8$&$  0.7$&$  4.7$&$ -3.5$&$ -2.4$&$ -2.0$&$  0.0$&$  0.0$\\ \hline
 \end{tabular}
 \end{center}
 \end{table}

\begin{table}[htb]
  \begin{center}
    \tiny
    \begin{tabular}{|*{2}{c|}|l|*{2}{c|}|*{2}{c|}|*{1}{c|}*{4}{r|}}
      \hline 
      $Q^2$  & $x$ & \multicolumn{1}{c|}{${\rm d} ^2 \sigma_{CC} /{\rm d} x {\rm d} Q^2$} & 
      $\delta_{tot}$ & $\delta_{sta}$ & $\delta_{unc}$ & $\delta^{h}_{unc}$ &
      $\delta_{cor}$ & $\delta^{V^{+}}_{cor}$ & 
      $\delta^{h^{+}}_{cor}$ & $\delta^{N^{+}}_{cor}$ & 
      $\delta^{B^{+}}_{cor}$ \vtab \\
      \tiny $(\rm GeV^2)$ & & \multicolumn{1}{c|}{\tiny $(\rm{pb/GeV^2})$} & 
      \tiny $(\%)$ &\tiny $(\%)$ &
      \tiny $(\%)$ &\tiny $(\%)$ &
      \tiny $(\%)$ &\tiny $(\%)$ &
      \tiny $(\%)$ &
      \tiny $(\%)$ &\tiny  $(\%)$ \\[1mm]
 \hline

$  300$ & $0.013$ & \phantom{C} $0.458\cdot 10^{ 0}$ & $57.6$ & $55.4$ &
$ 7.4$ & $ 2.2$ & $13.9$ & $ 12.1$ & $ -1.6$ & $  1.0$ & $ -6.5$ \\
$  300$ & $0.032$ & \phantom{C} $0.399\cdot 10^{ 0}$ & $27.3$ & $24.5$ &
$ 5.2$ & $ 2.3$ & $10.9$ & $ 10.0$ & $ -2.0$ & $  1.8$ & $ -3.3$ \\
$  300$ & $0.080$ & \phantom{C} $0.690\cdot 10^{-1}$ & $42.3$ & $40.7$ &
$ 8.7$ & $ 2.3$ & $ 7.6$ & $  4.7$ & $ -1.7$ & $  1.4$ & $ -5.6$ \\
\hline
$  500$ & $0.013$ & \phantom{C} $0.433\cdot 10^{ 0}$ & $39.9$ & $37.6$ &
$ 5.9$ & $ 2.8$ & $12.0$ & $ 11.6$ & $ -1.7$ & $  0.5$ & $ -2.3$ \\
$  500$ & $0.032$ & \phantom{C} $0.285\cdot 10^{ 0}$ & $21.0$ & $19.6$ &
$ 5.0$ & $ 3.0$ & $ 6.0$ & $  3.7$ & $ -1.8$ & $  0.4$ & $ -4.3$ \\
$  500$ & $0.080$ & \phantom{C} $0.790\cdot 10^{-1}$ & $22.4$ & $21.8$ &
$ 4.3$ & $ 3.0$ & $ 2.8$ & $  1.1$ & $ -1.4$ & $  2.1$ & $ -0.4$ \\
$  500$ & $0.130$ & \phantom{C} $0.551\cdot 10^{-1}$ & $29.9$ & $29.0$ &
$ 6.5$ & $ 3.0$ & $ 2.7$ & $  0.2$ & $ -1.3$ & $  2.2$ & $ -0.7$ \\
\hline
$ 1000$ & $0.032$ & \phantom{C} $0.186\cdot 10^{ 0}$ & $18.2$ & $17.5$ &
$ 4.4$ & $ 1.8$ & $ 2.2$ & $  1.5$ & $ -1.1$ & $  1.0$ & $ -0.6$ \\
$ 1000$ & $0.080$ & \phantom{C} $0.556\cdot 10^{-1}$ & $18.4$ & $17.9$ &
$ 4.0$ & $ 1.8$ & $ 1.6$ & $  0.5$ & $ -1.0$ & $  1.2$ & $  0.0$ \\
$ 1000$ & $0.130$ & \phantom{C} $0.310\cdot 10^{-1}$ & $24.5$ & $24.0$ &
$ 4.4$ & $ 1.8$ & $ 1.4$ & $  0.1$ & $ -1.0$ & $  1.0$ & $  0.0$ \\
$ 1000$ & $0.250$ & \phantom{C} $0.139\cdot 10^{-1}$ & $39.1$ & $37.6$ &
$10.5$ & $ 1.7$ & $ 1.4$ & $  0.0$ & $ -0.2$ & $ -1.4$ & $  0.0$ \\
\hline
$ 2000$ & $0.032$ & \phantom{C} $0.132\cdot 10^{ 0}$ & $16.2$ & $15.5$ &
$ 4.3$ & $ 2.0$ & $ 2.4$ & $  2.0$ & $ -1.3$ & $ -0.3$ & $ -0.1$ \\
$ 2000$ & $0.080$ & \phantom{C} $0.571\cdot 10^{-1}$ & $13.6$ & $13.0$ &
$ 3.8$ & $ 2.0$ & $ 1.1$ & $  0.6$ & $ -0.5$ & $  0.7$ & $ -0.1$ \\
$ 2000$ & $0.130$ & \phantom{C} $0.197\cdot 10^{-1}$ & $21.7$ & $21.2$ &
$ 4.4$ & $ 2.0$ & $ 1.1$ & $  0.1$ & $ -0.8$ & $  0.7$ & $ -0.2$ \\
$ 2000$ & $0.250$ & \phantom{C} $0.855\cdot 10^{-2}$ & $26.4$ & $25.6$ &
$ 6.5$ & $ 0.7$ & $ 0.8$ & $  0.0$ & $ -0.7$ & $ -0.3$ & $  0.0$ \\
\hline
$ 3000$ & $0.080$ & \phantom{C} $0.324\cdot 10^{-1}$ & $14.8$ & $14.0$ &
$ 4.6$ & $ 0.7$ & $ 1.4$ & $  0.7$ & $ -1.2$ & $  0.3$ & $  0.0$ \\
$ 3000$ & $0.130$ & \phantom{C} $0.250\cdot 10^{-1}$ & $15.2$ & $14.0$ &
$ 6.1$ & $ 0.7$ & $ 0.7$ & $  0.2$ & $  0.4$ & $  0.6$ & $  0.0$ \\
$ 3000$ & $0.250$ & \phantom{C} $0.749\cdot 10^{-2}$ & $20.1$ & $18.9$ &
$ 6.8$ & $ 0.8$ & $ 1.5$ & $  0.0$ & $ -0.7$ & $  1.3$ & $  0.0$ \\
$ 3000$ & $0.400$ & \phantom{C} $0.251\cdot 10^{-2}$ & $40.3$ & $35.2$ &
$17.1$ & $ 4.4$ & $ 9.7$ & $  0.0$ & $  1.7$ & $ -9.5$ & $  0.0$ \\
\hline
$ 5000$ & $0.080$ & \phantom{C} $0.213\cdot 10^{-1}$ & $19.2$ & $17.9$ &
$ 6.5$ & $ 4.4$ & $ 1.6$ & $  1.0$ & $  0.5$ & $  1.1$ & $  0.0$ \\
$ 5000$ & $0.130$ & \phantom{C} $0.108\cdot 10^{-1}$ & $18.2$ & $16.8$ &
$ 7.0$ & $ 4.4$ & $ 0.8$ & $  0.1$ & $  0.7$ & $  0.4$ & $  0.0$ \\
$ 5000$ & $0.250$ & \phantom{C} $0.550\cdot 10^{-2}$ & $16.9$ & $16.3$ &
$ 4.3$ & $ 0.7$ & $ 1.0$ & $  0.0$ & $ -0.1$ & $  1.0$ & $  0.0$ \\
$ 5000$ & $0.400$ & \phantom{C} $0.123\cdot 10^{-2}$ & $35.6$ & $33.1$ &
$12.4$ & $ 6.2$ & $ 4.3$ & $  0.0$ & $  3.6$ & $ -2.4$ & $  0.0$ \\
\hline
$ 8000$ & $0.130$ & \phantom{C} $0.722\cdot 10^{-2}$ & $21.1$ & $18.9$ &
$ 8.9$ & $ 6.2$ & $ 2.6$ & $  0.7$ & $  2.1$ & $  1.3$ & $  0.0$ \\
$ 8000$ & $0.250$ & \phantom{C} $0.342\cdot 10^{-2}$ & $17.4$ & $16.3$ &
$ 6.0$ & $ 2.5$ & $ 1.5$ & $  0.1$ & $  1.5$ & $  0.1$ & $  0.0$ \\
$ 8000$ & $0.400$ & \phantom{C} $0.946\cdot 10^{-3}$ & $30.4$ & $28.6$ &
$ 9.4$ & $ 6.7$ & $ 4.1$ & $  0.0$ & $  4.1$ & $  0.4$ & $  0.0$ \\
\hline
$15000$ & $0.250$ & \phantom{C} $0.139\cdot 10^{-2}$ & $27.3$ & $22.1$ &
$15.4$ & $ 7.1$ & $ 4.4$ & $  0.2$ & $  4.2$ & $  1.3$ & $  0.0$ \\
$15000$ & $0.400$ & \phantom{C} $0.419\cdot 10^{-3}$ & $29.5$ & $27.5$ &
$ 9.3$ & $ 7.9$ & $ 5.2$ & $  0.0$ & $  5.1$ & $  1.0$ & $  0.0$ \\

\hline
\end{tabular}
\caption{\sl  The CC double differential cross 
  section ${\rm d} ^2\sigma_{CC}/{\rm d} x {\rm d}Q^2$ with total
  error ($\delta_{tot}$), statistical error ($\delta_{sta}$),
  uncorrelated systematic error ($\delta_{unc}$), and its
  contributions from the hadronic energy error ($\delta^{h}_{unc}$).
  The effect of the other uncorrelated errors is included in
  $\delta_{unc}$.  In addition the correlated systematic error
  ($\delta_{cor}$), and its contributions from a positive variation of
  one standard deviation of the error coming from the
  anti-photoproduction cut ($\delta^{V^{+}}_{cor}$), of the hadronic
  energy error ($\delta^{h^{+}}_{cor}$), of the noise contribution
  ($\delta^{N^{+}}_{cor}$) and of the error due to background
  subtraction ($\delta^{B^{+}}_{cor}$).  The normalisation uncertainty
  of 1.8\% is not included in the errors.
\label{ccfull}
}
\end{center}
\end{table}

 \begin{table}[htb]
 \begin{center}
 \begin{tabular}{|r|c|c|c|c|c|}
 \hline
 $Q^2$  &$x$ & $x\Fz$ &
 $\delta_{sta}$ & $\delta_{sys}$ & $\delta_{tot}$ \\
 $(\rm GeV^2)$ & & & & & 
 \\\hline\hline
$ 1500$&$0.020$&$ 0.0529$&$ 0.0416$&$ 0.0284$&$ 0.0503$\\
$ 1500$&$0.032$&$ 0.0972$&$ 0.0433$&$ 0.0353$&$ 0.0558$\\
$ 1500$&$0.050$&$ 0.1028$&$ 0.0597$&$ 0.0442$&$ 0.0743$\\
$ 1500$&$0.080$&$ 0.0926$&$ 0.0840$&$ 0.0617$&$ 0.1042$\\ \hline
$ 5000$&$0.080$&$ 0.1778$&$ 0.0360$&$ 0.0238$&$ 0.0431$\\
$ 5000$&$0.130$&$ 0.2084$&$ 0.0506$&$ 0.0346$&$ 0.0613$\\
$ 5000$&$0.180$&$ 0.1667$&$ 0.0619$&$ 0.0377$&$ 0.0725$\\
$ 5000$&$0.250$&$ 0.1253$&$ 0.0787$&$ 0.0619$&$ 0.1002$\\
$ 5000$&$0.400$&$ 0.0374$&$ 0.0823$&$ 0.0661$&$ 0.1055$\\
$ 5000$&$0.650$&$-0.0120$&$ 0.0285$&$ 0.0152$&$ 0.0323$\\ \hline
$12000$&$0.180$&$ 0.1922$&$ 0.0851$&$ 0.0225$&$ 0.0880$\\
$12000$&$0.250$&$ 0.1352$&$ 0.0687$&$ 0.0234$&$ 0.0726$\\
$12000$&$0.400$&$-0.0057$&$ 0.0645$&$ 0.0338$&$ 0.0728$\\
$12000$&$0.650$&$-0.0141$&$ 0.0281$&$ 0.0163$&$ 0.0325$\\ \hline
 \end{tabular}
 \end{center}
 \caption[RESULT]
 {\sl \label{xf3tab} The structure function $x\Fz$~
 with statistical $(\delta_{sta})$,
 systematic $(\delta_{sta})$, and total 
 $(\delta_{tot})$ absolute errors. The 
 luminosity uncertainties of
 the $e^+p$ and $e^-p$ data sets
 are included in the systematic error.}

 \end{table}

\end{document}